\documentclass[aps, pre, a4paper, floatfix,  twocolumn, longbibliography, nofootinbib]{revtex4-1}

\usepackage[utf8]{inputenc}
\usepackage{epsfig}
\usepackage[T1]{fontenc}
\usepackage[english]{babel}
\usepackage[table]{xcolor}
\usepackage{t1enc}
\usepackage{graphicx}
\usepackage{amssymb}
\usepackage{amsmath}
\usepackage{relsize}
\usepackage[percent]{overpic}
\usepackage{bm}
\usepackage{hyperref}

\usepackage{siunitx}
\usepackage{accents}

\usepackage[normalem]{ulem}

\usepackage{pgf}
\usepackage{tikz}

\newcommand{\avg}[1]{{\left<#1\right>}}

\newcommand{\dd}{\mathrm{d}}
\newcommand{\ee}{\mathrm{e}}

\usepackage{mathtools}
\def\multiset#1#2{\ensuremath{\left(\kern-.3em\left(\genfrac{}{}{0pt}{}{#1}{#2}\right)\kern-.3em\right)}}

\usepackage{amsmath}

\newcommand{\A}{\bm{A}}
\newcommand{\G}{\bm{G}}
\newcommand{\bb}{\bm{b}}
\newcommand{\cc}{\bm{c}}
\newcommand{\x}{{\bm{x}}}
\newcommand{\y}{{\bm{y}}}
\newcommand{\z}{{\bm{z}}}
\newcommand{\w}{{\bm{w}}}

\usepackage{verbatim}
\usepackage{overpic}
\usepackage{booktabs}
\usepackage{placeins}

\usepackage{siunitx}
\usepackage{multirow}

\begin{document}

\title{Revealing consensus and dissensus between network partitions}

\author{Tiago P. Peixoto}
\email{peixotot@ceu.edu}
\affiliation{Department of Network and Data Science, Central European University, H-1051 Budapest, Hungary}
\affiliation{ISI Foundation, Via Chisola 5, 10126 Torino, Italy}
\affiliation{Department of Mathematical Sciences, University of Bath, Claverton Down, Bath BA2
  7AY, United Kingdom}

\begin{abstract}
  Community detection methods attempt to divide a network into groups of
  nodes that share similar properties, thus revealing its large-scale
  structure. A major challenge when employing such methods is that they
  are often degenerate, typically yielding a complex landscape of
  competing answers. As an attempt to extract understanding from a
  population of alternative solutions, many methods exist to establish a
  consensus among them in the form of a single partition ``point
  estimate'' that summarizes the whole distribution. Here we show that
  it is in general not possible to obtain a consistent answer from such
  point estimates when the underlying distribution is too
  heterogeneous. As an alternative, we provide a comprehensive set of
  methods designed to characterize and summarize complex populations of
  partitions in a manner that captures not only the existing consensus,
  but also the dissensus between elements of the population. Our
  approach is able to model mixed populations of partitions where
  multiple consensuses can coexist, representing different competing
  hypotheses for the network structure. We also show how our methods can
  be used to compare pairs of partitions, how they can be generalized to
  hierarchical divisions, and be used to perform statistical model
  selection between competing hypotheses.
\end{abstract}

\maketitle

\section{Introduction}

One of the most important tools in network analysis is the algorithmic
division of an unannotated network into groups of similar nodes --- a
task broadly known as network clustering or community
detection~\cite{fortunato_community_2010}. Such divisions allow
researchers to provide a summary of the large-scale structure of a
network, and in this way obtain fundamental insight about its function
and underlying mechanism of formation. Within this broad umbrella, many
community detection methods have been developed, based on different
mathematical definitions of the overall
task~\cite{fortunato_community_2016}. What most methods share in common
is that they are based on some objective function defined over all
possible partitions of the network, which if optimized yields the most
adequate partition for that particular network. Another universal
property of community detection methods is that when they are applied to
empirical networks, they exhibit at least some degree of degeneracy, in
that even if there exists a single partition with the largest score
among all others, there is usually an abundance of other solutions that
possess a very similar score, making a strict optimization among them
somewhat arbitrary~\cite{good_performance_2010}. This issue is
compounded with the fact that instances of the community detection
problem are generically computationally intractable, such that no known
algorithm that guarantees the correct solution can perform substantially
better than an exhaustive search over all
answers~\cite{brandes_modularity-np-completeness_2006,
decelle_asymptotic_2011}, which is not feasible for networks with more
than very few nodes. As a consequence, most available methods rely on
stochastic heuristics that give only approximations of the optimum, and
end up being specially susceptible to the degenerate landscape, yielding
different answers whenever they are employed.

In response to this inherent degeneracy, many authors have emphasized
the need to collectively analyse many outputs of any given community
detection method, not only the best scoring
result~\cite{guimera_missing_2009, clauset_hierarchical_2008,
calatayud_exploring_2019, riolo_consistency_2020}. In this direction,
one particularly interesting proposition is to recover the task of
detecting a single partition, but doing so in a manner that incorporates
the consensus over many different
alternatives~\cite{strehl_cluster_2002, topchy_clustering_2005,
clauset_hierarchical_2008, goder_consensus_2008,
lancichinetti_consensus_2012, zhang_scalable_2014,
riolo_consistency_2020, tandon_fast_2019}. If most results are aligned
with the same general solution, the consensus among them allow us in
fact to profit from the degeneracy, since small distortions due to
irrelevant details or statistical fluctuations are averaged out, leading
to a more robust answer than any of the individual solutions. However,
consensus clustering cannot provide a full answer to the community
detection problem. This is because any kind of approach based on
\emph{point estimates} possesses an Achilles' heel in situations where
the competing answers do not all point in a cohesive direction, and
instead amount to incompatible results. A consensus between diverging
answers is inconsistent in the same manner as the mean of a bimodal
distribution is not a meaningful representation of the corresponding
population. Therefore, extracting understanding from community detection
methods requires more than simply finding a consensus, as we need also
to characterize the \emph{dissensus} among the competing partitions. In
fact, we need robust methods that give us a complete picture of the
entire population of partitions.

Some authors have previously considered the problem of fully
characterizing the landscape of possible partitions. Good et
al~\cite{good_performance_2010} have used nonlinear dimensionality
reduction to project the space of partitions in two dimensions, thereby
revealing degeneracies. Closer to what is proposed in this work,
Calatayud et al~\cite{calatayud_exploring_2019} have used an \emph{ad
hoc} algorithm to cluster partitions, in order to determine how many
samples are necessary to better characterize a distribution. Although
these previous works effectively demonstrate the role of partition
heterogeneity in empirically relevant situations, the approaches so far
developed are implemented outside of a well-defined theoretical
framework, and rely on many seemingly arbitrary choices, such as
projection dimension, similarity function used, cluster forming
criterion, etc. Because of this, it is difficult to interpret in simple
terms the structures found by those methods, and also to evaluate if
they are meaningful and statistically significant, or are merely
artifacts of the provisional choices made.

In this work we develop a round set of methods to comprehensively
characterize a population of network partitions, in a manner that
reveals both the consensus and dissensus between them. Our methods start
from the formulation of interpretable probabilistic generative models
for arbitrary collections of partitions that are based on explicit
definitions of the notion of unique group labelings and clusters of
partitions. From these models, we are able to derive principled Bayesian
inference algorithms that are efficient and effective at characterizing
heterogeneous sets of partitions, according to their statistical
significance. Importantly, our methods are nonparametric, and do not
require \emph{a priori} choices to be made, such as distance thresholds
or even the number of existing clusters, with the latter being uncovered
by our method from the data alone. Our method also bypasses
dimensionality reduction~\cite{maaten_visualizing_2008,
mcinnes_umap_2018}, as required by some data clustering techniques, and
operates directly on a collection of partitions. Since it is grounded in
a broader statistical framework, our method also allows potential
generalizations, and principled comparison with alternative modelling
assumptions.

We approach our characterization task by first providing a solution to
the community label identification problem, which allows us to
unambiguously identify groups of nodes between partitions even when
their node compositions are not identical. This allows us to perform the
basic (but until now not fully solved) task of computing marginal
distributions of group memberships for each node in the network, and
also leads naturally to a way of comparing partitions based on the
maximum overlap distance, which has a series of useful properties that
we demonstrate. Our method yields a simple way to characterize the
consensus between a set of partitions, acting in a way analogous to a
maximum a posteriori estimation of a categorical distribution. We
highlight also the pitfalls of consensus estimation in community
detection, which fails when the ensemble of solutions is
heterogeneous. Finally, we provide a more powerful alternative,
consisting of the generalization of our method to the situation where
multiple consensuses are possible, such that groups of partitions can
align in different directions. The identification of these partitions
``modes'' yields a compact and understandable description of the
heterogeneous landscape of community detection results, allowing us to
assess their consistency and weigh the alternative explanations they
offer to the network data.

This work is divided as follows. We begin in Sec.~\ref{sec:label} with a
description of the label identification problem, which serves as a
motivation for our approach on consensus clustering developed in
Sec.~\ref{sec:random-label}, based on the inference of what we call the
random label model. In Sec.~\ref{sec:consensus} we discuss how we can
extract consensus from network partitions via ``point estimates,'' and
how this leads to inconsistencies in situations when the different
partitions disagree. We then show how we can find both consensus and
dissensus in Sec.~\ref{sec:modes}, by generalizing the random label
model, thus obtaining a comprehensive description of multimodal
populations of partitions, including how partitions may agree and
disagree with each other. In Sec.~\ref{sec:hierarchical} we show how our
ideas can be easily generalized to ensembles of hierarchical partitions,
and finally in Sec.~\ref{sec:evidence} we show how our methods allow us
to perform more accurate Bayesian model selection, which requires a
detailed depiction of the space of solutions that our approach is able
to provide. We end in Sec.~\ref{sec:conclusion} with a conclusion.

\section{The group identification problem in community detection}\label{sec:label}

In this work we will focus on the approach to community detection that
is based on the statistical inference of generative
models~\cite{peixoto_bayesian_2019}. Although our techniques can be used
with arbitrary community detection methods (or in fact for any data
clustering algorithm), those based on inference lend themselves more
naturally to our analysis, since they formally define a probability
distribution over partitions. More specifically, if we consider a
generative model for a network conditioned on a node partition $\bb =
\{b_i\}$, where $b_i$ is the group label of node $i$, such that each
network $\A$ occurs with a probability $P(\A|\bb)$, we obtain the
posterior distribution of network partitions by employing Bayes' rule,
\begin{equation}\label{eq:sbm_posterior}
  P(\bb|\A) = \frac{P(\A|\bb)P(\bb)}{P(\A)},
\end{equation}
where $P(\bb)$ the prior probability of partitions, and
$P(\A)=\sum_{\bb}P(\A|\bb)P(\bb)$ is the model evidence. There are many
ways to compute this probability, typically according to one of the many
possible parametrizations of the stochastic block model
(SBM)~\cite{holland_stochastic_1983} and corresponding choice of prior
probabilities for their parameters. Since our analysis will not depend
on any particular choice, we omit their derivations, and instead point
the reader to Ref.~\cite{peixoto_bayesian_2019} for a summary of the
most typical alternatives. To our present goal, it is sufficient to
establish that such a posterior distribution can be defined, and we have
mechanisms either to approximately maximize or sample partitions from
it.

The first central issue we seek to address is that for this class of
problems the actual numeric values of the group labels have no particular
significance, as we are interested simply in the division of the nodes
into groups, not in their particular placement in named categories. This
means that the posterior probability above is invariant to label
permutations. More specifically, if we consider a bijective mapping of
the labels $\mu(r)=s$, such that its inverse $\mu^{-1}(s)=r$ recovers
the original labels, then a label permutation $\cc = \{c_i\}$ where
$c_i=\mu(b_i)$, has the same posterior probability,
\begin{equation}
  P(\bb|\A) =  P(\cc|\A),
\end{equation}
for any choice of $\bm\mu$. Very often this is considered an unimportant
detail, since many inference methods break this label permutation
symmetry intrinsically. For example, if we try to find a partition that
maximizes the posterior distribution with a stochastic algorithm, we
will invariably find one of the many possible label permutations, in an
arbitrary manner that usually depends on the initial conditions, and we
can usually move on with the analysis from there. Methods like
belief-propagation~\cite{decelle_asymptotic_2011}, which can be employed
in the special case where the model parameters other than the partition
$\bb$ are known, yield marginal distributions over partitions that, due
to random initialization, also break the overall label permutation
symmetry, and yield a distribution centered around one particular group
labelling. The same occurs also for some Markov chain Monte Carlo (MCMC)
algorithms, for example those based on the movement of a single node at
a time~\cite{peixoto_efficient_2014,riolo_efficient_2017}, which will
often get trapped inside one particular choice of labels, since the swap
of two labels can only occur if the respective groups exchange all their
nodes one by one, a procedure that invariably moves the Markov chain
through low probability states, and thus is never observed in practice.
Although this spontaneous label symmetry breaking can be seen as a
helpful property in these cases, strictly speaking it is a failure of
the inference procedure in faithfully representing the overall label
symmetry that does exist in the posterior distribution. In fact, this
symmetry guarantees that the marginal posterior group membership
probability of any node must be the same for all $N$ nodes, i.e.
\begin{equation}\label{eq:trivial}
  \pi_i(r) = \sum_{\bb}\delta_{b_i,r}P(\bb|\A) = \sum_{B=r}^{N}\frac{P(B)}{B},
\end{equation}
where $P(B)$ is the marginal distribution of the number of labels
(nonempty groups), and we assume that the labels always lie in a
contiguous range from $1$ to $B$.  Because of this, the true answer to
the question ``what is the probability of a node belonging to a given
group?'' is always an unhelpful one, since it is the same one for every
node, and carries no information about the network structure. Far from
being a pedantic observation, this is a problem we encounter directly
when employing more robust inference methods such as the merge-split
MCMC of Ref.~\cite{peixoto_merge-split_2020}. In that algorithm, the
merge and split of groups are employed as direct move proposals which
significantly improve the mixing time and the tendency of the Markov
chain to get trapped in metastable states, when compared to single-node
moves. However, as a consequence, the merge and split of groups
result in the frequent sampling of the same partition where two group
labels have been swapped, after a merge and split. In fact, the
algorithm of Ref.~\cite{peixoto_merge-split_2020} also includes a joint
merge-split move, where the memberships of the nodes belonging to two
groups are redistributed in a single move, which often results in the
same exact partition, but with the labels swapped. Such an algorithm
will rapidly cycle through all possible label permutations leading to
the correct albeit trivial uniform marginal probabilities given by
Eq.~\ref{eq:trivial}.

\begin{figure}
  \begin{tabular}{ccc}
    \begin{overpic}[width=.33\columnwidth, trim=.8cm .8cm .8cm .8cm]{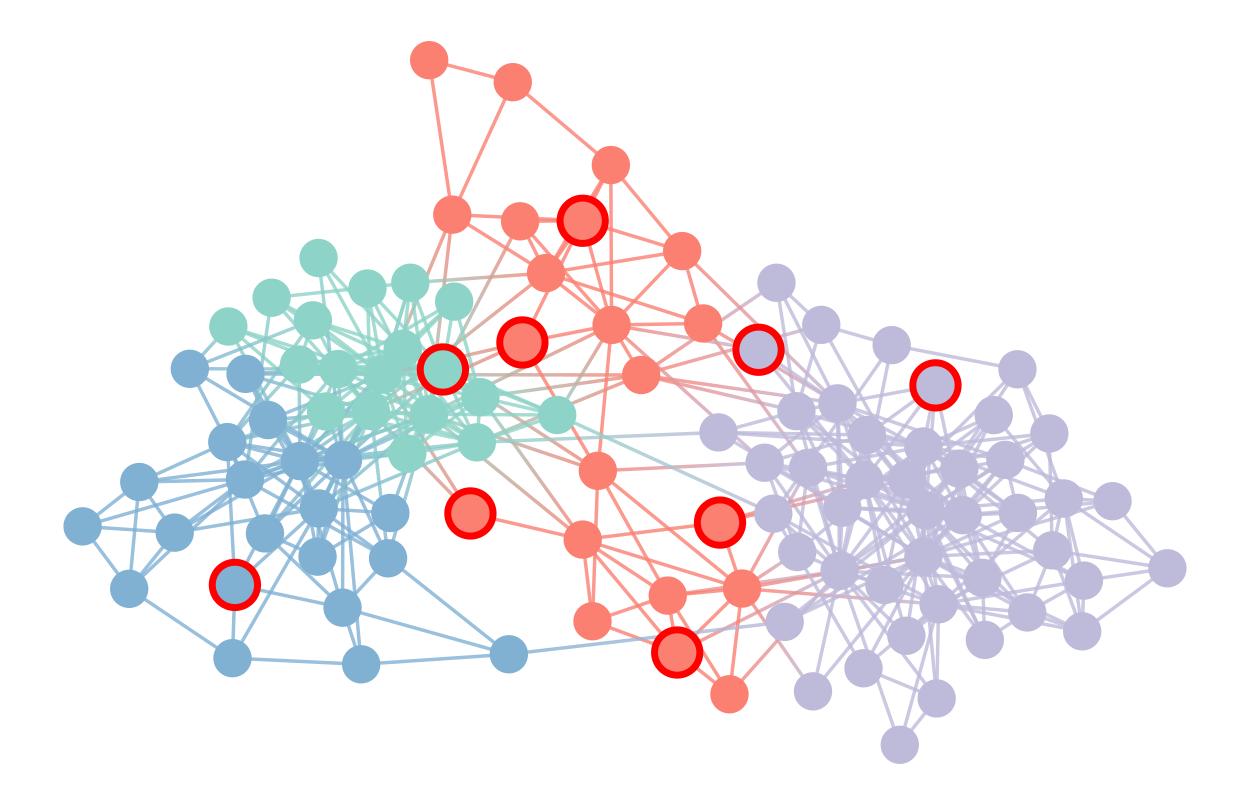}
      \put(0,50){(a)}
    \end{overpic}
    & \includegraphics[width=.33\columnwidth, trim=.8cm .8cm .8cm .8cm]{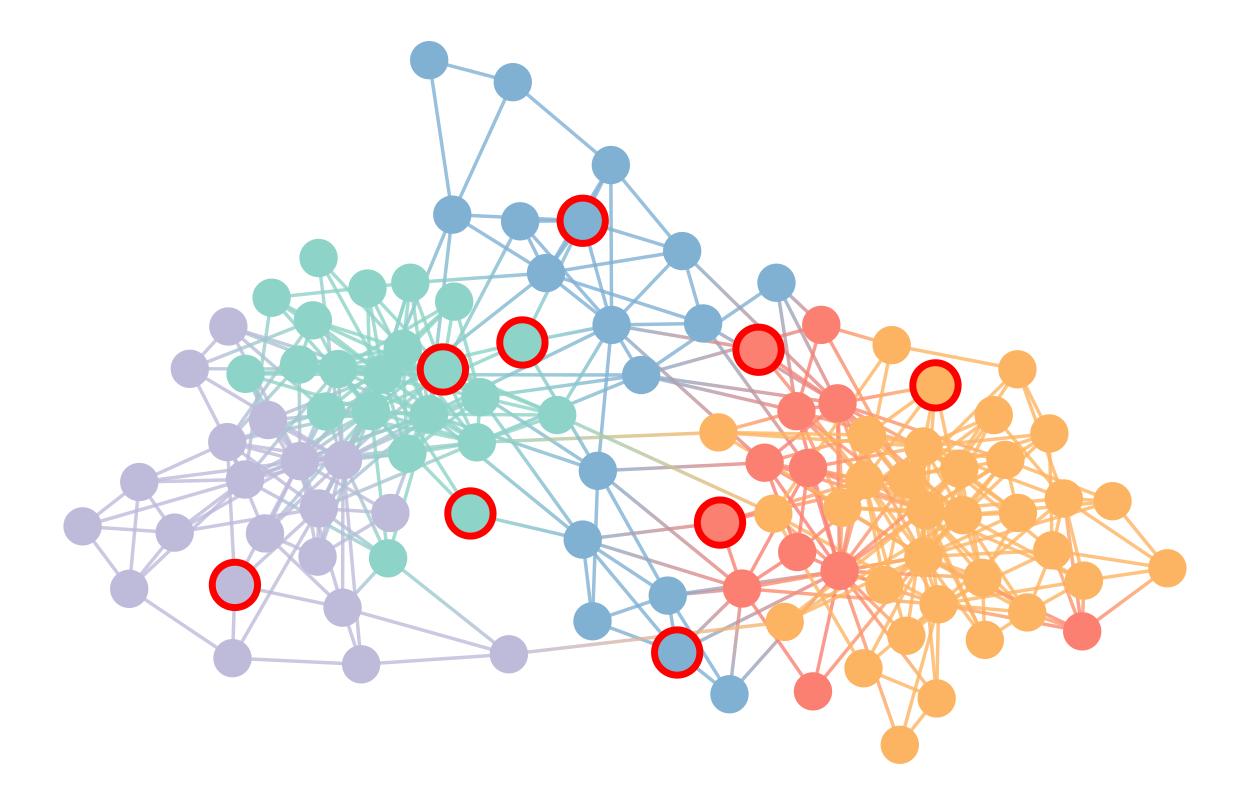}
    & \includegraphics[width=.33\columnwidth, trim=.8cm .8cm .8cm .8cm]{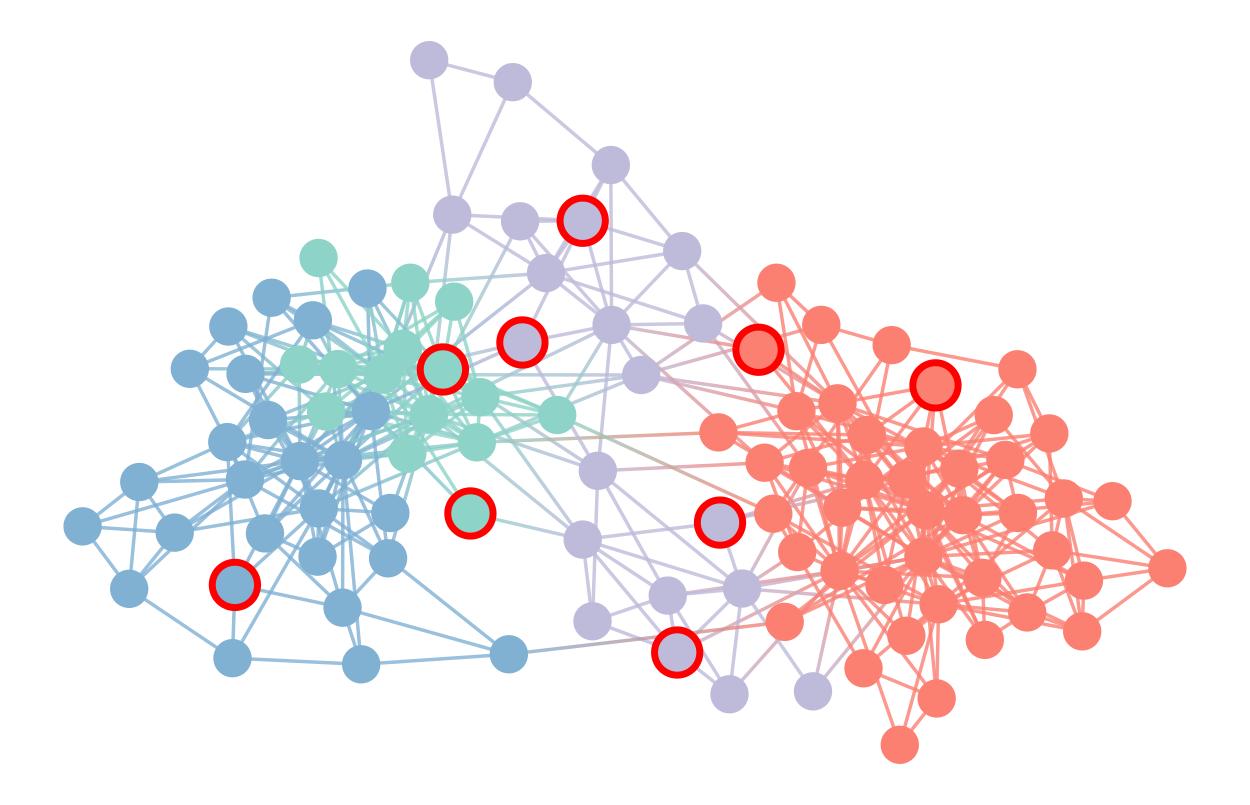}\\
    \includegraphics[width=.33\columnwidth, trim=.8cm .8cm .8cm .8cm]{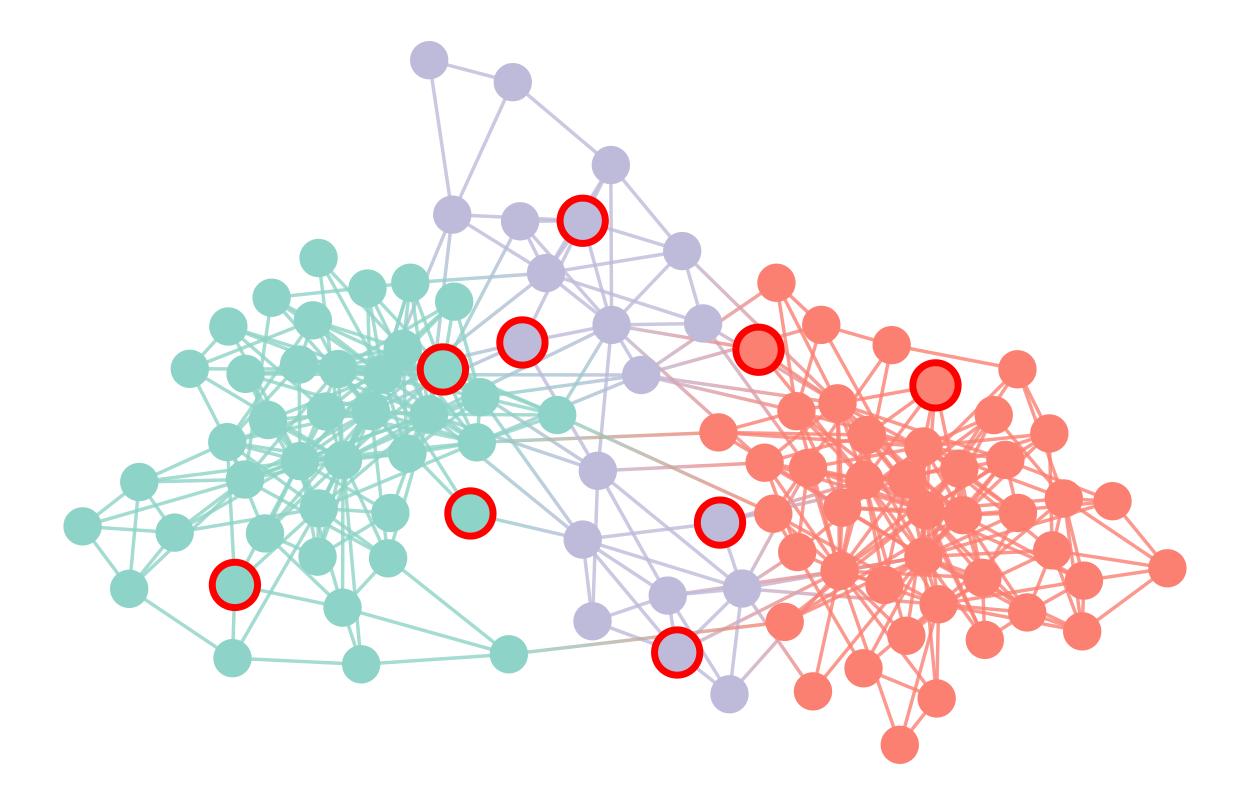} & \multicolumn{2}{c}{\multirow{2}{*}[4.5em]{
        \begin{overpic}[width=.6\columnwidth, trim=0 0 0cm 0]{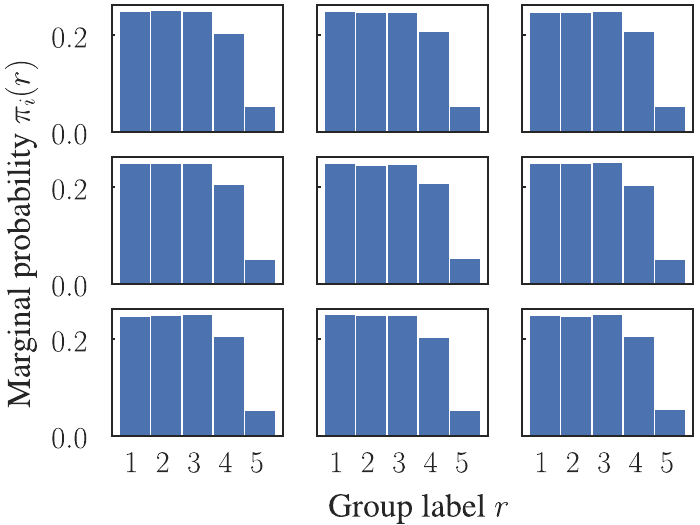}
          \put(-5,72){(b)}
        \end{overpic}}}\\
    \includegraphics[width=.33\columnwidth, trim=.8cm .8cm .8cm .8cm]{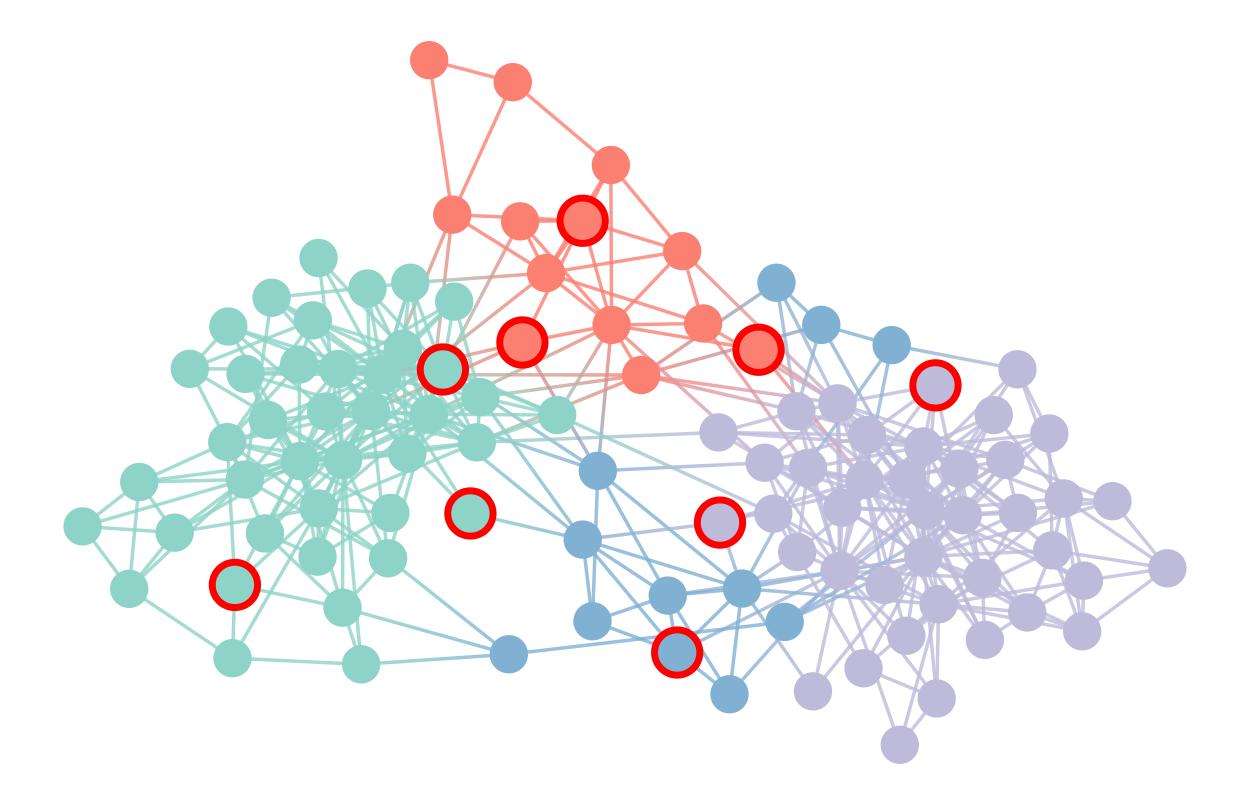} & \\
  \end{tabular} \caption{(a) Five sampled partitions from the posterior
  distribution of a network of political books, with the group labels
  represented as colors, using the Poisson DC-SBM and the MCMC algorithm
  of Ref.~\cite{peixoto_merge-split_2020} (b) Marginal posterior
  distribution of the group memberships of the nodes highlighted in red
  in (a), obtained for $10^5$ samples from the posterior
  distribution. The same asymptotic distribution is obtained for every
  single node in the network. \label{fig:label_cycle}}
\end{figure}

In Fig.~\ref{fig:label_cycle} we show how the label permutation
invariance can affect community detection for a network of co-purchases
of political books~\cite{krebs_political_nodate}, for which we used the
Poisson degree-corrected SBM (DC-SBM)~\cite{karrer_stochastic_2011},
using the parametrization of Ref.~\cite{peixoto_nonparametric_2017}, and
the merge-split MCMC of Ref.~\cite{peixoto_merge-split_2020}. Although
the individual partitions yield seemingly meaningful divisions, they are
observed with a random permutation of the labels, preventing an
aggregate statistics at the level of single nodes to yield useful
information.

At first we might think of a few simple strategies that can alleviate
the problem. For example, instead of marginal distributions, we can
consider the pairwise co-occurence probabilities $c_{ij} =
\sum_{\bb}\delta_{b_i,b_j}P(\bb|\A) \in [0,1]$, which quantify how often
two nodes belong to the same group, and thus is invariant with respect
to label permutations. However this gives us a large dense matrix of
size $N^2$ which is harder to interpret and manipulate than marginal
distributions --- indeed the usual approach is to try to cluster this
matrix~\cite{strehl_cluster_2002}, by finding groups of nodes that have
similar co-occurences with other nodes, but this just brings us back to
the same kind of problem. Another potential option is to choose a
canonical naming scheme for the group labels, for example by indexing
groups according to their size, such that $r < s$ if $n_r < n_s$, where
$n_r$ is the number of nodes with group label $r$. However this idea
quickly breaks down if we have groups of the same size, or if the group
sizes vary significantly in the posterior distribution. An alternative
canonical naming is one based on an arbitrary ordering of the nodes and
forcing the labels to be confined to a contiguous range, so that $b_j >
b_i$ for $j > i$ whenever $b_j$ corresponds to a group label previously
unseen for nodes $k\le i$. In this way every partition corresponds to a
single canonical labeling, which we can generate before collecting
statistics on the posterior distribution. Unfortunately, this approach
is not straightforward to implement, since the marginal distributions
will depend strongly on the chosen ordering of the nodes. For example,
if the first node happens to be one that can belong to two groups with
equal probability, whenever this node changes membership, it will incur
the relabeling of every other group, and thus spuriously causing the
marginal distribution of every other node to be broader, even if they
always belong to the ``same'' group. It seems intuitive therefore to
order the nodes according to the broadness of their marginal
distribution, with the most stable nodes first, but since determining
the marginal distribution depends on the ordering itself, it leads to a
circular problem.

In the following we will provide a different solution to this problem,
based on a generative model of labelled partitions, that is both
satisfying and easy to implement, and allows us in the end to obtain
marginal distributions in an unambiguous manner.

\section{Establishing consensus: The random label model}\label{sec:random-label}

If we have as an objective the estimation of the marginal probability
$\pi_i(r)$ of node $i$ belonging to group $r$, given $M$ partitions
$\{\bb\} = \{\bb^{(1)},\dots,\bb^{(M)}\}$ sampled from a posterior
distribution, this is done by computing the mean
\begin{equation}\label{eq:marginal}
  \pi_i(r) = \frac{1}{M}\sum_{m=1}^{M}\delta_{b_i^m,r}.
\end{equation}
This is fully equivalent to fitting a factorized ``mean-field'' model on
the same samples, given by
\begin{equation}
  P_{\text{MF}}(\bb|\bm p,B) = \prod_ip_i(b_i),
\end{equation}
where $p_i(r)$ is the probability of node $i$ belonging to group $r \in
\{1,\dots,B\}$. Given the same partitions, the maximum likelihood
estimate of the above model corresponds exactly to how we estimate
marginal distributions, i.e.
\begin{equation}
  \hat{p}_i(r) =
  \underset{p_i(r)}{\operatorname{argmax}}\prod_{m=1}^{M}P_{\text{MF}}(\bb^{(m)}|\bm p, B) = \pi_i(r).
\end{equation}
Although this computation is common practice, it is important to note
that this model is inconsistent with our posterior distribution of
Eq.~\ref{eq:sbm_posterior}, since it is in general not invariant to
label permutations, i.e. if we swap two labels $r$ and $s$ we have the
same distribution only if $p_i(r)=p_i(s)$ for every node $i$. Therefore,
in order to tackle the label symmetry problem, we may modify this
inference procedure, by making it also label symmetric. We do so by
assuming that our partitions are initially sampled from the above model,
but then the labels are randomly permuted. In other words, we have
\begin{equation}
  P(\bb|\bm p,B) = \sum_{\cc} P(\bb|\cc)P_{\text{MF}}(\cc|\bm p,B),
\end{equation}
where the intermediary partition $\cc$ is relabelled into $\bb$ with a
uniform probability
\begin{equation}
  P(\bb|\cc) = \frac{[\bb\sim\cc]}{q(\cc)!},
\end{equation}
where we make use of the symmetric indicator function
\begin{equation}
  [\bb\sim\cc] =
  \begin{cases}
    1 & \text{ if $\bb$ is a label permutation of $\cc$},\\
    0 & \text{ otherwise,}
  \end{cases}
\end{equation}
and where $q(\cc)$ is the number of labels actually present in partition
$\bm c$ [not to be confused with the total number of group labels $B$ in
the underlying model, since some groups may end up empty, so that
$q(\cc)\le B$], and $q(\cc)!$ in total number of label permutations of
$\cc$. Now, inferring the probabilities $\bm p$ from the model above
involves finding a single underlying canonical labelling that is erased
at each sample, but after it is identified allows us to obtain marginal
distributions. This canonical labeling itself is not unique, since every
permutation of its labels is equivalent, but we do not care about the
identity of the labels, just an overall alignment, which is what the
inference will achieve.

\begin{figure}
  \begin{tikzpicture}
    \node at (2.5, .5) {Group labels $r$};
    \node (r1) at (0.5, 0) [circle,draw] {};
    \node (r2) at (1.5, 0) [circle,draw] {};
    \node (r3) at (2.5, 0) [circle,draw] {};
    \node (r4) at (3.5, 0) [circle,draw] {};
    \node (r5) at (4.5, 0) [circle,draw] {};

    \node at (2.5, -2) {Group labels $s$};
    \node (s1) at (0, -1.5) [circle,draw] {};
    \node (s2) at (1, -1.5) [circle,draw] {};
    \node (s3) at (2, -1.5) [circle,draw] {};
    \node (s4) at (3, -1.5) [circle,draw] {};
    \node (s5) at (4, -1.5) [circle,draw] {};
    \node (s6) at (5, -1.5) [circle,draw] {};

    \draw[-, line width=1] (r1) to node[above,sloped] {$w_{rs}$} (s1);
    \draw[-, line width=4, color=green!70!black] (r1) to (s3);
    \draw[-, line width=3, color=green!70!black] (r2) to (s2);
    \draw[-, line width=3] (r2) to (s4);
    \draw[-, line width=1.2, color=green!70!black] (r3) to (s4);
    \draw[-, line width=1.4] (r3) to (s3);
    \draw[-, line width=1.1] (r4) to (s3);
    \draw[-, line width=3.2, color=green!70!black] (r4) to (s5);
    \draw[-, line width=1.6] (r4) to (s6);
    \draw[-, line width=3.5] (r5) to (s5);
    \draw[-, line width=3.2, color=green!70!black] (r5) to (s6);
  \end{tikzpicture}

  \caption{Relabeling a partition corresponds to finding the solution of
  a maximum bipartite weighted matching problem, where the partition
  labels are the nodes of a bipartite graph with weights $w_{rs}$ on the
  edges. The matching is a bijection $\mu(r)$ that needs to be chosen so
  that the total sum $\sum_rw_{r,\mu(r)}$ is maximized. In this
  illustration, the edge thickness corresponds to the weight $w_{rs}$,
  and the edges in green correspond to the maximum matching.
  \label{fig:matching}}
\end{figure}
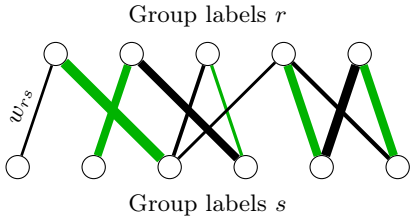

We proceed with the inference of the above model in the following
way. Suppose we observe $M$ partitions
$\{\bb\}=\{\bb^{(1)},\dots,\bb^{(M)}\}$ sampled from the posterior
distribution as before.  Our first step is to infer the hidden labels
$\{\cc\}=\{\cc^{(1)},\dots,\cc^{(M)}\}$ from the posterior
\begin{equation}
  P(\{\cc\},B | \{\bb\}) = \frac{P(\{\bb\}|\{\cc\})P(\{\cc\}|B)P(B)}{P(\{\bb\})}
\end{equation}
with the marginal likelihood integrated over all possible probabilities
$\bm p$, and given by
\begin{align}
  P(\{\cc\}|B) &= \int P_{\text{MF}}(\{\cc\}|\bm p)P(\bm p|B)\;\dd\bm p\\
  &= \prod_i \frac{(B-1)!}{(M+B-1)!}\prod_r n_i(r)!, \label{eq:random_label_marginal}
\end{align}
where
\begin{equation}
  n_i(r) = \sum_{m=1}^{M}\delta_{c_i^m,r}
\end{equation}
is the number of relabelled partitions where node $i$ has hidden label
$r$, and we have used an uninformative prior
\begin{align}
  P(\bm p|B) = \prod_i(B-1)!,
\end{align}
corresponding to a constant probability density for every node over a
$B$-dimensional simplex, each with volume $1/(B-1)!$, which is also
equivalent to a Dirichlet prior with unit hyperparameters. Therefore, up
to an unimportant multiplicative constant, we have that the posterior
distribution of hidden relabellings is given by
\begin{multline}\label{eq:posterior_c}
  P(\{\cc\}, B | \{\bb\}) \propto\\
  \left(\prod_{m=1}^{M}[\bb^{(m)}\sim\cc^{(m)}]\right)\prod_i\frac{(B-1)!}{(M+B-1)!}\prod_r n_i(r)!,
\end{multline}
where have assumed a uniform prior $P(B)=1/N$, which does not contribute
to the above.  We proceed by considering the conditional posterior
distribution of a single partition $\cc^{(m)}$,
\begin{multline}
  P(\cc^{(m)} | \{\bb\}, \{\cc^{(m'\ne m)}\}, B)\\
\begin{aligned}
  &\propto \prod_i\prod_r \left[n_i'(r)+\delta_{c_i^m,r}\right]!\\
  &\propto \prod_i\prod_r \left\{\left[n_i'(r)+1\right]!\right\}^{\delta_{c_i^m,r}}\left\{\left[n_i'(r)\right]!\right\}^{1-\delta_{c_i^m,r}}\\
  &\propto \prod_i\prod_r \left[n_i'(r)+1\right]^{\delta_{c_i^m,r}},
\end{aligned}
\end{multline}
where $n_i'(r)=\sum_{m'\ne m}\delta_{c_i^{(m')},r}$ is the label count
excluding $\cc^{(m)}$, and we have dropped the indicator function for
conciseness, but without forgetting that $[\cc^{(m)}\sim\bb^{(m)}]=1$
must always hold. If we seek to find the most likely hidden labelling
$\cc^{(m)}$ we need to maximize the above probability, or equivalently
its logarithm, which is given by
\begin{multline}
  \ln P(\cc^{(m)} | \{\bb\}, \{\cc_{m'\ne m}\}, B) = \sum_{i,r} \delta_{c_i^{m},r}\ln\left[n_i'(r)+1\right],
\end{multline}
up to an unimportant additive constant. The maximization involves
searching through all $q(\bb^{(m)})!$ possible relabellings of
$\bb^{(m)}$. Unfortunately, this number grows too fast for an exhaustive
search to be feasible, unless the number of labels is very
small. Luckily, as we now show, it is possible to re-frame the
optimization, in a manner that exposes its feasibility. We begin by
representing the mapping between the labels of $\bb^{(m)}$ and
$\cc^{(m)}$ via the bijective function $\mu(r)$, chosen so that
\begin{equation}
  \mu(b_i^m) = c_i^m, \quad\forall i.
\end{equation}
Now, by introducing the matrix
\begin{equation}
  w_{rs}=\sum_i\delta_{b_i,r}\ln\left[n_i'(s)+1\right],
\end{equation}
we can express the log-likelihood as
\begin{equation}\label{eq:bip_matching}
  \ln P(\cc^{(m)} | \{\bb\}, \{\cc_{m'\ne m}\}, B) = \sum_rw_{r,\mu(r)}.
\end{equation}
Therefore, if we consider the matrix $w_{rs}$ as the weighted adjacency
matrix of a bipartite graph, where the group labels of $\bb^{(m)}$ and
$\cc^{(m)}$ form the nodes on each partition (see
Fig.~\ref{fig:matching}), the above log-likelihood corresponds to the
sum of the weights of the edges selected by $\bm\mu$. Finding such a
bijection is an instance of a very well known combinatorial optimization
problem called \emph{maximum bipartite weighted matching}, also known as
the \emph{assignment problem}, which corresponds to finding a
``matching'' on a bipartite graph, defined as a subset of the edges that
share no common nodes, such that the sum of the weights of the edges
belonging to the matching is maximized. This corresponds precisely to
the sum given in Eq.~\ref{eq:bip_matching}, where a given choice of
$\bm\mu$ corresponds to a particular matching. In particular we are
interested in the unbalanced and imperfect version of the matching
problem, where the number of groups on both sides might be different,
and groups on either side might be left
unmatched~\cite{ramshaw_minimum-cost_2012}, in which case for each
unmatched group we give it a label of a new group. Luckily, fast
polynomial algorithms for this problem have been long known. For example
using the ``Hungarian'' or Kuhn–Munkres
algorithm~\cite{kuhn_hungarian_1955,munkres_algorithms_1957} this
problem can be solved with a worst-case running time of $O(q(\bb)^3)$,
which is substantially better than an exhaustive search, rendering our
approach not only feasible but efficient.

Having found the maximum of Eq.~\ref{eq:bip_matching}, we are still left
with inferring the value of $B$ according to
Eq.~\ref{eq:posterior_c}. But, as it is easy to verify, the likelihood
is a monotonically decreasing function of $B$. Therefore, since
$q(\cc)\le B$, this step amounts simply to choosing $B$ so that
\begin{equation}\label{eq:B_post}
  B = \underset{m}{\max}\; q(\cc^{(m)}).
\end{equation}
Equipped with the above, we can summarize our whole inference algorithm
as follows:
\begin{enumerate}
  \item We sample $M$ partitions $\bb^{(1)},\dots,\bb^{(M)}$ from the
    posterior distribution $P(\bb|\A)$.
  \item We initialize $\cc^{(m)} = \bb ^{(m)}$ for every sample $m$.
  \item For each sample $m$, in random order, we obtain a new
        relabelling $\cc^{(m)}$ such that
        Eq.~\ref{eq:bip_matching} is maximized.
  \item If any value of $\cc^{(m)}$ is changed during the last
        step, we repeat it, otherwise we stop and return $\{\cc\}$.
  \item We update the inferred value of $B$ according to
        Eq.~\ref{eq:B_post}.
\end{enumerate}
By the end of this algorithm, we are guaranteed to find a local maximum
of Eq.~\ref{eq:posterior_c}, but not a global one, hence we need to run
it multiple times and obtain the result with the largest posterior
probability. However, we found that repeated runs of the algorithm give
the same result the vast majority of cases we tried.\footnote{We offer a
freely available reference C++ implementation of every algorithm
described in this work as part of the \texttt{graph-tool}
Python library~\cite{peixoto_graph-tool_2014}.}

\begin{figure*}
  \begin{tabular}{ccccc}
    \multicolumn{4}{c}{Sampled}&\\
    \includegraphics[width=.2\textwidth, trim=.8cm .8cm .8cm .2cm]{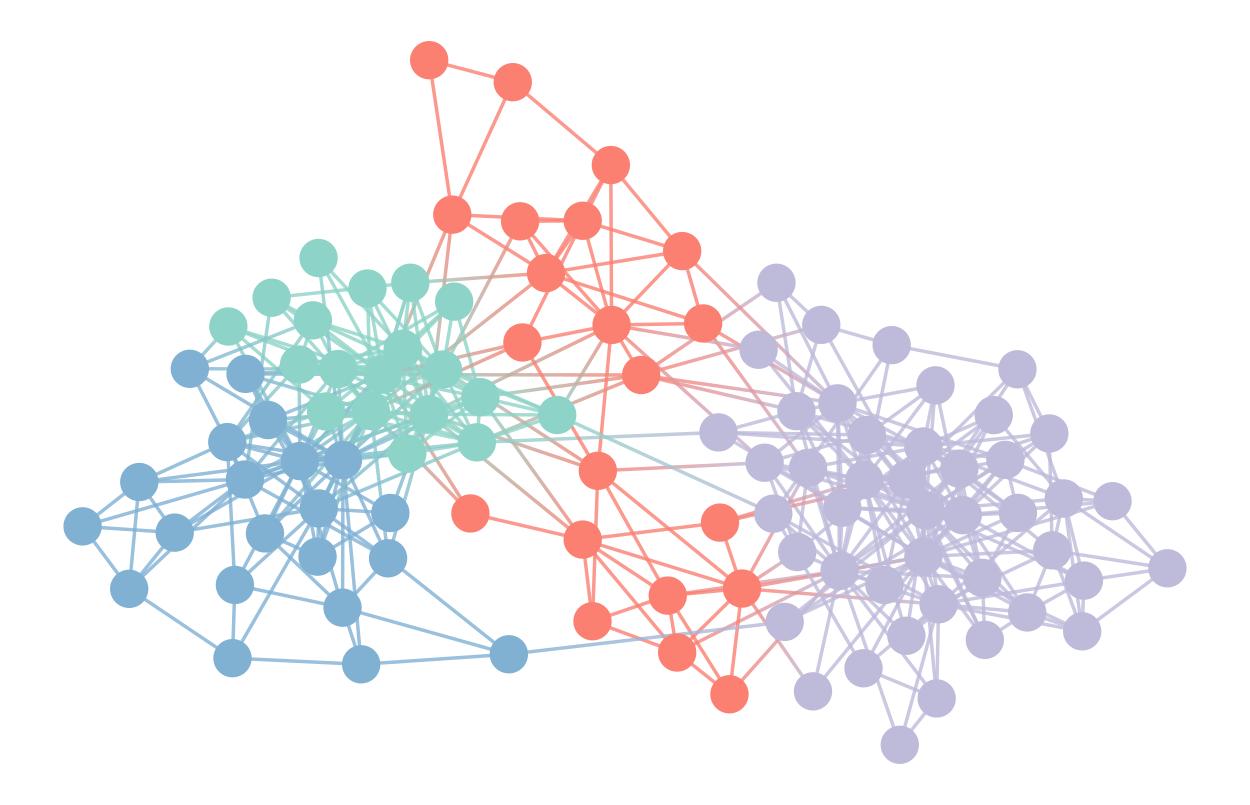}
    & \includegraphics[width=.2\textwidth, trim=.8cm .8cm .8cm .2cm]{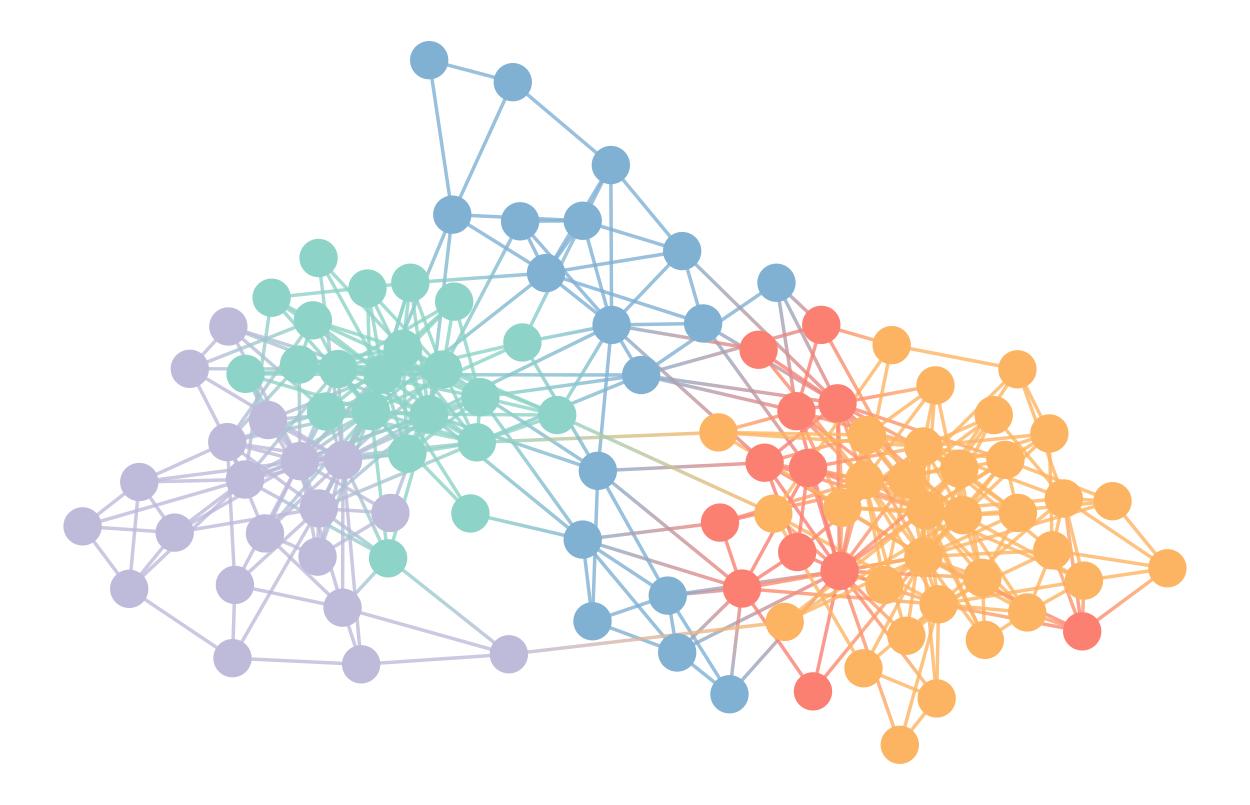}
    & \includegraphics[width=.2\textwidth, trim=.8cm .8cm .8cm .2cm]{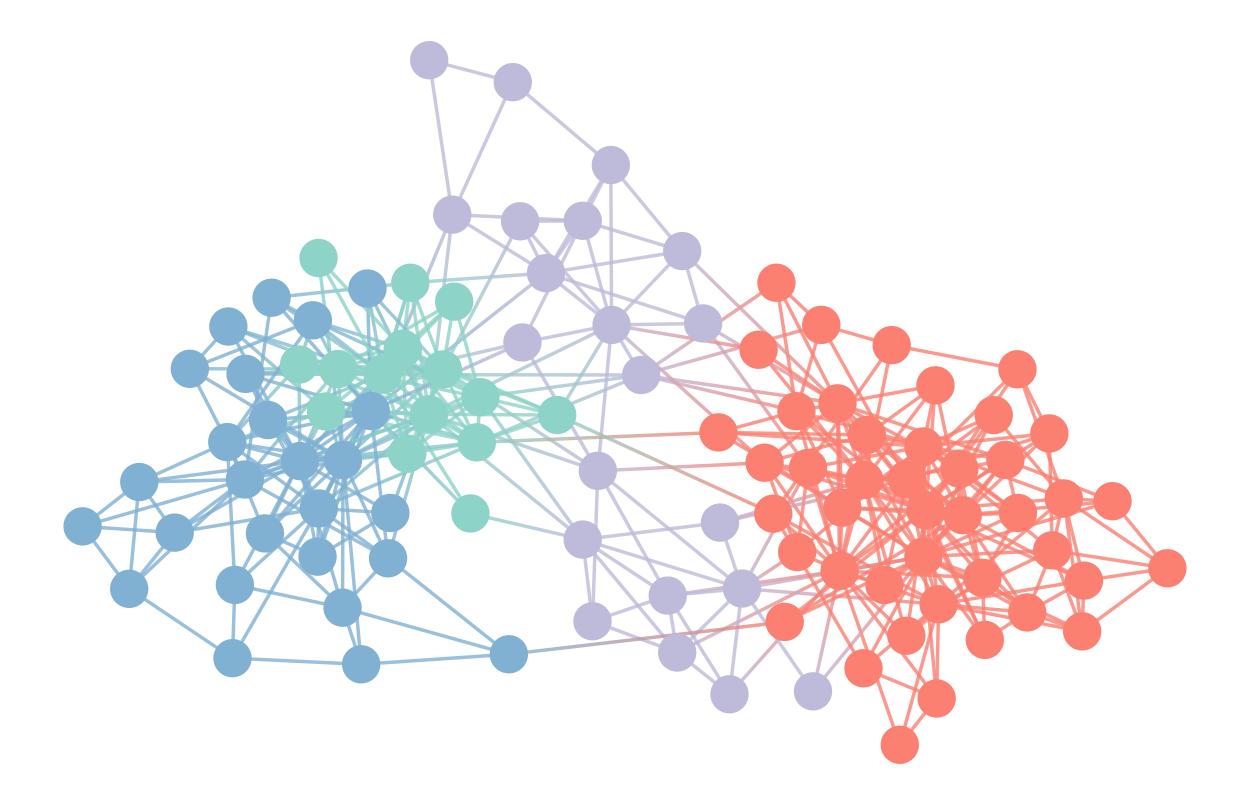}
    & \includegraphics[width=.2\textwidth, trim=.8cm .8cm .8cm .2cm]{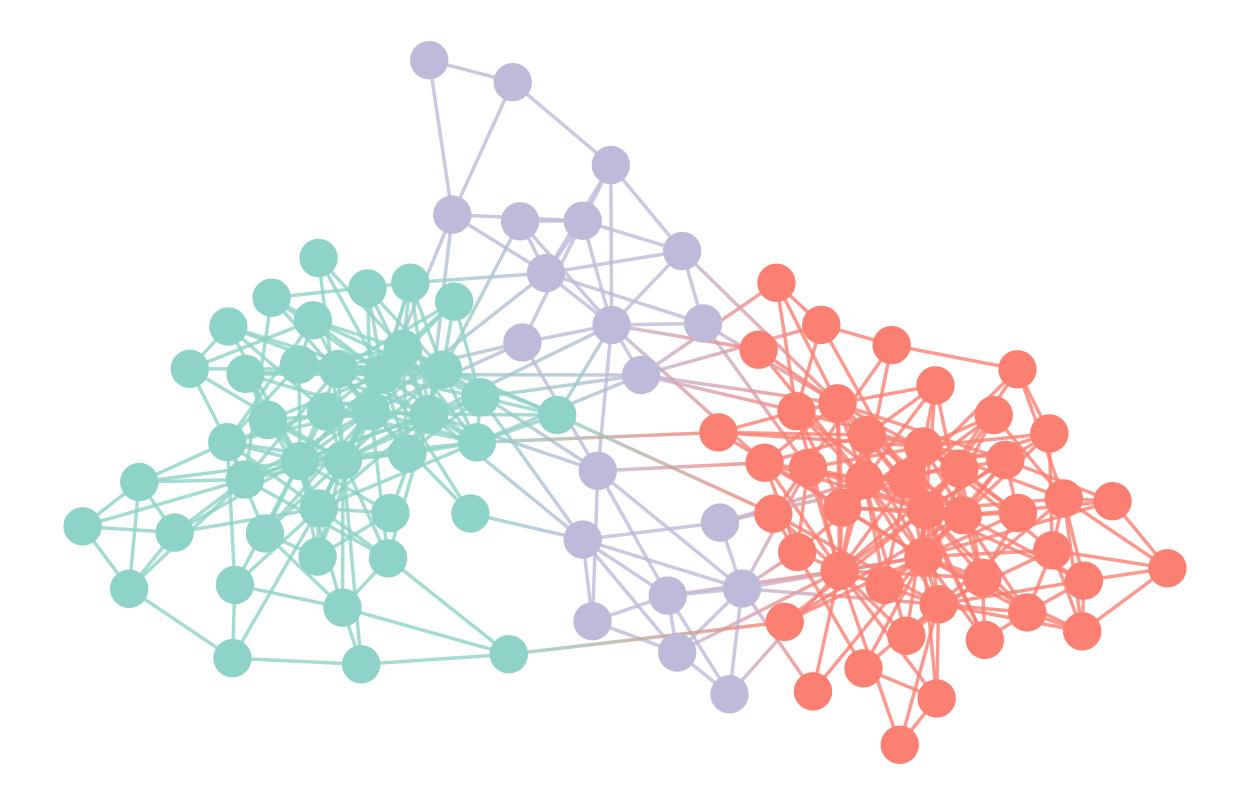}
    \includegraphics[width=.2\textwidth, trim=.8cm .8cm .8cm .5cm]{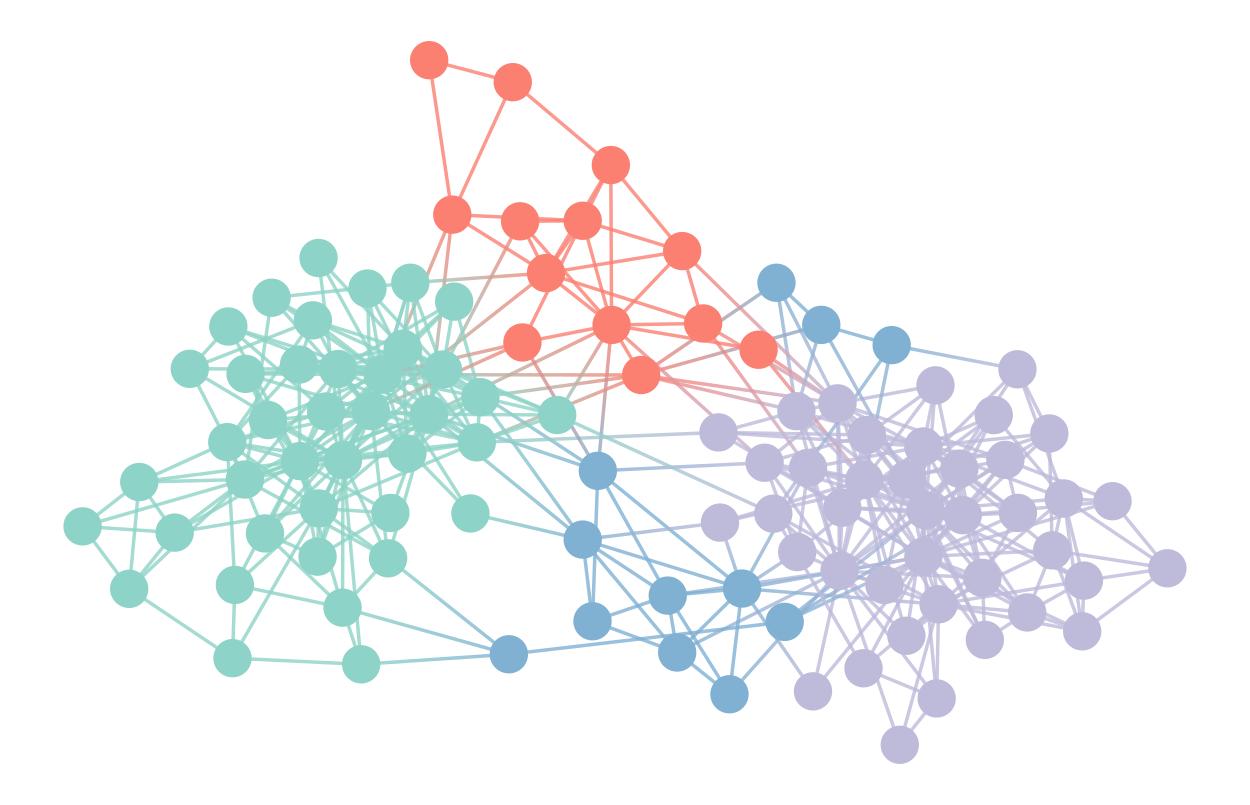}\\
    \multicolumn{4}{c}{Relabelled}&\\
    \includegraphics[width=.2\textwidth, trim=.8cm .8cm .8cm .2cm]{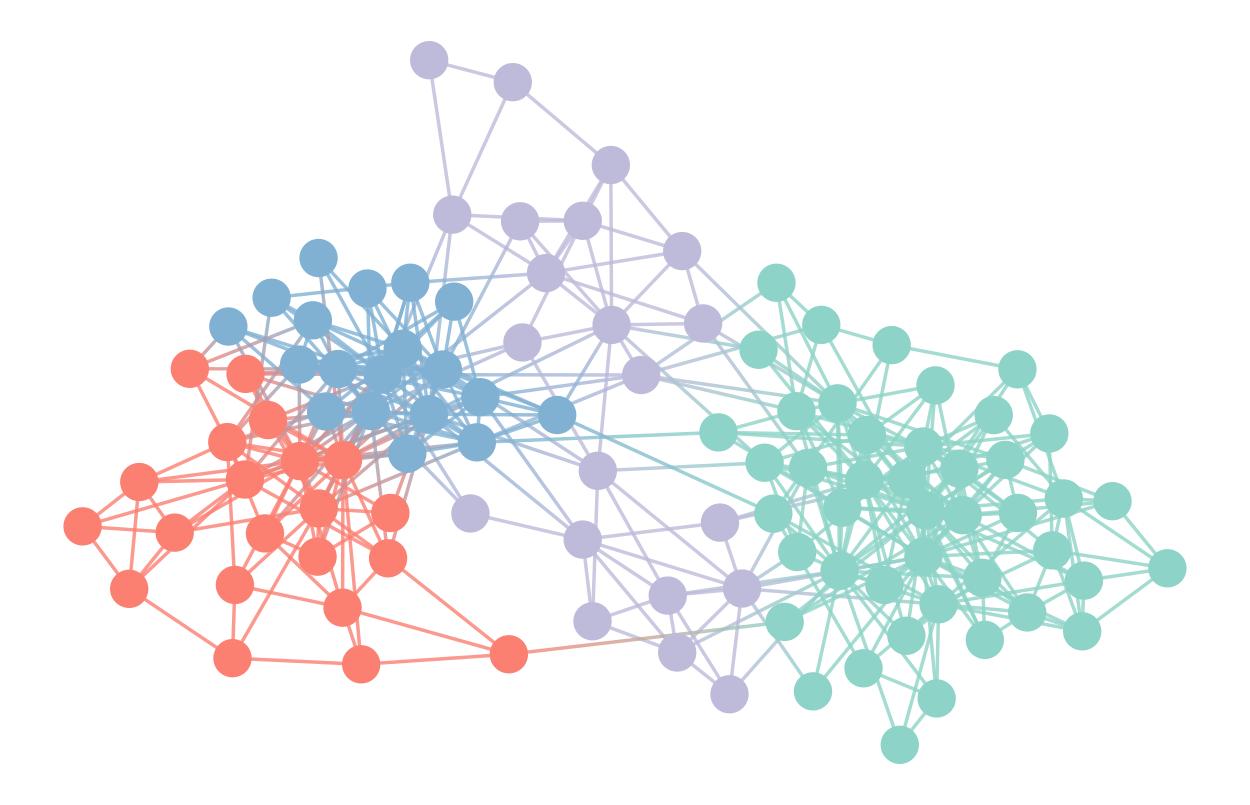}
    & \includegraphics[width=.2\textwidth, trim=.8cm .8cm .8cm .2cm]{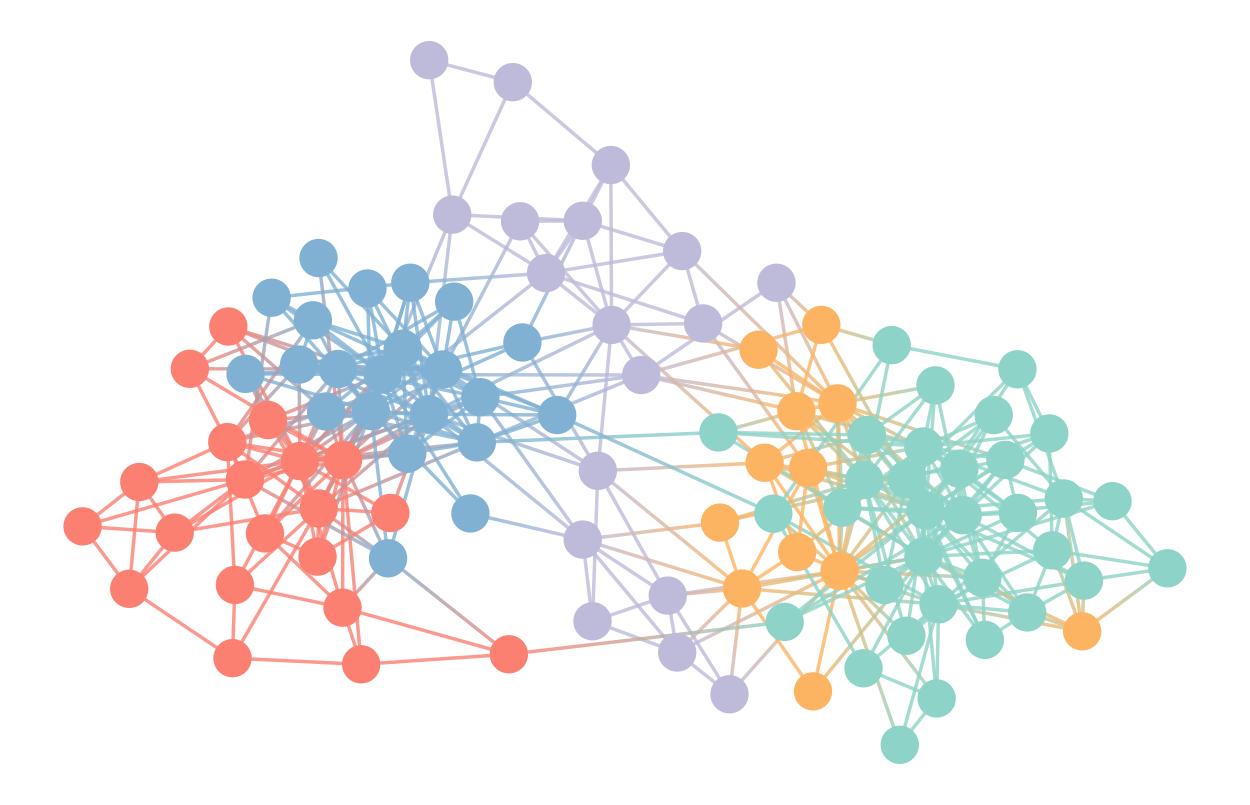}
    & \includegraphics[width=.2\textwidth, trim=.8cm .8cm .8cm .2cm]{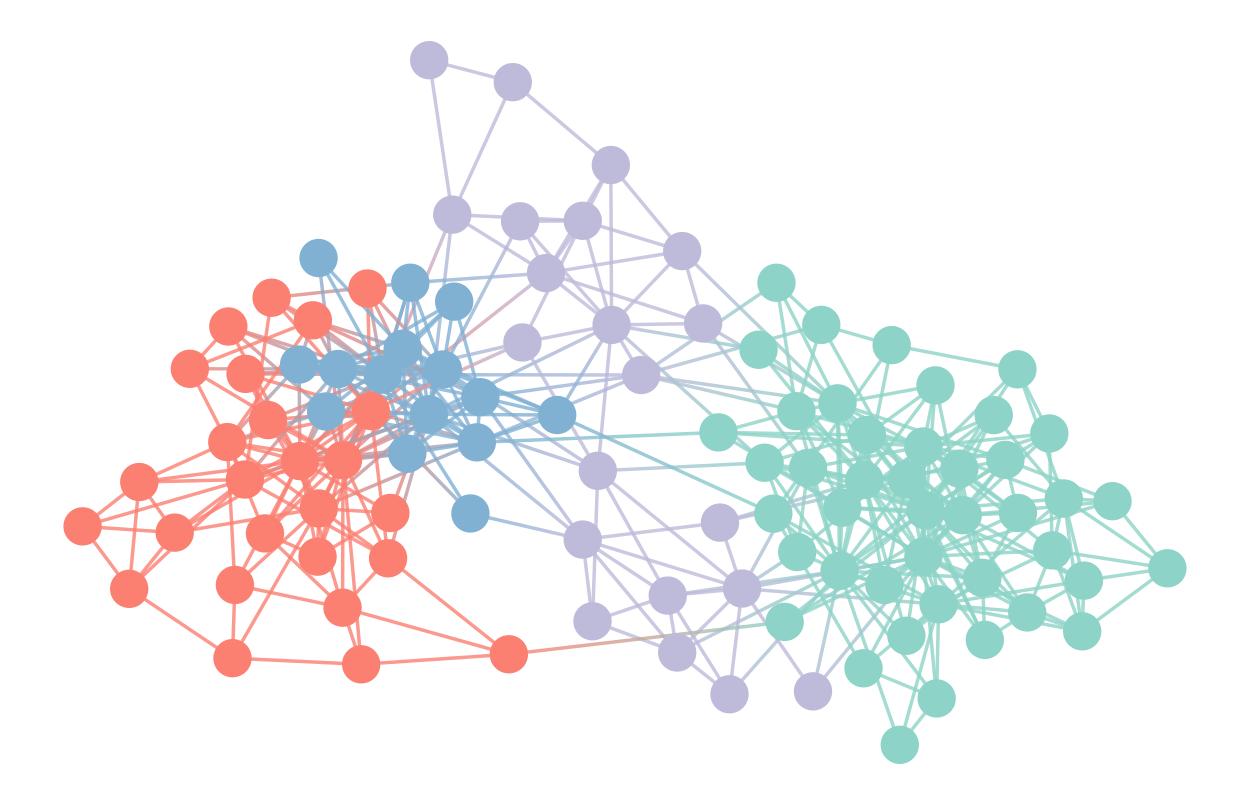}
    & \includegraphics[width=.2\textwidth, trim=.8cm .8cm .8cm .2cm]{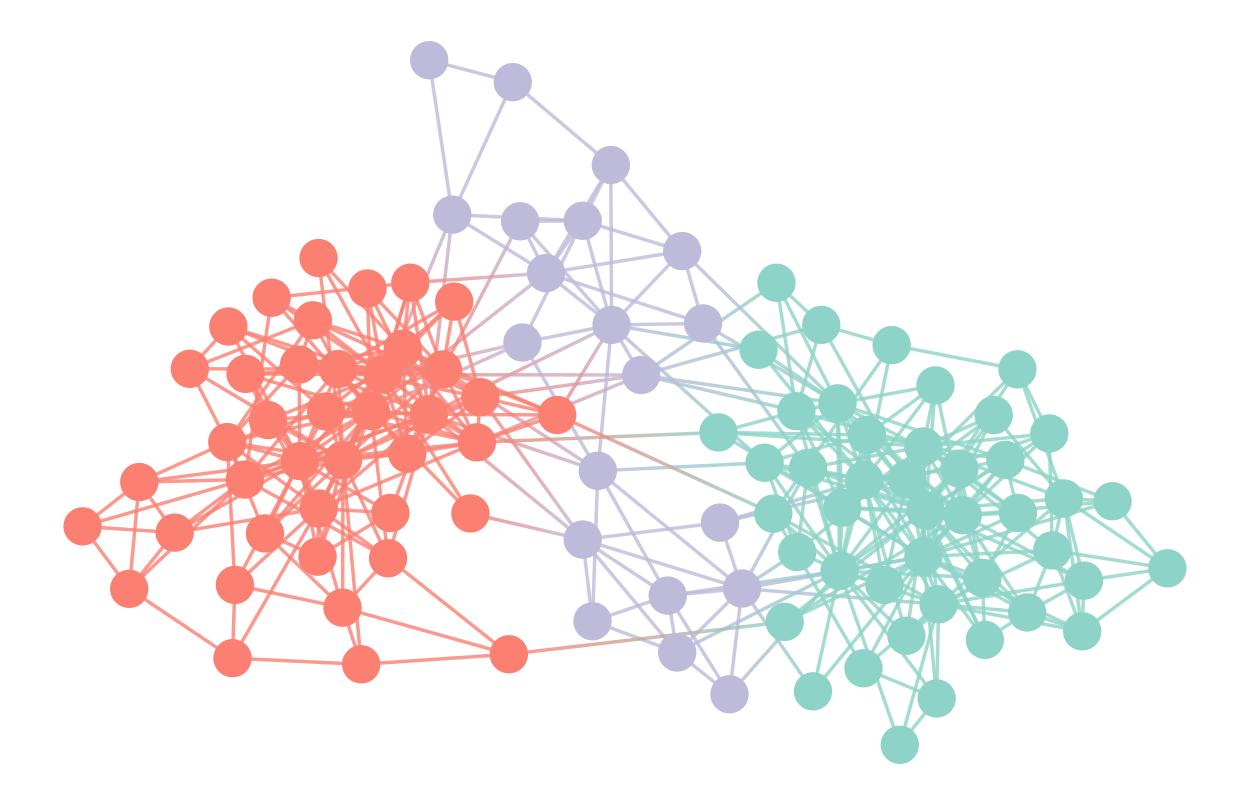}
    \includegraphics[width=.2\textwidth, trim=.8cm .8cm .8cm .2cm]{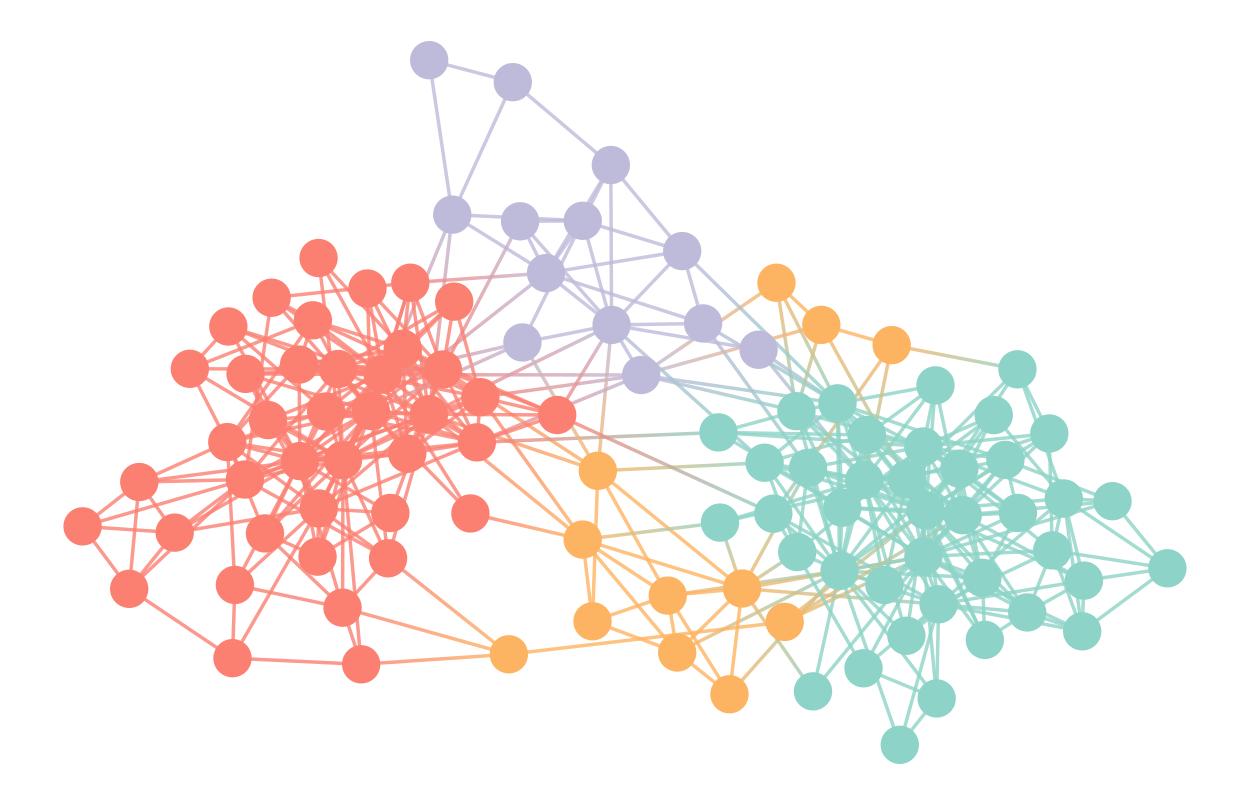}\\
  \end{tabular}

  \caption{Five sampled partitions from Fig.~\ref{fig:label_cycle}, on
  the top panel, with their relabelled counterparts on the bottom panel,
  using the algorithm described in the text, where it becomes possible
  to identify groups consistently according to their
  label (color).\label{fig:polbooks_relabel}}
\end{figure*}

\begin{figure}
  \begin{tabular}{c}
    \begin{overpic}[width=\columnwidth]{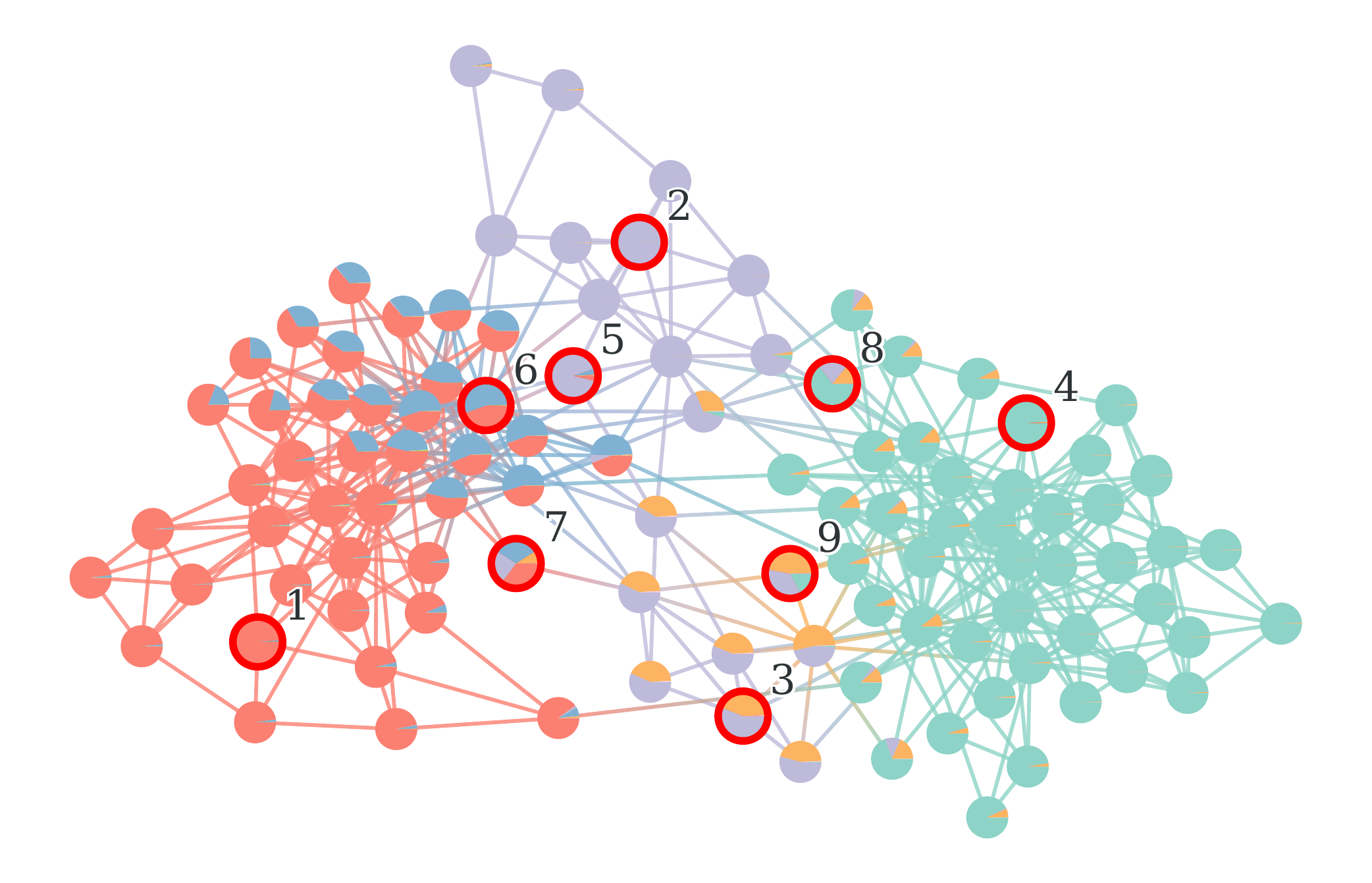}
      \put(0,55){(a)}
    \end{overpic} \\
    \begin{overpic}[width=\columnwidth]{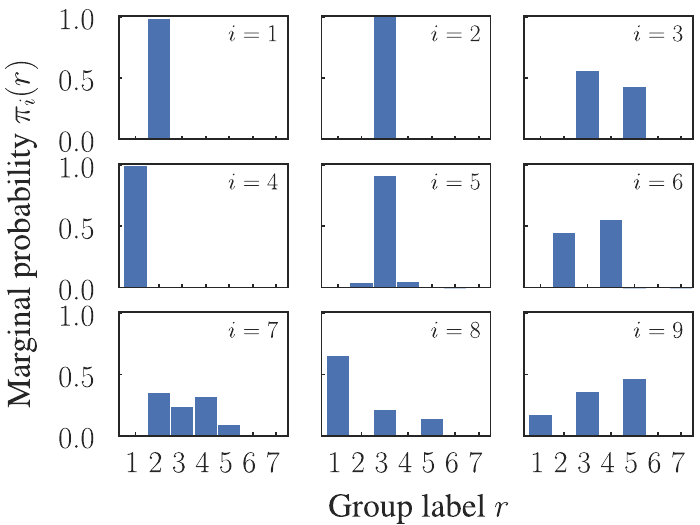}
      \put(0,75){(b)}
    \end{overpic}
  \end{tabular}

  \caption{(a) Marginal posterior group membership distribution on the
  nodes obtained from relabelled partitions for a network of political
  books, the same as in Fig~\ref{fig:label_cycle}, obtained with the
  algorithm described in the text for with $M=10^5$ samples, represented
  as pie diagrams on the nodes. (b) The same distributions for the nodes
  highlighted in red in (a). \label{fig:polbooks_marginal}}
\end{figure}

Computationally, step 3 is the heart of the above algorithm, as it
corresponds to the alignment of each partition with the rest. It takes
time $O[M(N+B^3)]$ in the worst case, where $B$ is the total number of
labels used, since for each partition we need time $O(N)$ to compute the
weights $w_{rs}$, and time $O(B^3)$ to solve the maximum bipartite
weighted matching problem. We can then use the final values of $\{\cc\}$
to easily obtain the marginal probabilities, via
\begin{equation}
  \hat p_i(r) = \underset{p_i(r)}{\operatorname{argmax}}\; P(\bm p|\{\cc\})
  = \frac{1}{M}\sum_{m=1}^{M}\delta_{c_i^m,r}.
\end{equation}
Note that the above procedure is not much more computationally intensive
than obtaining the marginals in the naive way, i.e. directly from
the originally labelled partitions $\bb$, which requires a time $O(MN)$ to
record the label counts. It does, however, require more memory, with a
total $O(MN)$ storage requirement, as we need to keep all $M$ partitions
for the whole duration of the algorithm. In practice, however, we do not
need to perform the whole procedure above for all $M$ partitions, as it
is often sufficient to choose a relatively small subset of them,
provided they give a good representation of the ensemble, and then we
run steps 1 to 4 only on this subset. Based on that, we can simply
process each remaining partition by simply finding its relabelling
$\cc^{(m)}$, updating the global label counts $n_i(r)$, and then discard
the partition. Although this gives only an approximation of the
optimization procedure, we find it works very well in practice, yielding
results that are often indistinguishable from what is obtained with the
full algorithm, while requiring less memory.

In Fig.~\ref{fig:polbooks_relabel} we show the partitions of the
political books network considered in Fig.~\ref{fig:label_cycle}, but
now relabelled according to the algorithm above. Despite groups changing
size and composition, and the appearance and disappearance of groups,
the unique labelling allows us to identify them clearly across
partitions. In Fig.~\ref{fig:polbooks_marginal} these relabellings are
used to obtain marginal distributions on the nodes, where we can say
unambiguously with each frequency a node belongs to a given group.

\subsection{The maximum overlap distance}\label{sec:max-overlap}

The method described in this section serves as a principled way
to disambiguate group labels in an ensemble of partitions, but the ideas
articulated in its derivation also lead us to a way of comparing two
partitions with each other in a general and meaningful way. Consider the
situation where we employ the model above, but we have only $M=2$
partitions. In this case, without loss of generality, we can set one of
them arbitrarily to correspond to the canonical labelling, and we seek
to relabel the second one, by maximizing Eq.~\ref{eq:bip_matching},
which in this case simplifies to
\begin{equation}\label{eq:overlap_ln2}
  \sum_rm_{r,\mu(r)}\ln 2,
\end{equation}
where
\begin{equation}
  m_{rs} = \sum_i\delta_{b_i^{(1)},r}\delta_{b_i^{(2)},s}
\end{equation}
is the so-called contingency table between partitions $\bb^{(1)}$ and
$\bb^{(2)}$, which quantifies how many nodes in group $r$ of $\bb^{(1)}$
belong to group $s$ of $\bb^{(2)}$. Therefore, maximizing
Eq.~\ref{eq:overlap_ln2} is equivalent to finding the bijection $\bm\mu$
so that $\x$ with $x_i=\mu(b_i^{(1)})$ and $\y = \bb^{(2)}$ maximize the
partition overlap
\begin{equation}
  \omega(\x, \y) = \sum_i\delta_{x_i,y_i},
\end{equation}
which counts how many nodes share the same label in both partitions.
Therefore, incorporating our inference procedure leads to the
\emph{maximum overlap distance}
\begin{equation}\label{eq:overlap_dist}
  d(\x, \y) = N - \underset{\bm\mu}{\operatorname{max}}\;\sum_i\delta_{\mu(x_i),y_i}.
\end{equation}
This quantity has a simple interpretation as the minimal classification
error, i.e. the smallest possible number of nodes with an incorrect
group placement in one partition if the other is assumed to be the
correct one. This measure has been considered before in
Refs.~\cite{meila_experimental_2001, meila_comparing_2005,
meila_comparing_2007}, but here we see its derivation based on a
probabilistic generative model. In appendix~\ref{app:max-overlap} we
review some of its useful properties.

\section{Consensus as point estimates}\label{sec:consensus}

The explicit objective of community detection, like any data clustering
method, is to find a partition of the nodes of a network, in a manner
that captures its structure in a meaningful way. However, instead of a
single partition, the inference approach gives us a distribution of
partitions, which ascribes to every possible division of the network a
plausibility, reflecting both our modelling assumptions as well as the
actual structure of the network. In order to convert this information
into a single partition ``point estimate'' we have to be more specific
about what we would consider a successful outcome, or more precisely how
we define the error of our estimate. A consistent scenario is to assume
that our observed network is indeed generated from our model
$P(\A|\bb^*)$ where $\bb^*$ is the true partition we are trying to
find. In order to quantify the quality of our inference we need to
specify an error function $\epsilon(\x,\y)$ that satisfies
\begin{equation}
  \bb^* = \underset{\bb}{\operatorname{argmin}}\;\epsilon(\bb,\bb^*).
\end{equation}
Based on a choice for this function, and since we do not really have
access to the true partition $\bb^*$, our best possible estimate
$\hat\bb$ from the posterior distribution is the one which minimizes the
average error over all possible answers, weighted according to their
plausibility, i.e.
\begin{equation}\label{eq:min-loss}
  \hat\bb =
  \underset{\bb}{\operatorname{argmin}}\;\sum_{\bb'}\epsilon(\bb,\bb')P(\bb'|\A).
\end{equation}
Therefore, it is clear that our final estimate will depend on our choice
of error function $\epsilon(\x,\y)$, and hence is not a property of the
posterior distribution alone. In statistics and optimization literature
the function $\epsilon(\x,\y)$ is called a ``loss function,'' and it
determines the ultimate objective of the inference procedure.

In addition to producing a point estimate $\hat\bb$, it is also useful
for our inference procedure to yield an uncertainty value
$\sigma_{\hat\bb}$, which quantifies how sure we are about the result,
with $\sigma_{\hat\bb}=0$ indicating perfect certainty. Such choices are
not unique, as there is often multiple ways to characterize the
uncertainty or how broad is a distribution. But as we will see, the
choice of the error function allows us to identify what are arguably the
simplest and most direct options.

In the following we to consider simple choices of the error function,
and investigate how they compare to each other in the inference results
they produce.

\subsection{Maximum a posteriori (MAP) estimation}

Arguably the simplest error function we can use is the indicator
function (also called the ``zero-one'' or ``all-or-nothing'' loss)
\begin{equation}
  \epsilon(\x,\y) = 1-\prod_i\delta_{x_i,y_i},
\end{equation}
which would separate the true partition completely from any other,
without differentiating among wrong ones. Inserting this in
Eq.~\ref{eq:min-loss}, we obtain the maximum a posteriori (MAP)
estimator
\begin{equation}
  \hat\bb =
  \underset{\bb}{\operatorname{argmax}}\;P(\bb|\A),
\end{equation}
which is simply the most plausible partition according to the posterior
distribution. The corresponding uncertainty for this estimate is simply
$\sigma_{\hat\bb}=1-P(\bb|\A)$, such that if $\sigma_{\hat\bb}=0$ we are
maximally certain about the result. Despite its simplicity, there are
several problems with this kind of estimation. Namely, the drastic
nature of the error function completely ignores partitions which may be
almost correct, with virtually all nodes correctly classified, except
very few or in fact even one node placed in the incorrect group. We
therefore rely on a very strong signal in the data, where the true
partition is given a plausibility that is larger than any small
perturbation around it, in order to be able to make an accurate
estimation. This puts us in a precarious position in realistic
situations where our data are noisy and complex, and does not perfectly
match our modelling assumptions. Furthermore the uncertainty
$\sigma_{\hat\bb}$ is in most cases difficult to compute, as it involves
determining the intractable sum $P(\A)=\sum_{\bb}P(\A,\bb)$ which serves
as a normalization constant for $P(\bb|\A)$ (although we will consider
approximations for this in Sec.~\ref{sec:evidence}). Even if computed
exactly, typically we will have $\sigma_{\hat\bb}$ approaching the
maximum value of one, since very few networks have a single partition
with a dominating posterior probability.

\subsection{Maximum overlap consensus (MOC) estimation}

As an alternative to the MAP estimation, we may consider a more relaxed
error function given by the overlap distance
\begin{equation}
  \epsilon(\x,\y) = N-\sum_i\delta_{x_i,y_i},
\end{equation}
which counts the number of nodes correctly classified when compared to
the true partition. With this function, from Eq.~\ref{eq:min-loss} we
obtain the maximum marginal estimator
\begin{equation}
  \hat b_i = \underset{r}{\operatorname{argmax}}\;\pi_i(r),
\end{equation}
with
\begin{equation}
  \pi_i(r) = \sum_{\bb}\delta_{b_i,r}P(\bb|\A),
\end{equation}
being the marginal posterior distribution for node $i$. The uncertainty
in this case is then simply the average of the uncertainty for each
node, $\sigma_{\hat\bb}=1-\sum_i\pi_i(\hat b_i)/N$. Since this estimator
considers the average over all partitions instead of simply its maximum,
it incorporates more information from the posterior
distribution. Nevertheless, we encounter again the same problem we have
described before, namely that due to label permutation invariance the
marginal distribution will be identical for every node, and this
estimator will yield in fact useless results. We can fix this problem by
employing instead the \emph{maximum} overlap distance of
Eq.~\ref{eq:overlap_dist} as an error function
$\epsilon(\x,\y)=d(\x,\y)$, leading to the estimator
\begin{equation}
\hat \bb =
  \underset{\bb}{\operatorname{argmax}}\;\sum_{\bb} \underset{\bm\mu}{\operatorname{max}}\;\sum_i\delta_{\hat{b}_i,\mu(b_i)}P(\bb|\A).
\end{equation}
Performing the maximization yields now a set of self-consistent
equations,
\begin{equation}
  \hat b_i = \underset{r}{\operatorname{argmax}}\;\pi'_i(r|\{\mu_{\bb}\}),
\end{equation}
with the marginal distributions obtained over the relabeled partitions,
\begin{equation}\label{eq:center_marginals}
  \pi_i'(r|\{\mu_{\bb}\}) = \sum_{\bb}\delta_{\mu_{\bb}(b_i),r}P(\bb|\A),
\end{equation}
where the relabeling is done in order to maximize the overlap with $\hat\bb$
\begin{equation}
  \bm\mu_{\bb} = \underset{\bm\mu}{\operatorname{argmax}} \sum_i\delta_{\hat b_i, \mu(b_i)}.
\end{equation}
Like before, the uncertainty is given by
$\sigma_{\hat\bb}=1-\sum_i\pi_i'(\hat{b}_i|\{\mu_{\bb}\})$. In practice
we implement this estimator by sampling a set of $M$ partitions
$\{\bb\}$ from the posterior distribution and then performing the double
maximization
\begin{align}
  \hat b_i &= \underset{r}{\operatorname{argmax}}\;\sum_m \delta_{\mu_m(b^m_i), r}\label{eq:oc_marginal}\\
  \bm\mu_m &= \underset{\bm\mu}{\operatorname{argmax}} \sum_r\hat m_{r,\mu(r)}^{(m)}, \label{eq:oc_matching}
\end{align}
where in the last equation we have that $\hat m_{rs}^{(m)} =
\sum_i\delta_{b_i^m,r}\delta_{\hat b_i,s}$ is the contingency table
between $\bb^{(m)}$ and $\hat\bb$. The solution of
Eq.~\ref{eq:oc_marginal} is obtained by simply counting how often each
label appears for each node and then extracting the label with the
largest count, and Eq.~\ref{eq:oc_matching} is once more an instance of
the maximum bipartite weighted matching problem. The overall solution
can be obtained by simple iteration, starting from an arbitrary choice
of $\hat\bb$, and then alternating between the solution of equation
Eq.~\ref{eq:oc_matching} and using its result to solve
Eq.~\ref{eq:oc_marginal}, until $\hat\bb$ no longer changes. This
guarantees a local optimum of the optimization problem, but not
necessarily a global one, therefore this algorithm needs to be repeated
multiple times with different initial conditions, and the best result
kept. Since it involves the relabelling over all $M$ partitions, the
overall algorithmic complexity of a single iteration is $O(MNB
+ MB^3)$.

Note that the marginal distributions obtained via
Eq.~\ref{eq:center_marginals} with the MOC estimator are not necessarily
the same as those obtained by inferring the random label model
considered previously. This is because while the MOC calculation
attempts to find a single partition with a maximum overlap to all
samples, inferring the random label model amounts to finding the most
likely marginal distribution compatible with all samples, irrespective
of its maximum. Although in many cases these two calculations will give
similar answers, they are not equivalent.

\subsection{Error functions based on the contingency table}

In principle, we can make a variety of other choices for error
functions. A particular class of them are those based on the contingency
table between partitions, using concepts from information theory. These
error functions are not based on an explicit labeling or alignment of
partitions, but instead focus on the joint probability of labels in both
partitions being compared. A popular function of this kind is the
variation of information (VI)~\cite{meila_comparing_2003}, which is defined as
\begin{equation}
  \text{VI}(\x,\y) = -\frac{1}{N}\sum_{rs}m_{rs}\left[\ln\frac{m_{rs}}{n_r} + \ln\frac{m_{rs}}{n_s'}\right],
\end{equation}
with $m_{rs}=\sum_i\delta_{x_i,r}\delta_{y_i,s}$ being the contingency
table between $\x$ and $\y$, $n_r=\sum_sm_{rs}$ and $n'_s=\sum_rm_{rs}$
are the group sizes in both partitions. We can use VI as an error
function by setting
\begin{equation}
  \epsilon(\x,\y) = \text{VI}(\x,\y).
\end{equation}
As detailed in
Ref.~\cite{meila_comparing_2003}, VI is a dissimilarity function that
fulfills many desirable formal properties, including triangle
inequality, making it a proper metric distance (like the maximum overlap
distance). Another possible alternative consists of using the reduced
mutual information (RMI)~\cite{newman_improved_2020}, as done by Riolo
and Newman~\cite{riolo_consistency_2020}, with
\begin{equation}
  \epsilon(\x,\y) = -\text{RMI}(\x,\y)
\end{equation}
where
\begin{equation}
  \text{RMI}(\x,\y) = \frac{1}{N}\left[\ln \frac{N!\prod_{rs}m_{rs}!}{\prod_rn_r!\prod_sn_s'!}-\ln\Omega(\bm n, \bm n')\right],
\end{equation}
with $\Omega(\bm n, \bm n')$ being the total number of contingency
tables with fixed row and column sums, which we omit here for brevity
(see Ref.~\cite{newman_improved_2020} for asymptotic
approximations). The negative sign used in the definition of
$\epsilon(\x,\y)$ is because RMI is a similarity function, which takes
its maximum value when $\x$ and $\y$ are identical, unlike VI, which is
a dissimilarity that takes its minimum value of zero in the same
case. RMI can be seen as a correction to mutual information, which fails
as an appropriate similarity function in key cases. It is based on a
nonparametric minimum description length encoding of both partitions,
which quantifies the amount of information required to describe them if
the contingency table is known, together with the necessary information
required to describe the contingency table itself.

In either of the above cases, our point estimate $\hat\bb$ consists of
minimizing the sum of the error function over $M$ samples from the
posterior distribution, according to Eq.~\ref{eq:min-loss}. Unlike the
indicator and the maximum overlap distance, the above loss functions are
more cumbersome to optimize, with the overall optimization
itself amounting to a nonconvex clustering problem of its own. Therefore
we can use some of the same algorithms we use to perform community
detection in the first place, with a good choice being the merge-split
MCMC of Ref.~\cite{peixoto_merge-split_2020}, which we have used in our
analysis.

\subsection{Consensus point estimates are inconsistent for heterogeneous distributions}

Our aim is not to list or perform an exhaustive comparison between all
possible error functions, but instead to focus on the fact that they do
not always yield the same answer. Although there is only one way with
which all partitions in an ensemble can be identical, there are many
ways in which they can be different. While the various error functions
allow us to extract a form of consensus between differing partitions,
they each achieve this based on different features of the
population. Therefore, for partition ensembles with sufficiently strong
heterogeneity, the different estimators may give conflicting answers.
Such a disagreement can signal an arbitrariness in the inference
procedure, and our inability of summarizing the population in a simple
manner. We illustrate this problem with a few simple examples.

\begin{figure}
  \includegraphics[width=\columnwidth]{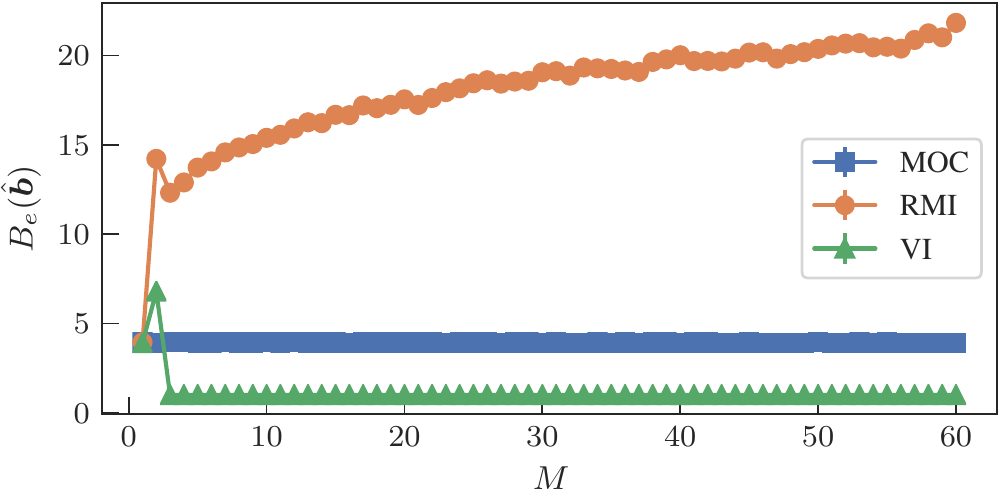}

  \caption{Effective number of groups $B_e(\hat\bb)$ for the consensus
  estimate $\hat\bb$ obtained for $M$ random partitions of $N=100$ nodes
  into into $B=4$ groups, according to the different error functions as
  indicated in the legend. The results were obtained by averaging over
  $50$ realizations.\label{fig:random-consensus}}
\end{figure}

We consider first a simple artificial scenario with strong
heterogeneity, composed of $M$ independently sampled partitions of $N$
nodes, where to each node is sampled a group label uniformly at random
from the interval $[1,B]$. Indeed, in this example there is no real
consensus at all between partitions. Intuitively, we might expect the
estimated consensus between such fully random partitions to be a sort of
``neutral'' partition, in the same way the average of a fully isotropic
set of points in Cartesian space will tend towards the origin. However,
all consensus estimators considered previously behave very differently
from each other in this example. In Fig.~\ref{fig:random-consensus} we
compare the effective number of groups $B_e(\hat\bb)=\ee^{S}$ obtained
for each point estimate, with
\begin{equation}
  S = -\sum_r\frac{n_r}{N}\ln\frac{n_r}{N},
\end{equation}
being the group label entropy. Arguably, the estimator that behaves the
closest to the intuitive expectation just mentioned is VI, which for
$M>2$ yields a consensus partition composed of a single group,
$B_e(\hat\bb)=1$. The MOC estimator yields instead a partition into
$B_e(\hat\bb)=4$ groups, which itself is hard to distinguish from a
random partition sampled from the original ensemble. This is because the
marginal distributions obtained by Eq.~\ref{eq:center_marginals} will be
close to uniform, even after the label alignments of
Eq.~\ref{eq:oc_matching} are achieved, such that the maximum chosen by
Eq.~\ref{eq:oc_marginal} will be determined by small quenched
fluctuations in the partition ensemble. Finally, the RMI estimate yields
consensus partitions with a number of groups that increases with the
number of samples $M$. This is because the RMI estimate tends to find
the overlaps between partitions, i.e. sets of nodes that tend to occur
together in the same group across many
partitions~\cite{riolo_consistency_2020}. In our random case, two nodes
belong to the same group due to pure coincidence, therefore the
probability of this happening for a large set of nodes decreases for
larger $M$, thus making the overlapping sets progressively smaller, and
leading to a larger number of groups in the consensus. Inspecting any of
the obtained point estimates in isolation, it would be difficult to get
a coherent picture of the underlying ensemble, since none of them allow
us to distinguish between an ensemble concentrated on the point
estimate, or the maximally heterogeneous situation we have just
considered. If we would consider instead the uncertainty of the MOC
estimate (which yields $\sigma_{\hat\bb}\approx .69$ for $M\to\infty$),
or even more explicitly the marginal distributions of
Eq.~\ref{eq:center_marginals} (or those of the inferred random label
model of Sec.~\ref{sec:random-label}), we would see that they are very
broad, matching closely the true random distribution. But nevertheless,
none of the point estimates by themselves can reveal this information.

\begin{figure}
  \begin{tabular}{c}
    \includegraphics[width=\columnwidth]{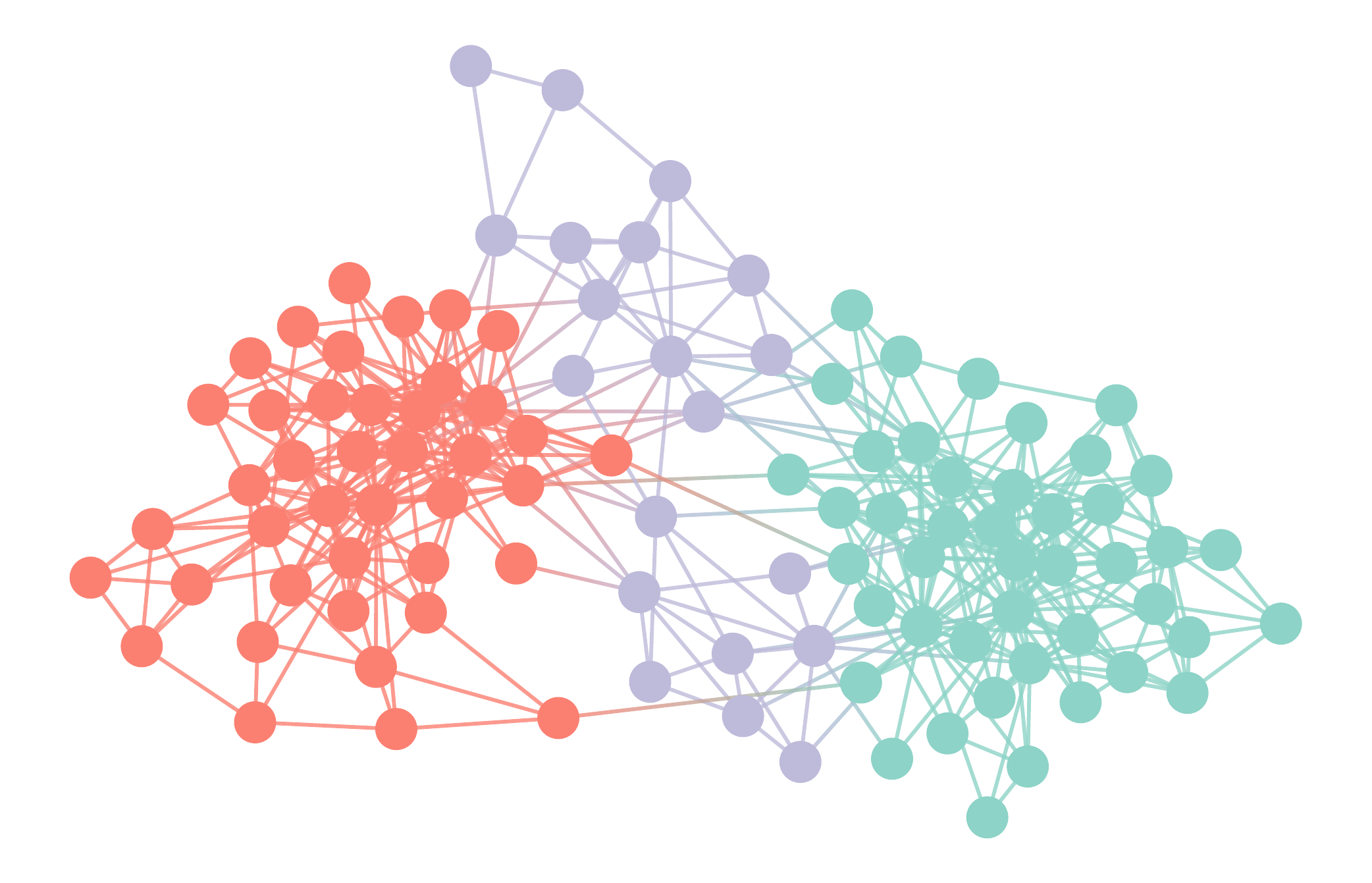}\\
    MAP ($\sigma_{\hat\bb}=0.99988$) and VI estimates\\
    \includegraphics[width=\columnwidth]{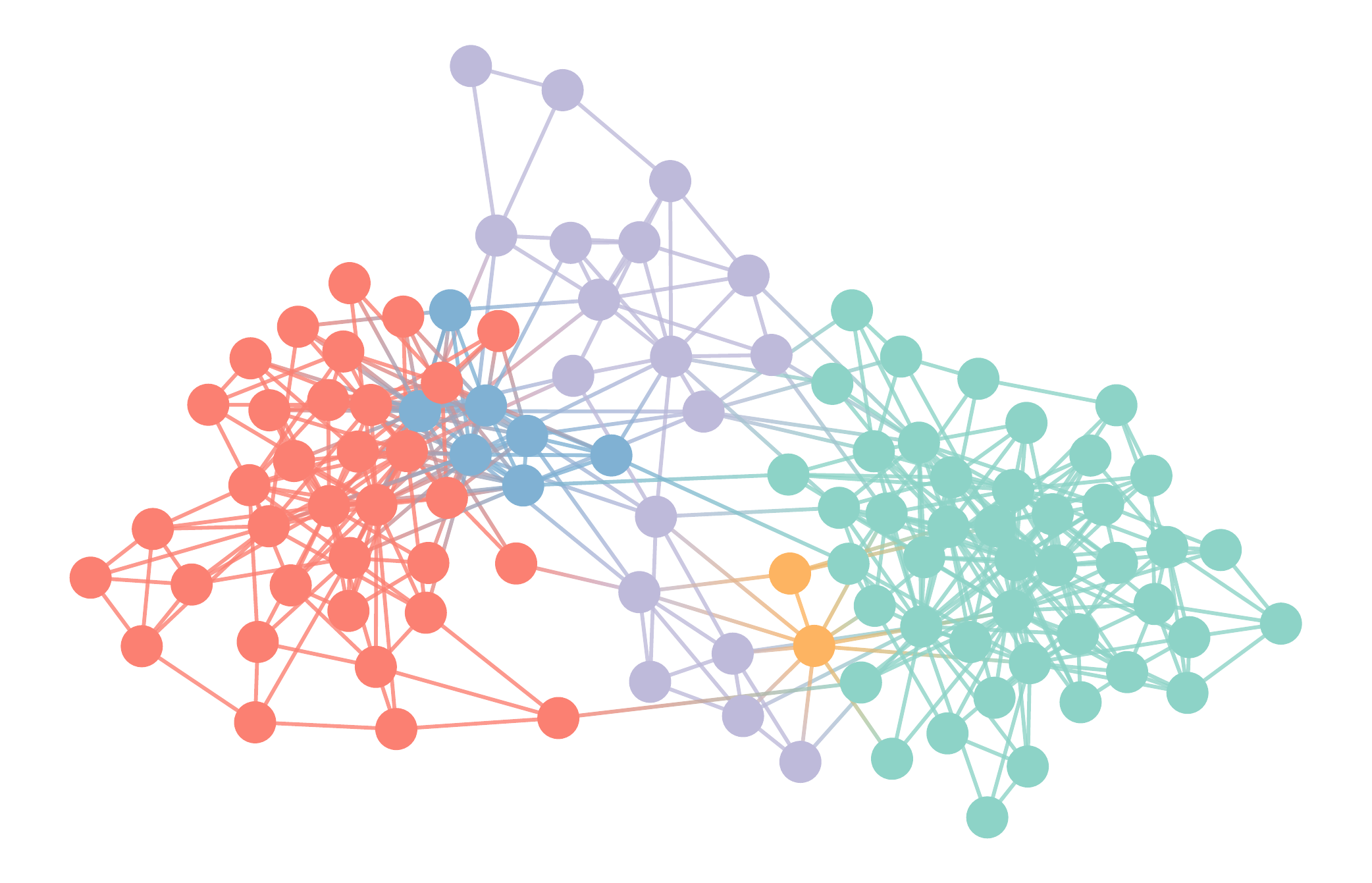}\\
    MOC estimate ($\sigma_{\hat\bb}=0.15$)\\
    \includegraphics[width=\columnwidth]{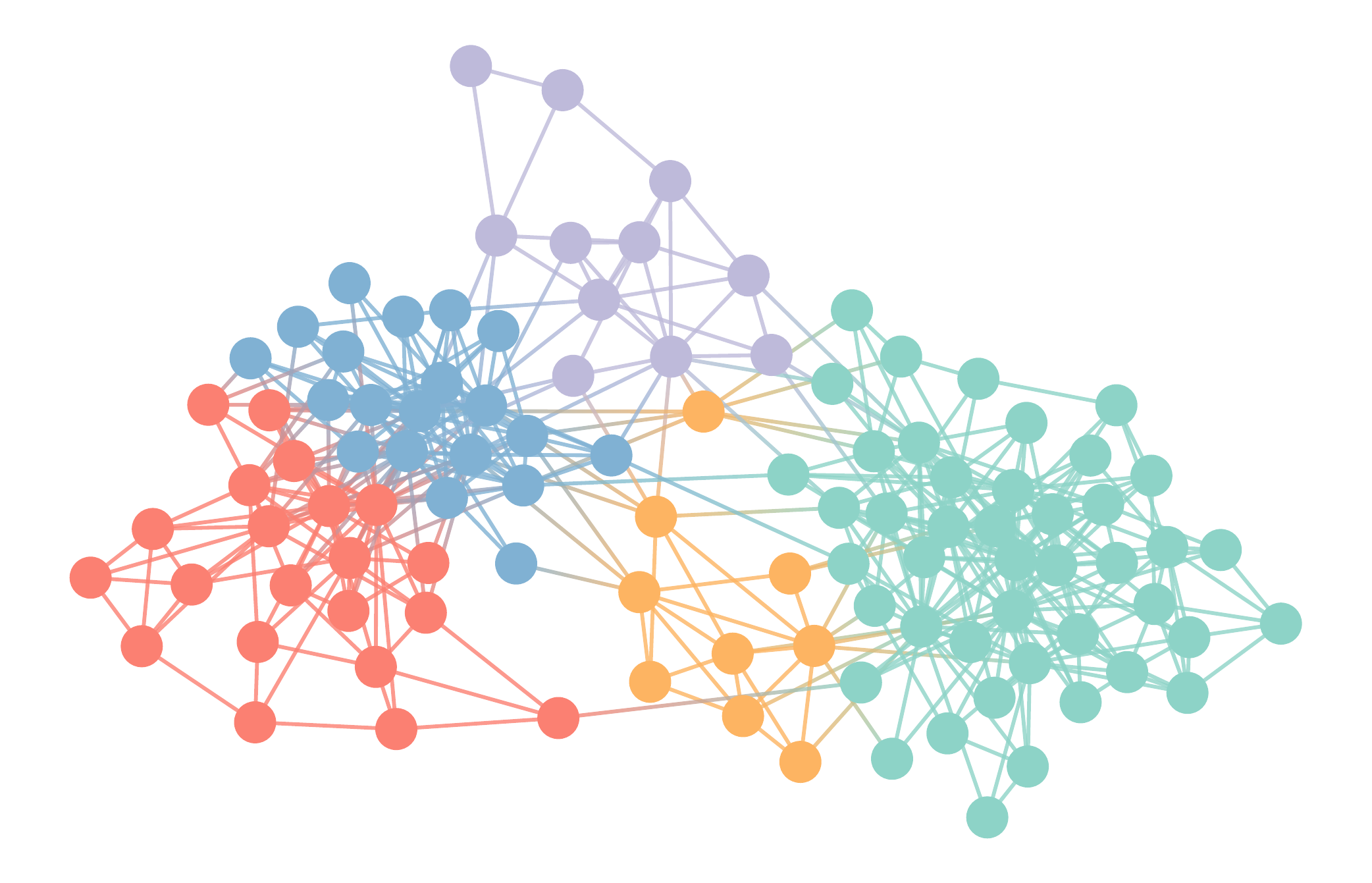}\\
    RMI estimate
  \end{tabular} \caption{Inference of the community structure of the
  political books network, according to the the DC-SBM and using the
  different estimators as shown in the legend. For the VI/MAP estimate
  (top panel), the three groups can be interpreted, from left to right,
  as ``liberal'', ``neutral'' and
  ``conservative''.\label{fig:polbooks-centers}}
\end{figure}

We further illustrate the discrepancy issue with a more realistic
example where we can see both agreements and disagreements between the
different estimates. In Fig.~\ref{fig:polbooks-centers} we show the
estimates obtained for the same political books network considered
previously, again using the DC-SBM to obtain a posterior distribution of
partitions. We observe, rather curiously, that the MAP estimate
coincides perfectly with the VI estimate, but gives a different result
from the MOC and RMI estimates. The MAP/VI estimates separate the
network into three groups, which in this context can be understood as
types of books describing ``liberal'' and ``conservative'' politics, and
``neutral'' books not taking any side. The MOC estimate further divides
the ``liberal'' category into a new subgroup, and somewhat strangely at
first, singles out two ``neutral'' books into their own category. As can
be seen in Fig.~\ref{fig:polbooks_marginal}, the reason for this is that
the posterior distribution exhibits a possible subdivision of the
neutral group into two, however it is only these two nodes that happen
to belong to this subdivision with the highest probability. The RMI
estimate also yields a division into five groups, but the two extra
groups have a larger size when compared to the MOC result. In view of
the behavior seen for the fully random example considered earlier, the
discrepancies raise some doubts about what is the most faithful
division. Are the MOC and RMI arbitrary divisions due to the randomness
of the posterior distribution or do they point to a meaningful summary?
Are the MAP/VI estimates being too conservative about the structure of
the posterior distribution?

\begin{figure}
  \begin{tabular}{cc}
    \includegraphics[width=.5\columnwidth]{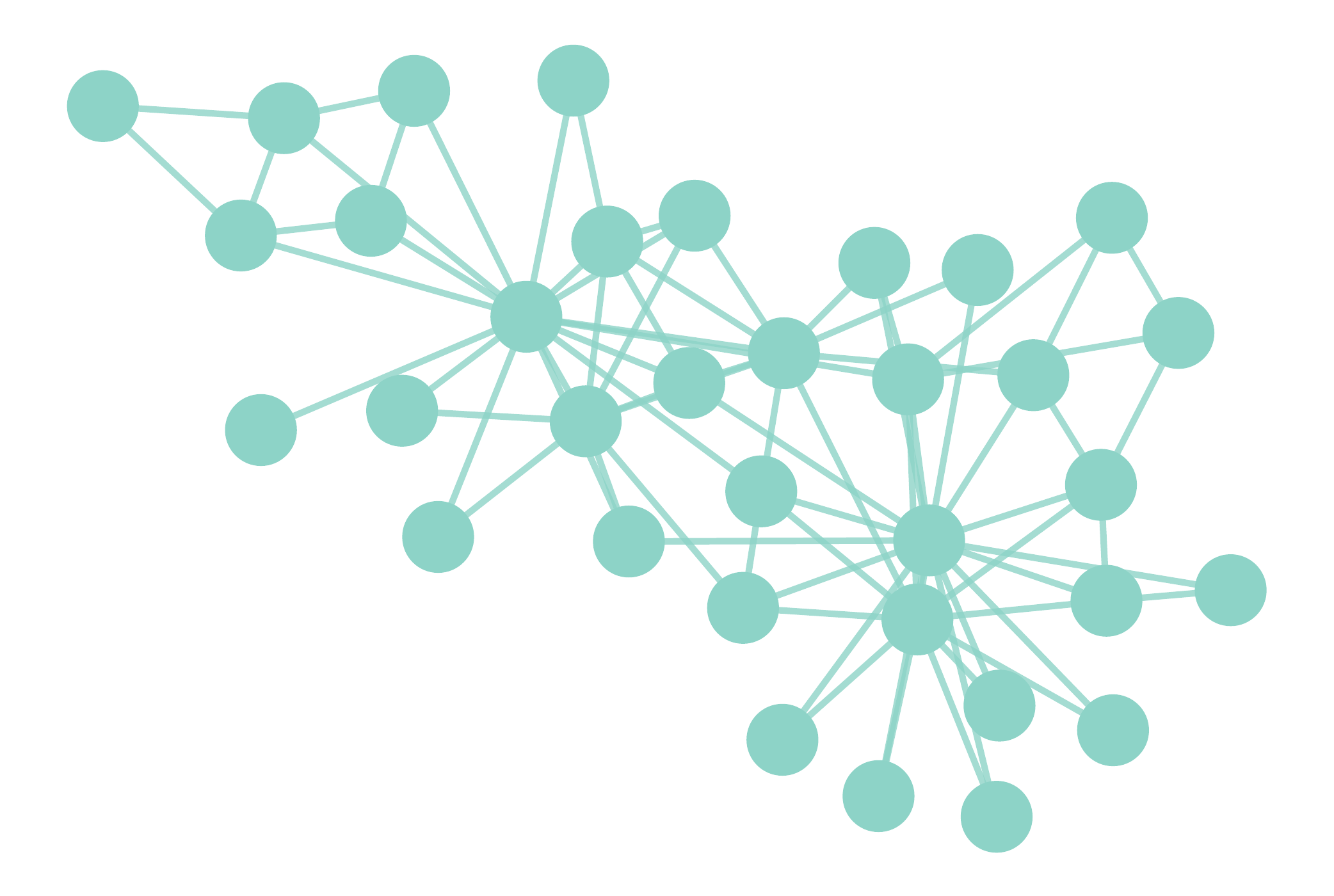}&
    \includegraphics[width=.5\columnwidth]{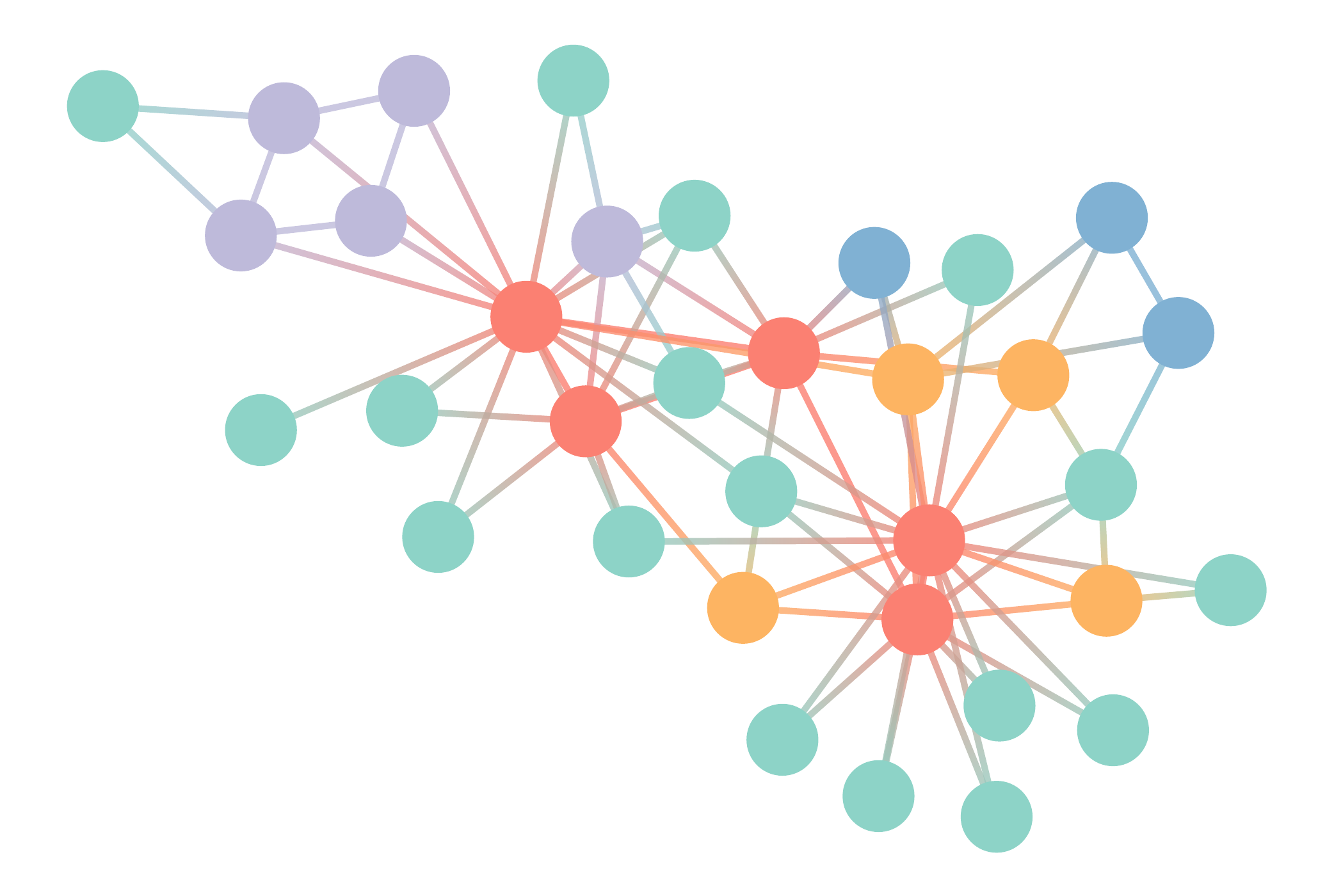}\\
    \begin{minipage}{.5\columnwidth}
      MAP ($\sigma_{\hat\bb}=0.51$), MOC ($\sigma_{\hat\bb}=0.18$) and VI estimates
    \end{minipage} &
    RMI estimate
  \end{tabular} \caption{Inference of the community structure of
  Zachary's karate club network, according to the DC-SBM and using the
  different estimators as shown in the
  legend.\label{fig:karate-centers}}
\end{figure}

With some other networks, the discrepancy between estimators can be even
stronger, making such questions even harder to answer. In
Fig.~\ref{fig:karate-centers} we show the results for Zachary's karate
club network~\cite{zachary_information_1977}, again using the DC-SBM. In
this case, the MAP, MOC and VI estimators yield the same division of the
network into a single group, whereas the RMI estimate yields a partition
into five groups, following no clear pattern. None of the estimates
resemble the putative division for this network in two
assortative communities.

Despite the partial agreement between some of the estimates in the
examples above, the disagreements still raise obvious interpretation
questions. Here we argue that this discrepancy cannot be resolved simply
by trying alternative ways to form a consensus, since trying to
summarize a whole distribution with a point estimate is in general an
impossible task, and therefore we need instead a way to also
characterize the \emph{dissensus} between partitions, by exposing also
the existing heterogeneity of the posterior distribution.

To some extent, the characterization of dissensus is already achieved by
the random label model of Sec.~\ref{sec:random-label}, since it attempts
to describe the posterior distribution via marginal probabilities,
rather than just a point estimate, and therefore can convey how
concentrated it is. However, because this model assumes the group
membership of each node to be independent, it still hides a significant
fraction of the potential heterogeneity in the ensemble, which can come
from the correlation between these memberships. In the next section we
will generalize this approach to the situation where the posterior
distribution is multimodal, so that multiple consensuses are
simultaneously possible. We will see how this allows us to extract a
more complete and coherent picture of distributions of partitions.

\section{Extracting dissensus between partitions}\label{sec:modes}

We aim to characterize the discrepancy between partitions by considering
the possibility of several possible consensuses that only exist between
a subset of the partitions. This corresponds to the situation where the
inference procedure can yield substantially different explanations for
the same network. We do so by modelling the posterior distribution of
partitions with a mixture model, where each partition can belong to one
of $K$ clusters --- which we call ``modes'' to differentiate from the
groups of nodes in the network. Inside each mode the partitions are
generated according to the same random label model considered before,
but with different parameters. More specifically, a partition $\bb$ is
sampled according to
\begin{equation}
  P(\bb|\bm p, \w) = \sum_k P(\bb|\bm p, k)P(k|\w)
\end{equation}
where
\begin{equation}
  P(k|\w) = w_k
\end{equation}
is the relative size of mode $k$, with $\sum_kw_k=1$, and inside a mode
$k$ the partitions are sampled according to the random label model,
\begin{equation}
  P(\bb|\bm p, k) = \sum_{\cc}P(\bb|\cc) P_{\text{MF}}(\cc|\bm p, k)
\end{equation}
with the hidden labels generated according to
\begin{equation}
  P_{\text{MF}}(\cc|\bm p, k) = \prod_i p_i^{(k)}(c_i),
\end{equation}
where $p_i^{(k)}(r)$ is the probability that a node $i$ has group label
$r$ in mode $k$, and finally a random label permutation chosen uniformly
at random,
\begin{equation}
  P(\bb|\cc) = \frac{[\bb\sim\cc]}{q(\bb)!}.
\end{equation}
Naturally, we recover the original random label model for $K=1$.

We perform the inference of the above model by considering the mode
label $k$ as a latent variable, which yields a joint probability
together with the original and relabelled partitions
\begin{equation}
  P(\bb,\cc, k|\bm p, \w) = P(\bb|\cc)P(\cc|\bm p, k)P(k|\w).
\end{equation}
If we now observe $M$ partitions $\{\bb\} =
\{\bb^{(1)},\dots,\bb^{(M)}\}$ sampled from the SBM posterior
distribution, we assume that each one has been sampled from one of the
$K$ modes, so that for each observed partition $\bb_{m}$ we want to
infer its relabelled counterpart together with its originating mode,
i.e. $(\cc^{(m)},k)$. The joint posterior distribution for these pairs,
together with the total number of modes $K$, and the number of groups
$\bm B=\{B_k\}$ in each mode, is given by
\begin{multline}
  P(\{\cc,k_m\}, \bm B, K|\{\bb\}) =\\
  \frac{P(\{\bb\}|\{\cc\})P(\{\cc\}|\bm k,\bm B)P(\bm B)P(\bm k|K)P(K)}{P(\{\bb\})},
\end{multline}
where the relabelling probability is given by
\begin{equation}
  P(\{\bb\}|\{\cc\}) = \prod_mP(\bb^{(m)}|\cc^{(m)})
\end{equation}
and with the marginal likelihood obtained by integrating over all possible
probabilities $\bm p$ for each mode,
\begin{align}
  P(\{\cc\}|\bm k,\bm B) &= \prod_k\int \left[\prod_m P(\cc^{(m)}|\bm p, k_m)^{\delta_{k_m,k}}\right]P(\bm p)\;\dd\bm p\\
  &= \prod_k\prod_i \frac{(B_k-1)!}{(M_k+B_k-1)!}\prod_r n_i^{(k)}(r)!,
\end{align}
with $M_k = \sum_m \delta_{k_m,k}$ being the number of samples that
belong to mode $k$, $B_k$ the total number of group labels in mode $k$,
and $n_i^{(k)}(r) = \sum_m \delta_{c_i^m,r}\delta_{k_m,k}$ are the
marginal label counts in mode $k$, and finally the prior mode
distribution is obtained by integrating over all possible mode mixtures
$\w$,
\begin{align}
  P(\bm k|K) &= \int \left[\prod_m P(k_m|\w)\right]P(\w|K)\;\dd\w,\\
  &=\frac{(K-1)!}{(M + K -1)!}\prod_k M_k!.
\end{align}
where we used once more an uninformative prior
\begin{align}
  P(\w|K) = (K-1)!.
\end{align}
For the total number of modes $K$ we use a uniform prior $P(K)\propto
1$, which has no effect in resulting inference.  With this posterior in
place, we can find the most likely mode distribution with a clustering
algorithm that attempts to maximize it. We do so by starting with an
arbitrary initial placement of the $M$ partitions into modes, and
implementing a greedy version of the merge-split algorithm of
Ref.~\cite{peixoto_merge-split_2020} that chooses at random between the
following steps, and accepting it only if increases the posterior
probability:
\begin{enumerate}
  \item A random partition $\bb^{(m)}$ is moved from its current mode to a
    randomly chosen one, including a new mode.
  \item Two randomly chosen modes are merged into one, reducing the
        total number of modes.
  \item A randomly chosen mode is split into two, increasing the total
        number of modes. The division itself is chosen by a surrogate
        greedy algorithm, which tries one of the following strategies at
        random:
        \begin{enumerate}
          \item Start with a random split of the modes into two, and
                attempt to move each sample in random sequence between
                the two modes if the move increases the posterior
                probability, and stop when no improvement is possible.
         \item Start with each of the samples in their own modes, with a
               single sample each, and place them in sequence in two new
               modes that are initially empty, according to the choice
               with the largest posterior probability.
         \item Start with all samples in a single mode, and proceed
               like in strategy (b).
        \end{enumerate}
   \item Two randomly chosen modes are merged into one, and then split
         like in option 3, preserving the total number of modes.
\end{enumerate}
The algorithm stops whenever further improvements to the posterior
cannot be made. In the above, whenever a sample $m$ is placed into a
mode $k$, its hidden labelling $\cc^{(m)}$ is obtained by maximizing the
conditional posterior probability,
\begin{equation}
  P(\cc^{(m)} | \{\bb\}, \{\cc^{(m'\ne m)}\}, B_k, k) \propto \prod_i\prod_r \left[n_i'(r|k)+1\right]^{\delta_{c_i^m,r}},
\end{equation}
where $n_i'(r|k)=\sum_{m'\ne m}\delta_{k_m',k}\delta_{c_i^{m'},r}$ is
the label count of node $i$ considering all samples belonging to mode
$k$, excluding $\cc^{(m)}$. Like in the original random label model,
this maximization is performed by solving the corresponding maximum
bipartite weighted matching problem with the Kuhn–Munkres algorithm in
time $O(N + B^3)$, where $B$ is the number of partition labels
involved. Overall, a single ``sweep'' of the above algorithm, where each
sample has been moved once, is achieved in time $O[M(N+B^3)]$. For the
choice of $M$ itself, this in general will depend on the structure of
the data. The general guideline is that $M$ should be large enough so
that if it is increased the inference results (i.e. number of modes and
their composition) no longer change. A good strategy is to make $M$ as
large as the initial computational budget allows, and then compare the
results with a \emph{smaller} choice of $M$, and then evaluate if the
results are the same. In terms of practical speed, when compared e.g. to
sampling partitions from the SBM posterior via MCMC, we find that
performing the overall clustering algorithm is most often substantially
faster than generating the partitions in the first place.

After we have found the mode memberships $\bm k$, the mode fractions can
be estimated as
\begin{align}
  w_k = \frac{M_k}{M}.
\end{align}
This is interpreted as the relative posterior plausibility of each mode
serving as an alternative explanation for the data.

\begin{figure*}
  \setlength{\tabcolsep}{0em}\renewcommand{\arraystretch}{0}
  \begin{tabular}{cccc}
    \includegraphics[width=.24\textwidth]{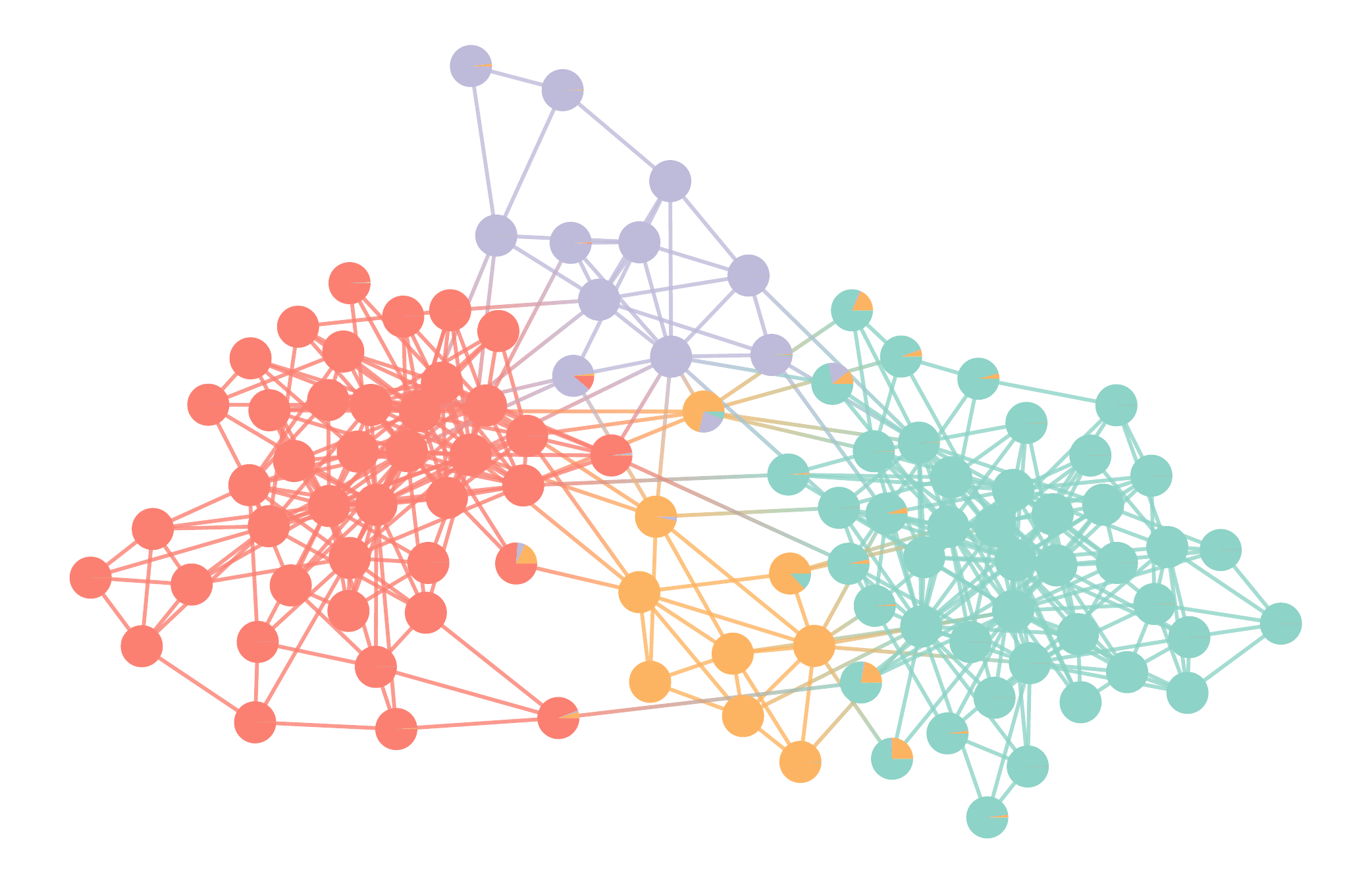}
    &
    \includegraphics[width=.24\textwidth]{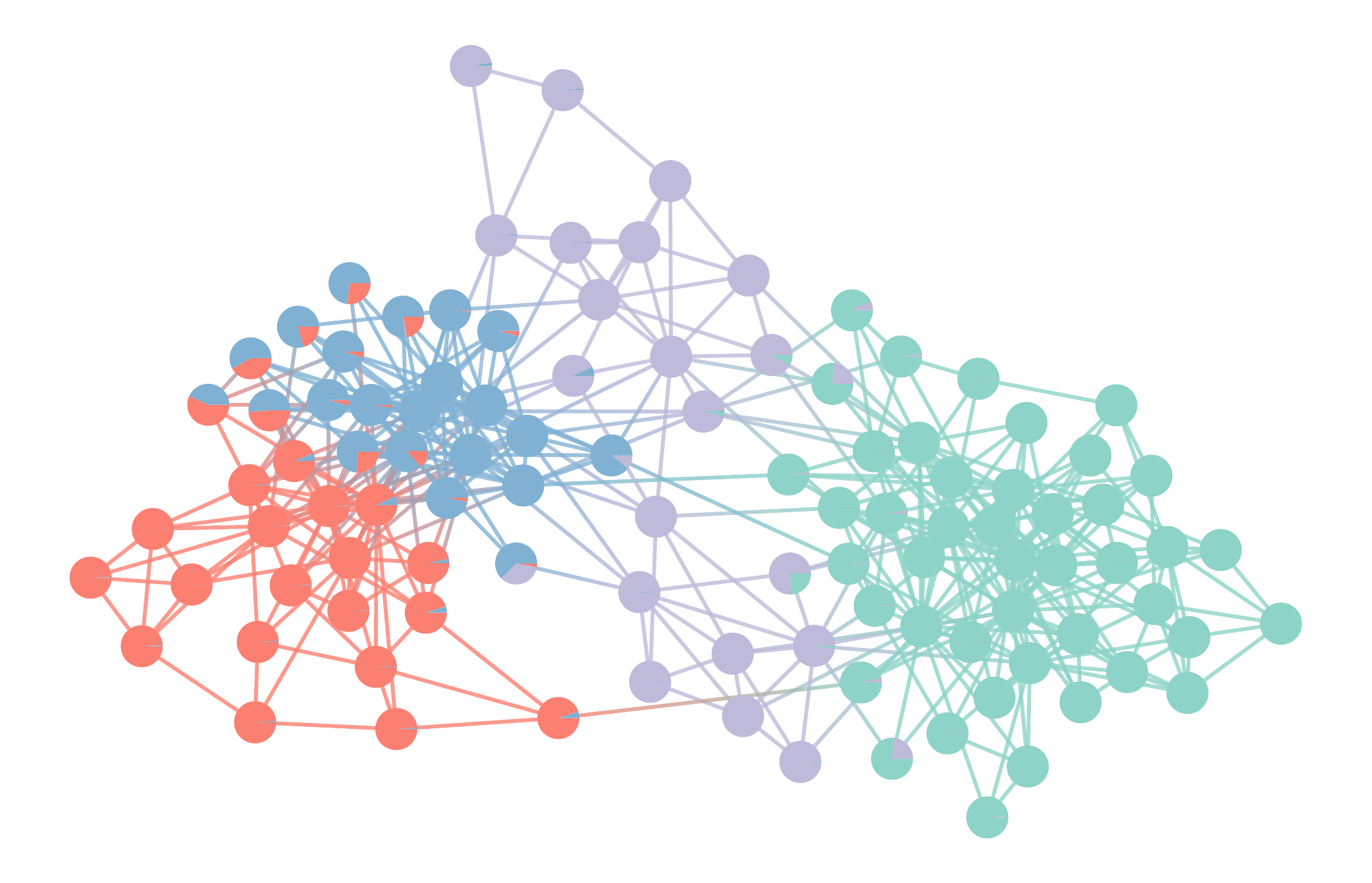}
    &
    \includegraphics[width=.24\textwidth]{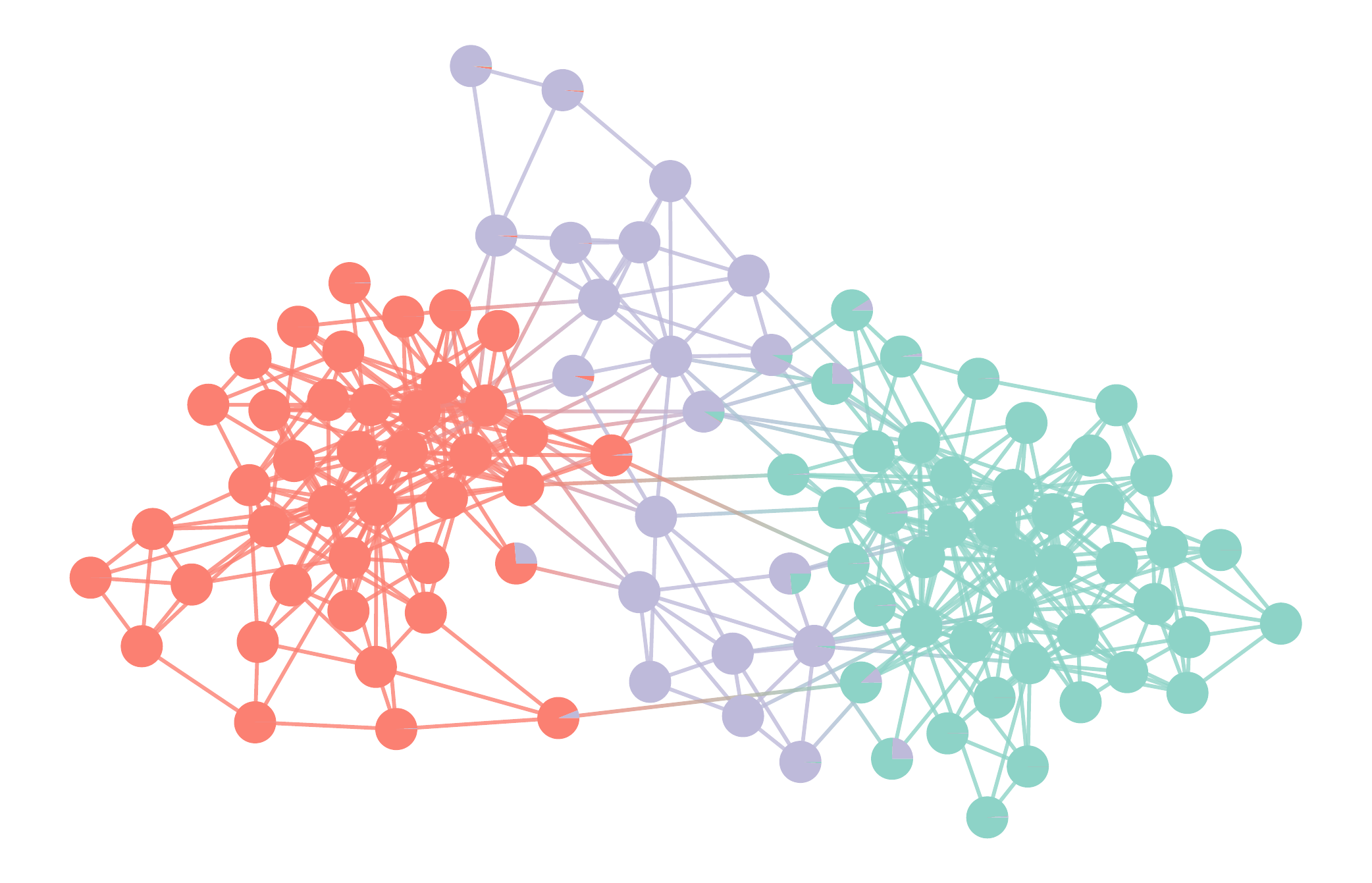}
    &
    \includegraphics[width=.24\textwidth]{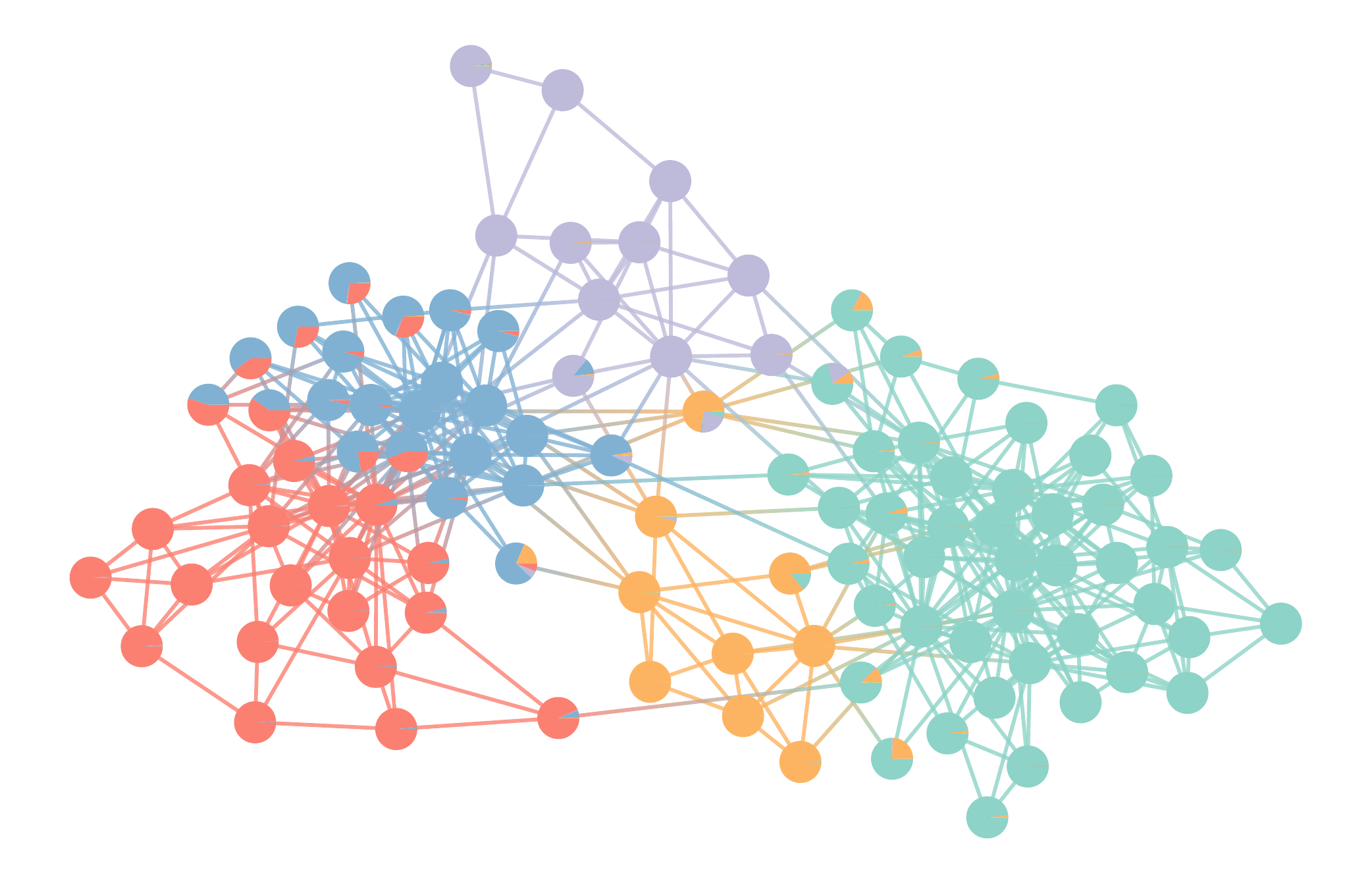}\\
    (a) $w_1 = 0.232$, $\sigma_{\hat\bb}=0.021$ &
    (b) $w_2 = 0.223$, $\sigma_{\hat\bb}=0.045$&
    (c) $w_3 = 0.134$, $\sigma_{\hat\bb}=0.016$&
    (d) $w_4 = 0.132$, $\sigma_{\hat\bb}=0.053$\\
    \includegraphics[width=.24\textwidth]{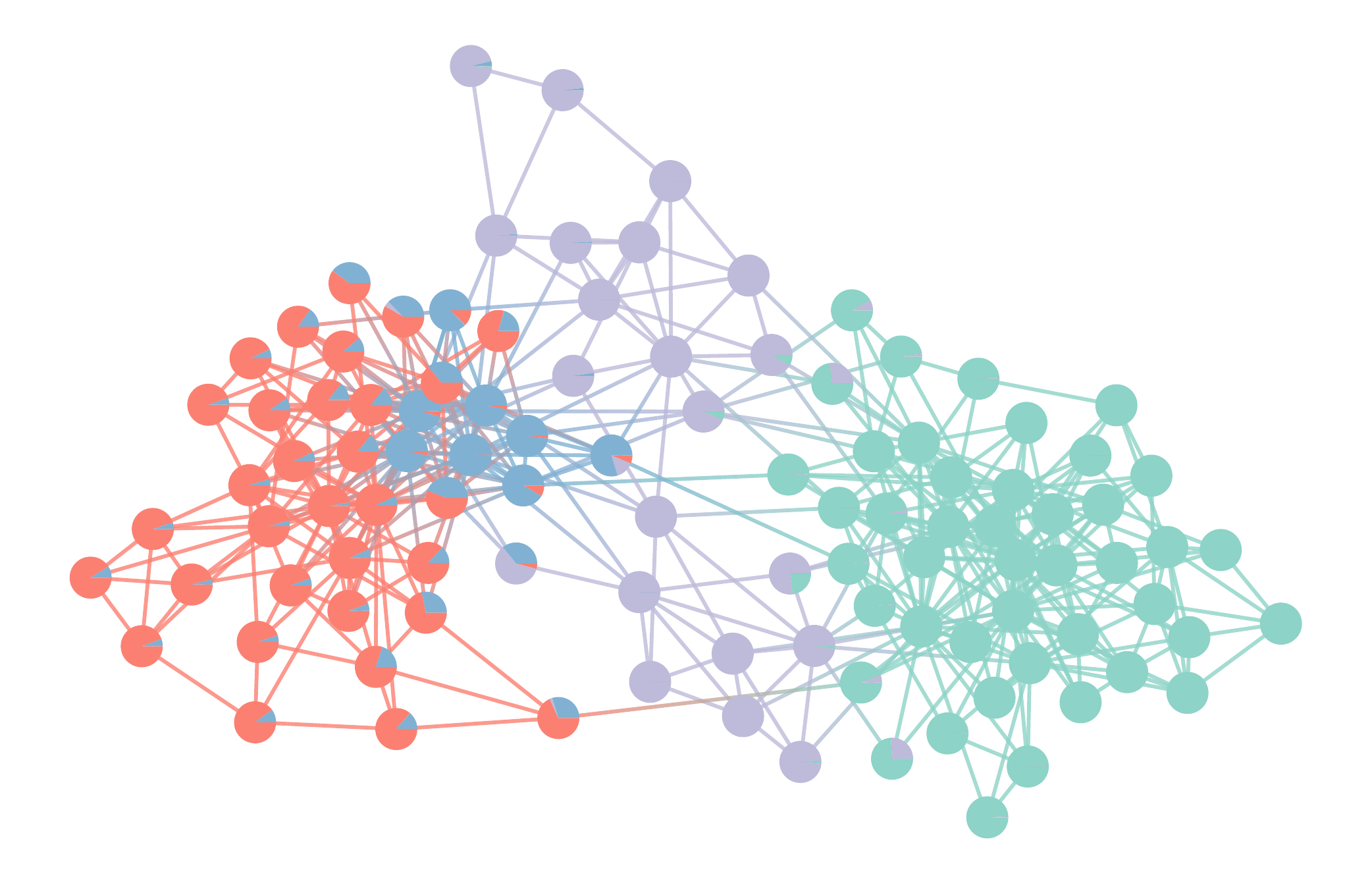}
    &
    \includegraphics[width=.24\textwidth]{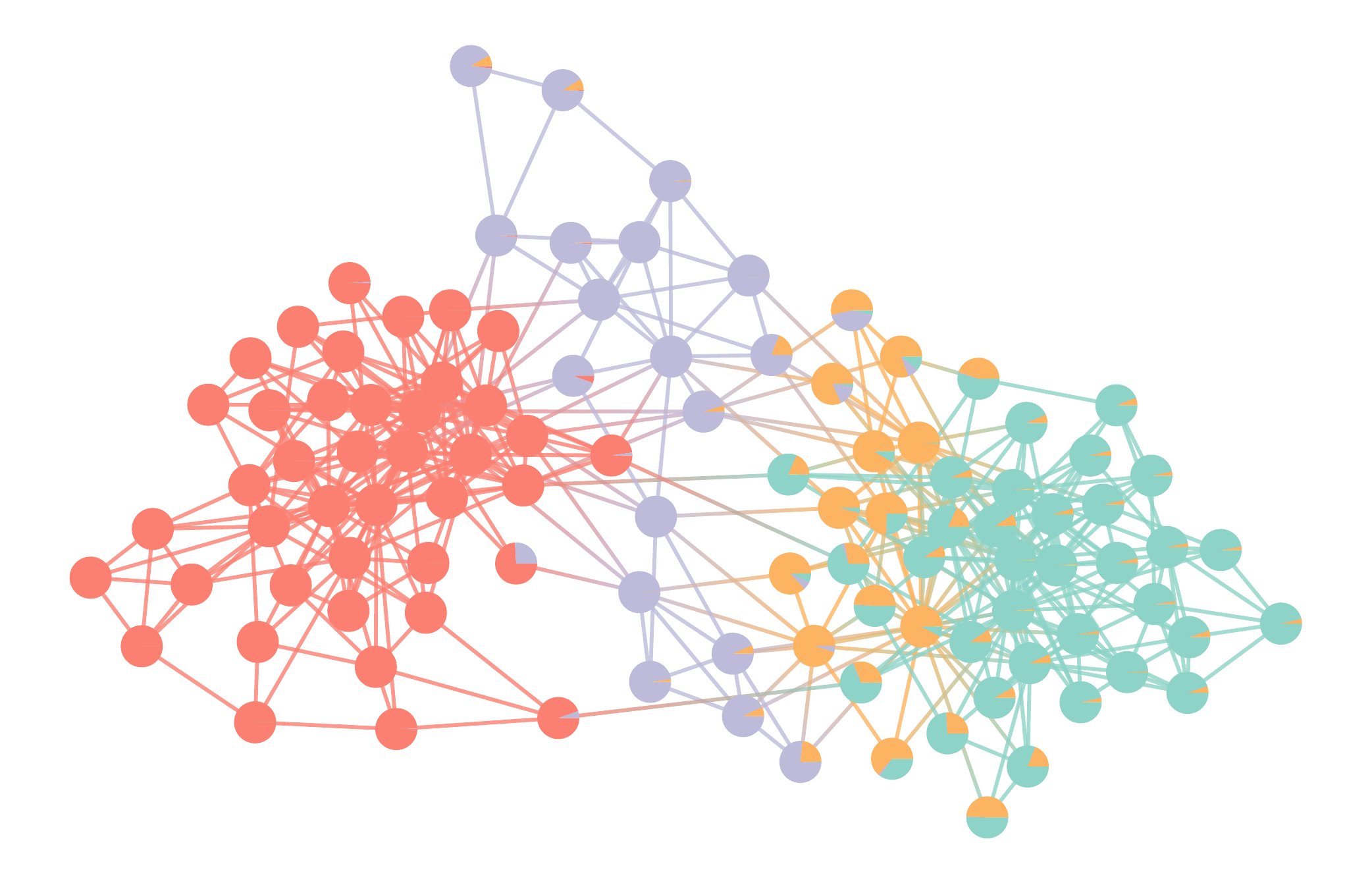}
    &

    \multicolumn{2}{c}{\multirow{4}{*}[4.5em]{
        \begin{overpic}[width=.5\textwidth, trim=2cm 1.2cm .5cm 3.2cm,clip]{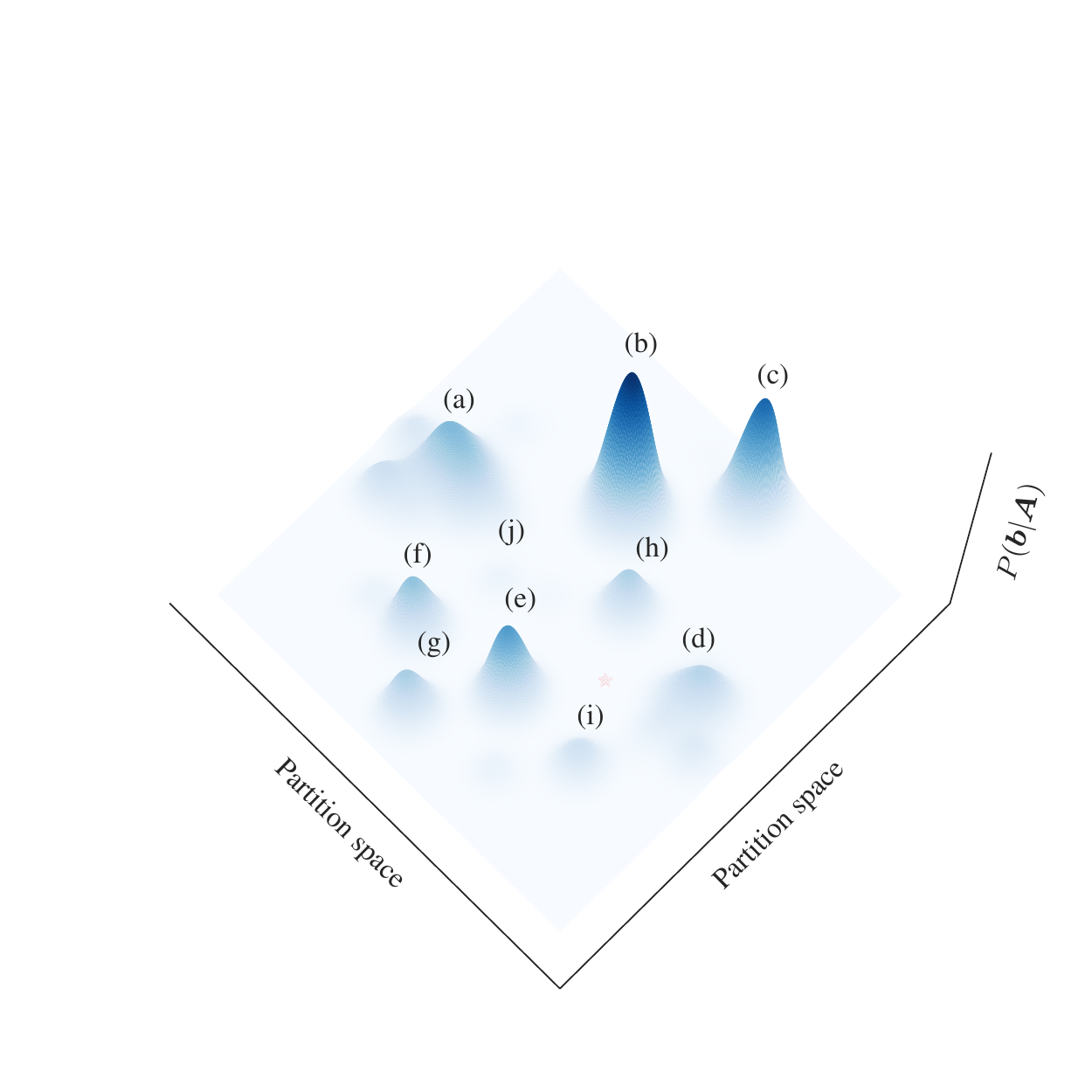}
          \put(48.1,34.1){\color{red}\smaller$\bigstar$}
          \put(66.6,66.4){\color{red}$\blacklozenge$}
          \put(59,35.5){\color{red}$\blacktriangle$}
        \end{overpic}
    }}\\
    (e) $w_5 = 0.0960$, $\sigma_{\hat\bb}=0.066$&
    (f) $w_6 = 0.0588$, $\sigma_{\hat\bb}=0.067$\\
     \includegraphics[width=.24\textwidth]{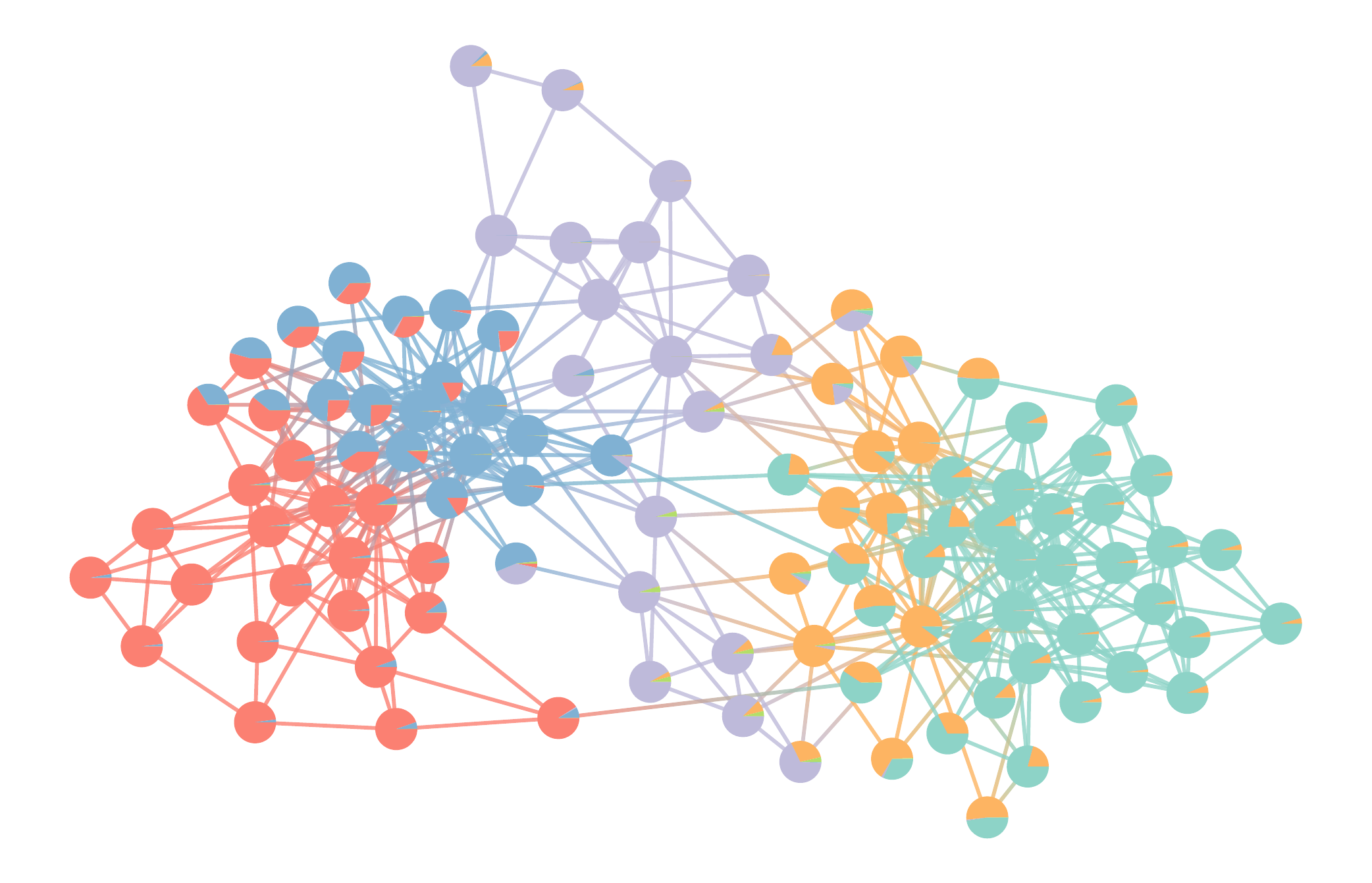}
    &
    \includegraphics[width=.24\textwidth]{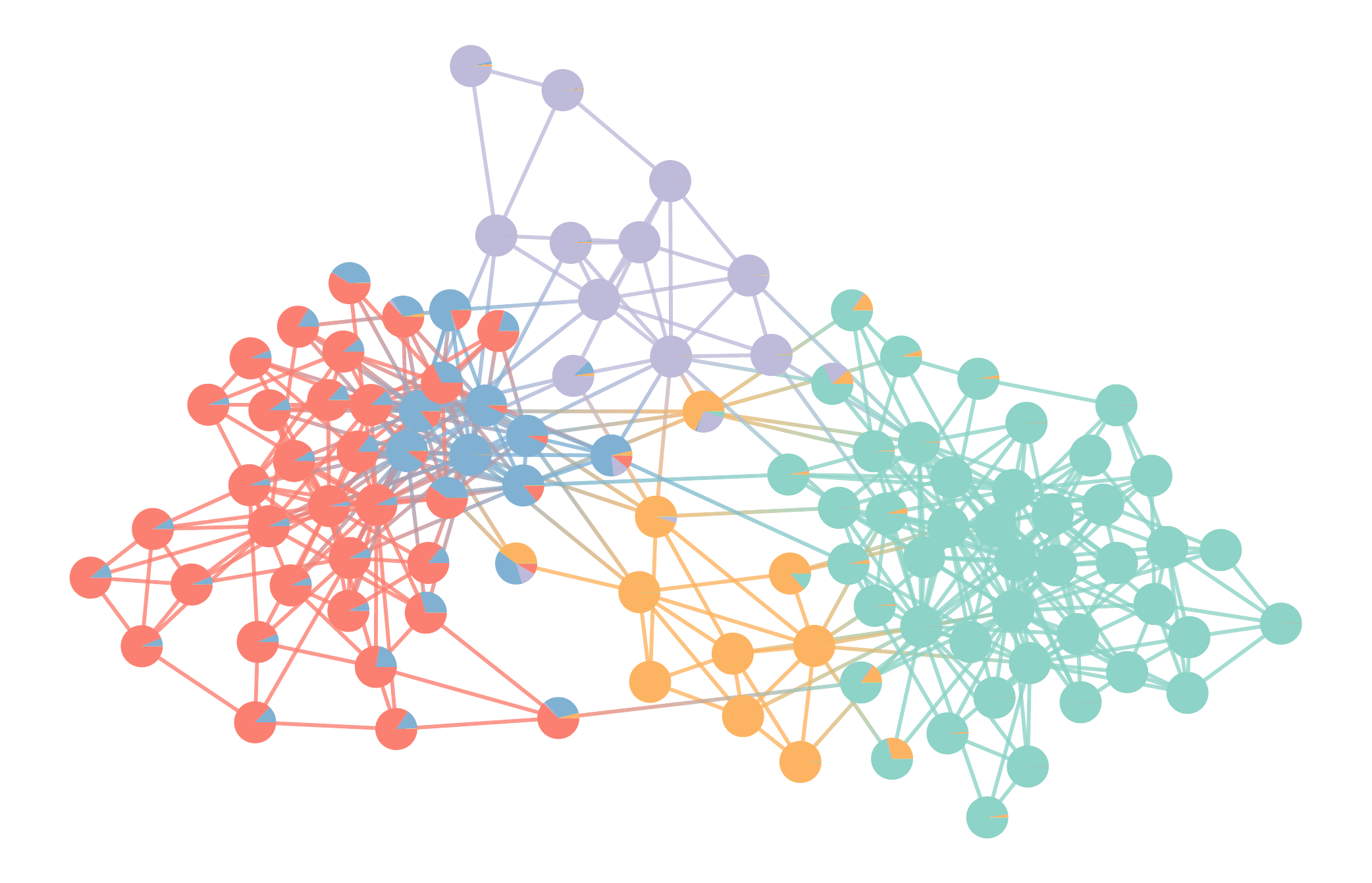}\\
    (g) $w_7 = 0.0528$, $\sigma_{\hat\bb}=0.13$&
    (h) $w_8 = 0.0500$, $\sigma_{\hat\bb}=0.080$\\
    \includegraphics[width=.24\textwidth]{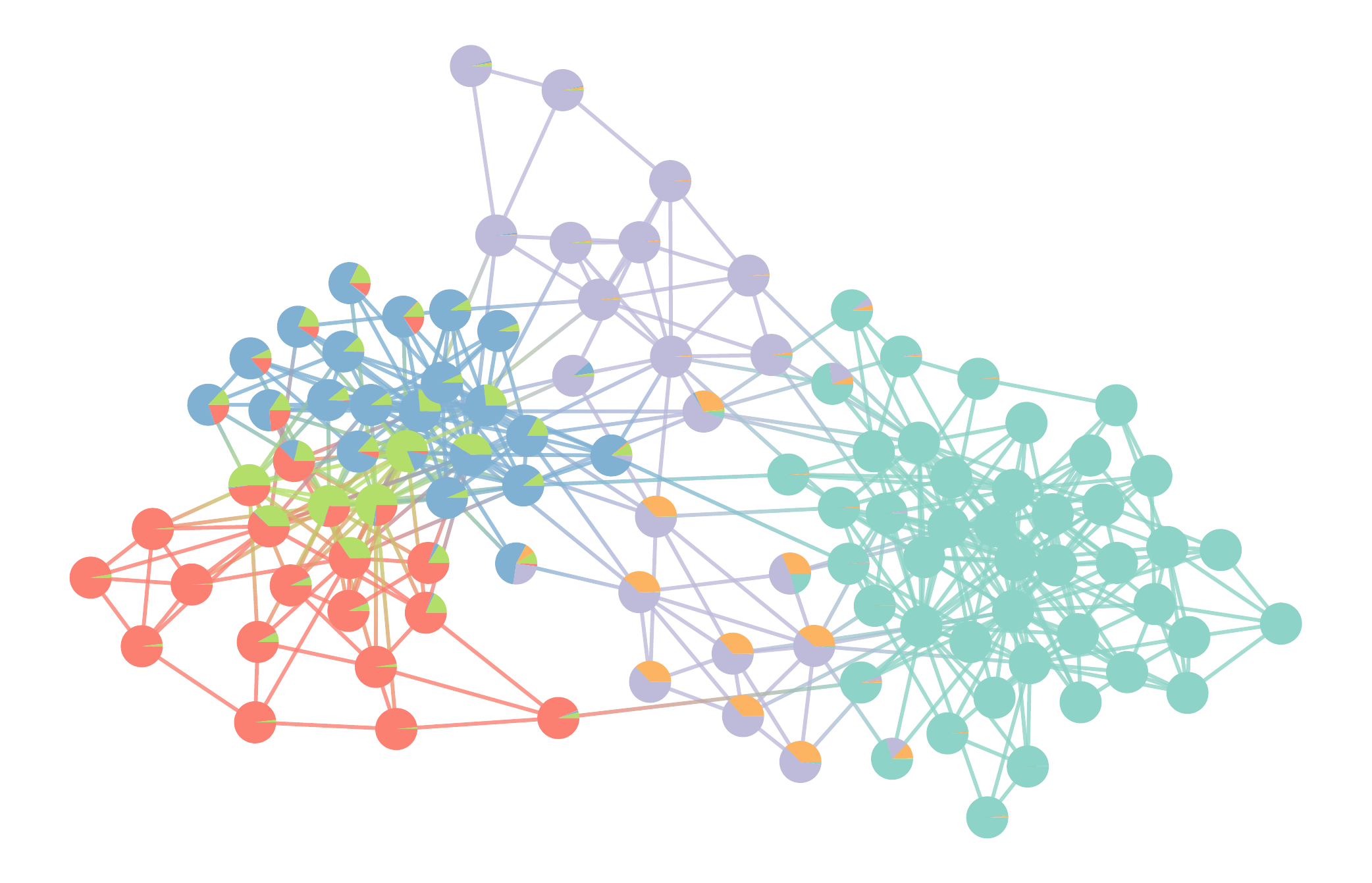}
    &
    \includegraphics[width=.24\textwidth]{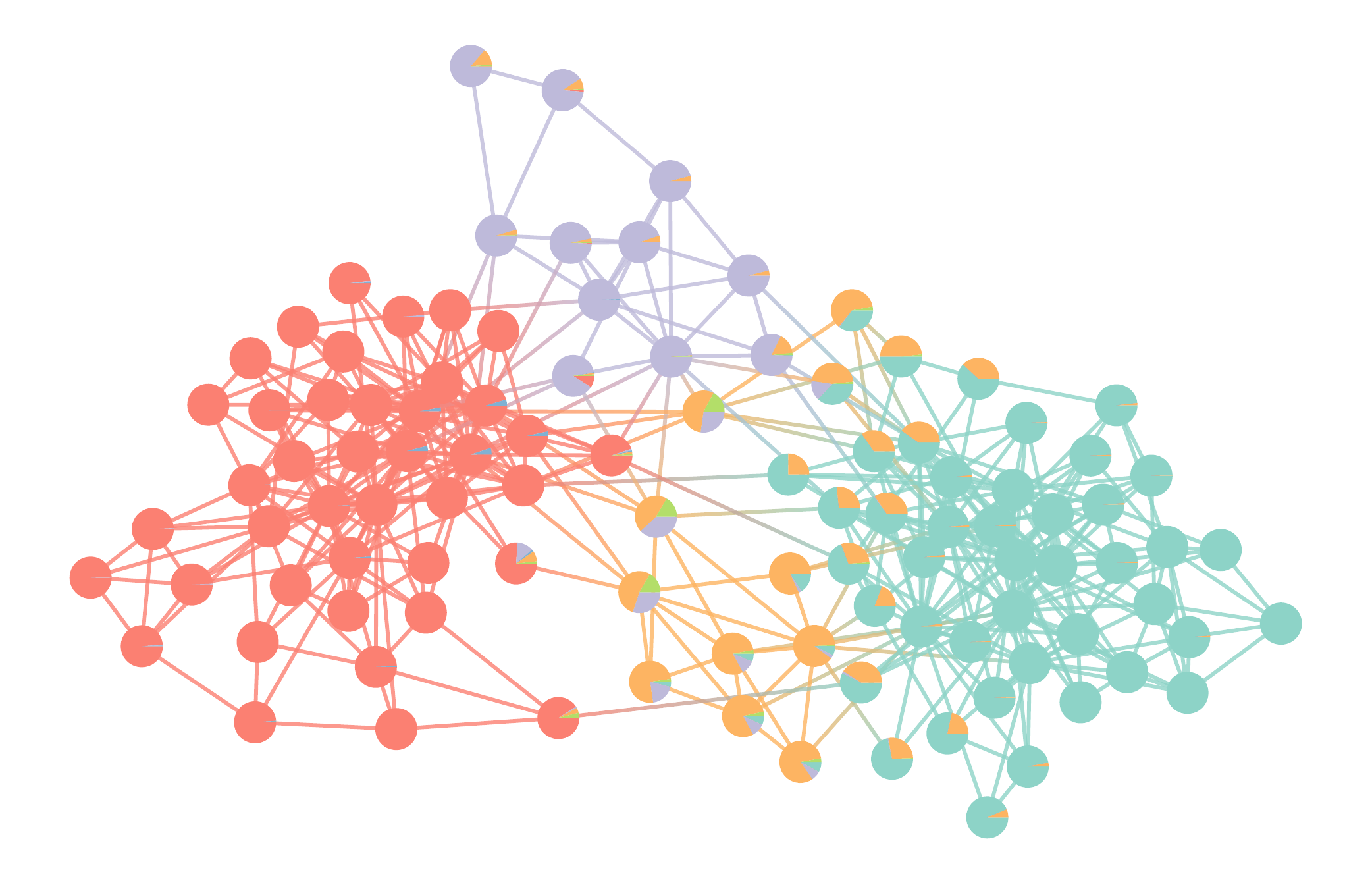}\\
    (i) $w_9 = 0.0124$, $\sigma_{\hat\bb}=0.12$&
    (j) $w_{10} = 0.00829$, $\sigma_{\hat\bb}=0.087$&
  \end{tabular}

  \caption{Inferred partition modes from $M=10^5$ samples of the DC-SBM
  posterior distribution for the political books network. Panels (a) to
  (j) show the marginal distributions for each identified mode as pie
  diagrams on the nodes of the network, with the legend specifying the
  relative mode fraction $w_k$ and the uncertainty $\sigma_{\hat\bb}$ of
  the maximum for each mode. The bottom right panel shows the projection
  of the partition distribution in two dimensions according to the UMAP
  dimensionality reduction algorithm~\cite{mcinnes_umap_2018}, where the
  different modes can be identified as local peaks of the
  distribution. The star symbol ($\bigstar$) shows the location of the MOC
  estimate, the diamond symbol ($\blacklozenge$) the position of the
  MAP/VI estimate, and the triangle ($\blacktriangle$) the
  position of the RMI estimate. \label{fig:polbooks_modes}}
\end{figure*}

In the following, we consider a simple example that illustrates how the
method above can characterize the structure of a distribution of
partitions, and we proceed to investigate how the multimodal nature of
the posterior distribution can be used to assess the quality of fit of
the network model being used.

\subsection{Simple example}

\begin{figure*}
  \begin{tabular}{ccc}
    \multirow{4}{*}[7em]{\includegraphics[width=\columnwidth]{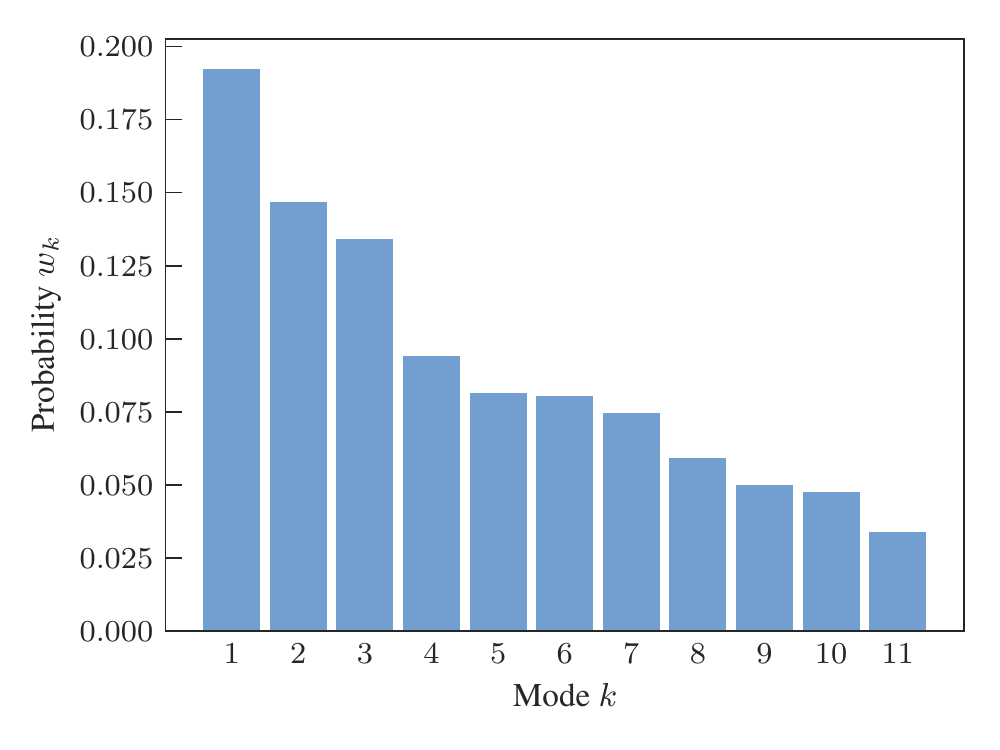}} &
    \includegraphics[width=.5\columnwidth, trim=.8cm .8cm .8cm .2cm]{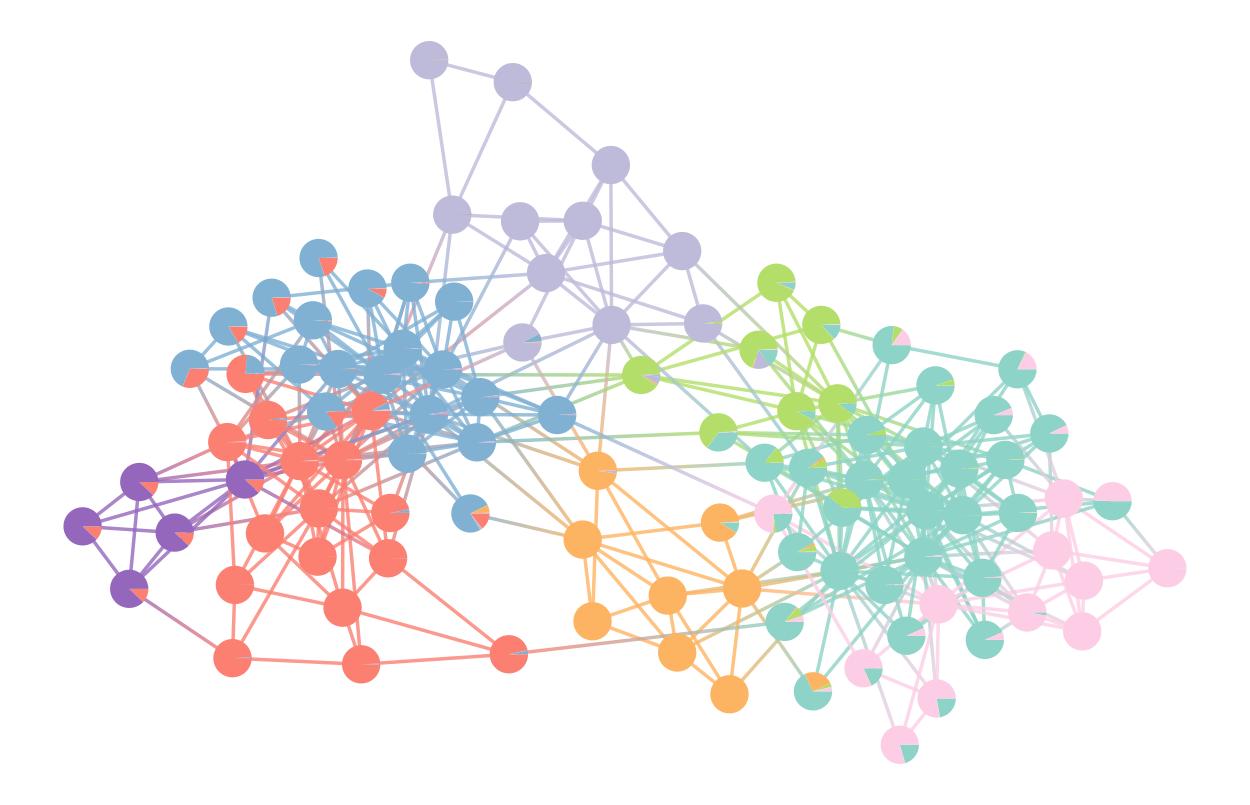}
      &
    \includegraphics[width=.5\columnwidth, trim=.8cm .8cm .8cm .2cm]{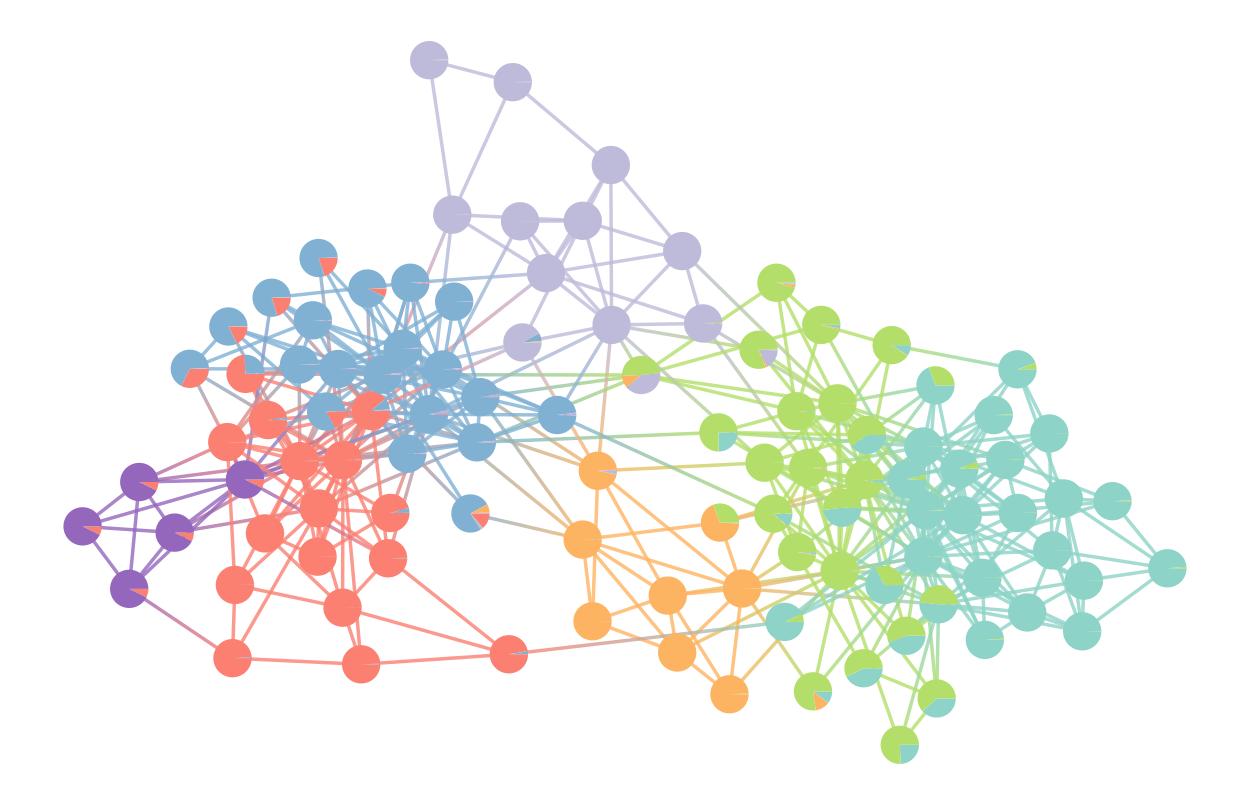}\\
    &
    (a) $k=1$ & (b) $k=2$ \\
    &
    \includegraphics[width=.5\columnwidth, trim=.8cm .8cm .8cm .2cm]{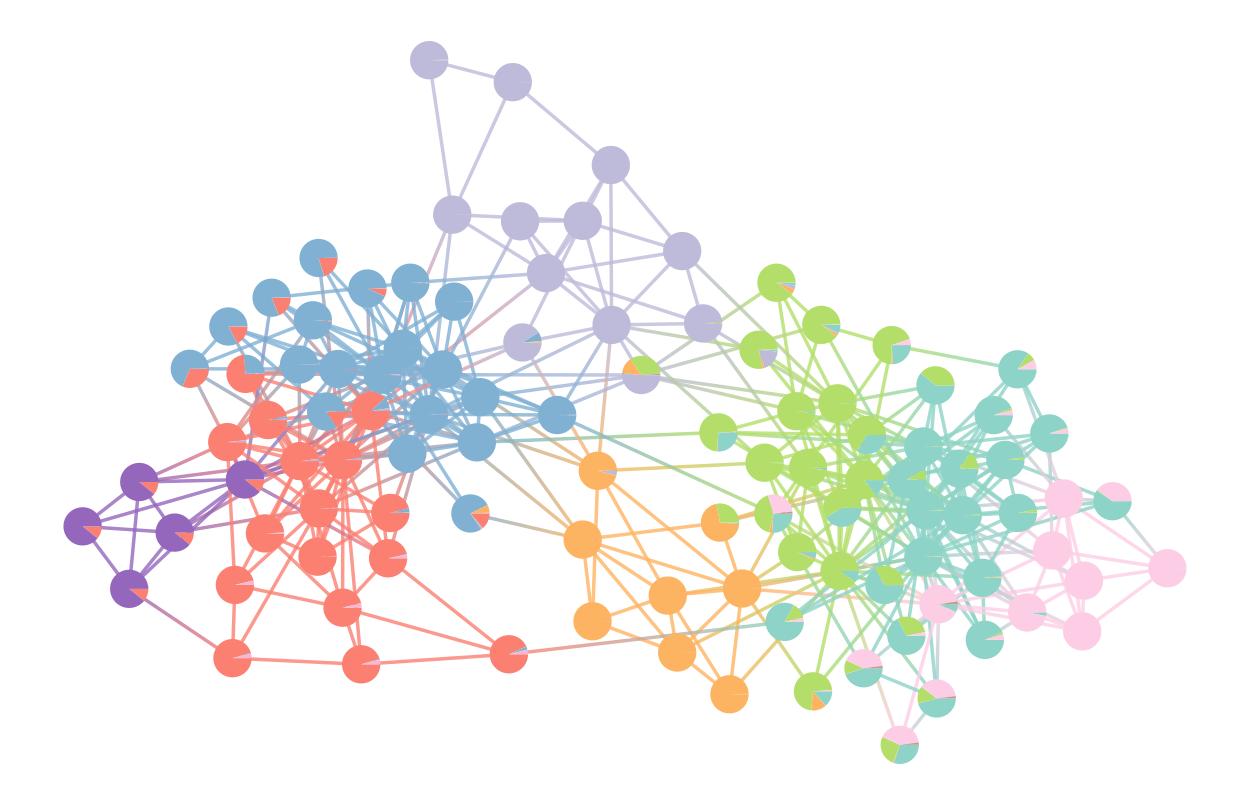}
    &
    \includegraphics[width=.5\columnwidth, trim=.8cm .8cm .8cm .2cm]{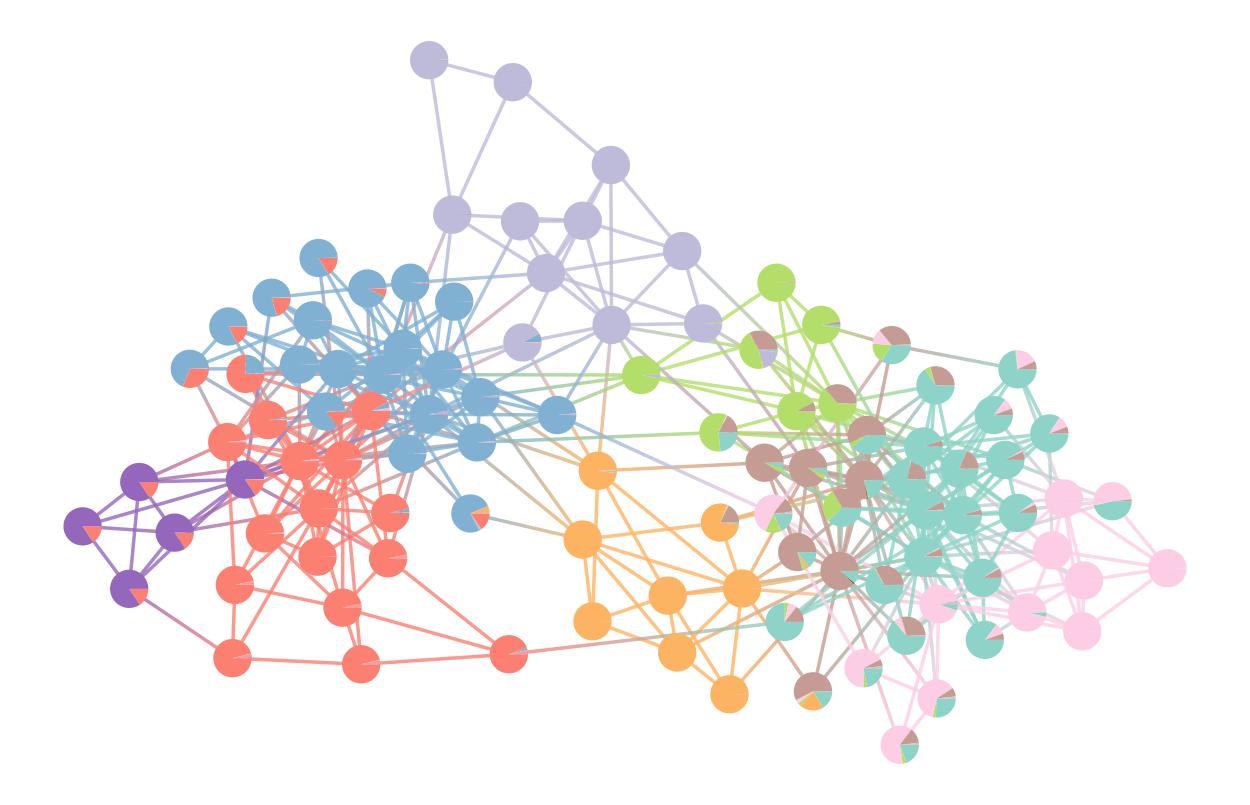}\\
   &  (c) $k=3$ & (d) $k=4$
  \end{tabular}
  \caption{Inferred partition modes from $M=10^5$ samples of the latent Poisson DC-SBM
    posterior distribution for the political books network. The left
    panel shows the mode fractions $w_k$, and the right panel the four
    largest modes, with the marginal distributions shown as pie diagrams
    on the nodes of the network.
    \label{fig:polbooks-multi-modes}}
\end{figure*}

In Fig.~\ref{fig:polbooks_modes} we show the result of the above
algorithm for the posterior distribution obtained for the same political
books network considered previously, where in total $K=11$ modes are
identified. For each mode we show the corresponding marginal
distribution of the relabeled partitions, and the uncertainty
$\sigma_{\hat\bb}=1-\sum_ip_i(\hat b_i)$ of its maximum $\hat\bb$, which
serves as a quantification of how broadly distributed are the individual
modes. As a means of illustration, in Fig.~\ref{fig:polbooks_modes} we
show also a two-dimensional projection of the distribution of
partitions, obtained using the UMAP dimensionality reduction
algorithm~\cite{mcinnes_umap_2018} using the maximum overlap distance as
the dissimilarity metric (similar results can also be found with other
dissimilarity functions, as shown in Appendix~\ref{app:dimred}). This
algorithm attempts to project the distribution of partitions in two
dimensions, while preserving the relative distances between partitions
in the projection. As a result we see that each mode is clearly
discernible as a local concentration of partitions, much like we would
expect of a heterogeneous mixture of continuous variables. We note here
that we have not informed the UMAP algorithm of the modes we have found
with the algorithm above, and therefore this serves an additional
evidence for the existence of the uncovered heterogeneity in the
posterior distribution. The most important result of this analysis is
that no single mode has a dominating fraction of the distribution, with
the largest mode corresponding only to around $23\%$ of the posterior
distribution, and with the second largest mode being very close to
it. This means that there is no single cohesive picture that emerges
from the distribution, and therefore our attempt at summarizing it with
a single partition seems particularly ill-suited.

In view of this more detailed picture of the ensemble of partitions, it
is worth revisiting the consensus results obtained previously with the
various error functions. As shown in Fig.~\ref{fig:polbooks_modes}, the
MAP/VI estimates correspond to the most likely partition of mode (c),
which is overall only the third most plausible mode with
$w_3=0.134$. From the point of view of the MAP estimator, this serves to
illustrate how choosing the most likely partition may in fact run
counter to intuition: Although the single most likely partition belongs
to mode (c), collectively, the partitions in mode (a) and (b) have a
larger plausibility. This means that, if we are forced to choose a
single explanation for the data, it would make more sense instead to
choose mode (a), despite the fact that it does not contain the single
most likely partition. More concretely, when comparing modes (a), (b),
and (c), we see that the network does in fact contain more evidence for
a division of either the ``neutral'' or the ``liberal'' groups into
subgroups than the MAP estimate implies, however not both, as mode (d),
corresponding to the simultaneous subdivisions, has a smaller
plausibility than the other options. The VI estimate also points to mode
(c), but it is unclear why. This is indeed a problem with using VI,
since despite its strong formal properties, it lacks a clear
interpretability.

\begin{figure*}
  \begin{tabular}{cc}
  \begin{tikzpicture}
    \node at (0,0) {\includegraphics[width=\columnwidth]{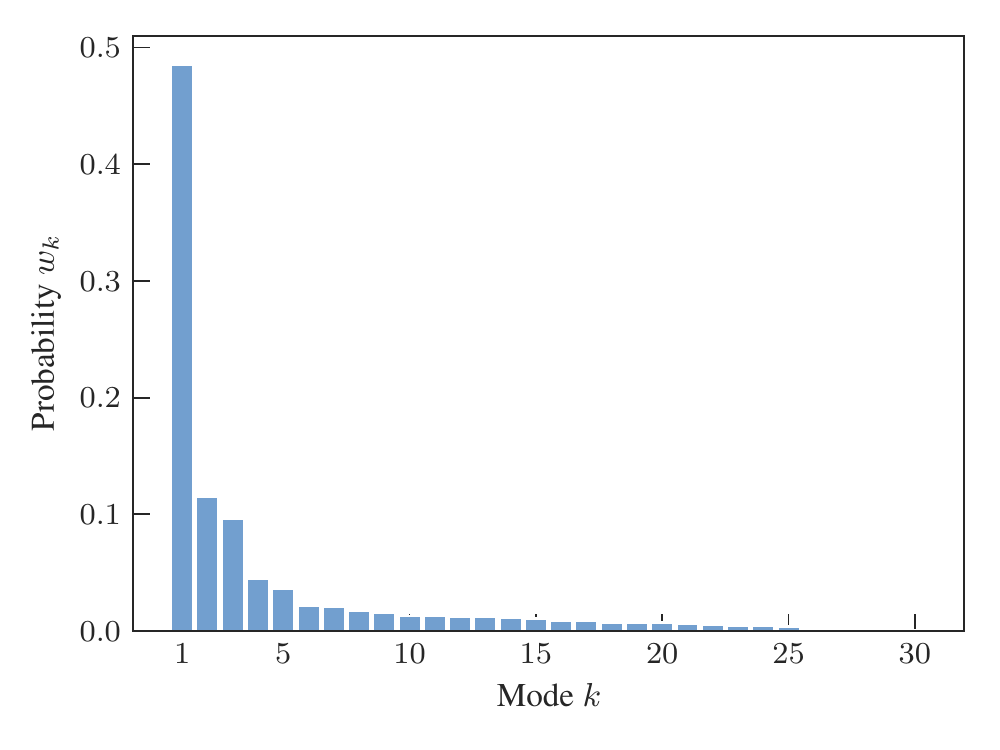}};
    \node at (-.75,1.5) {\includegraphics[width=.4\columnwidth]{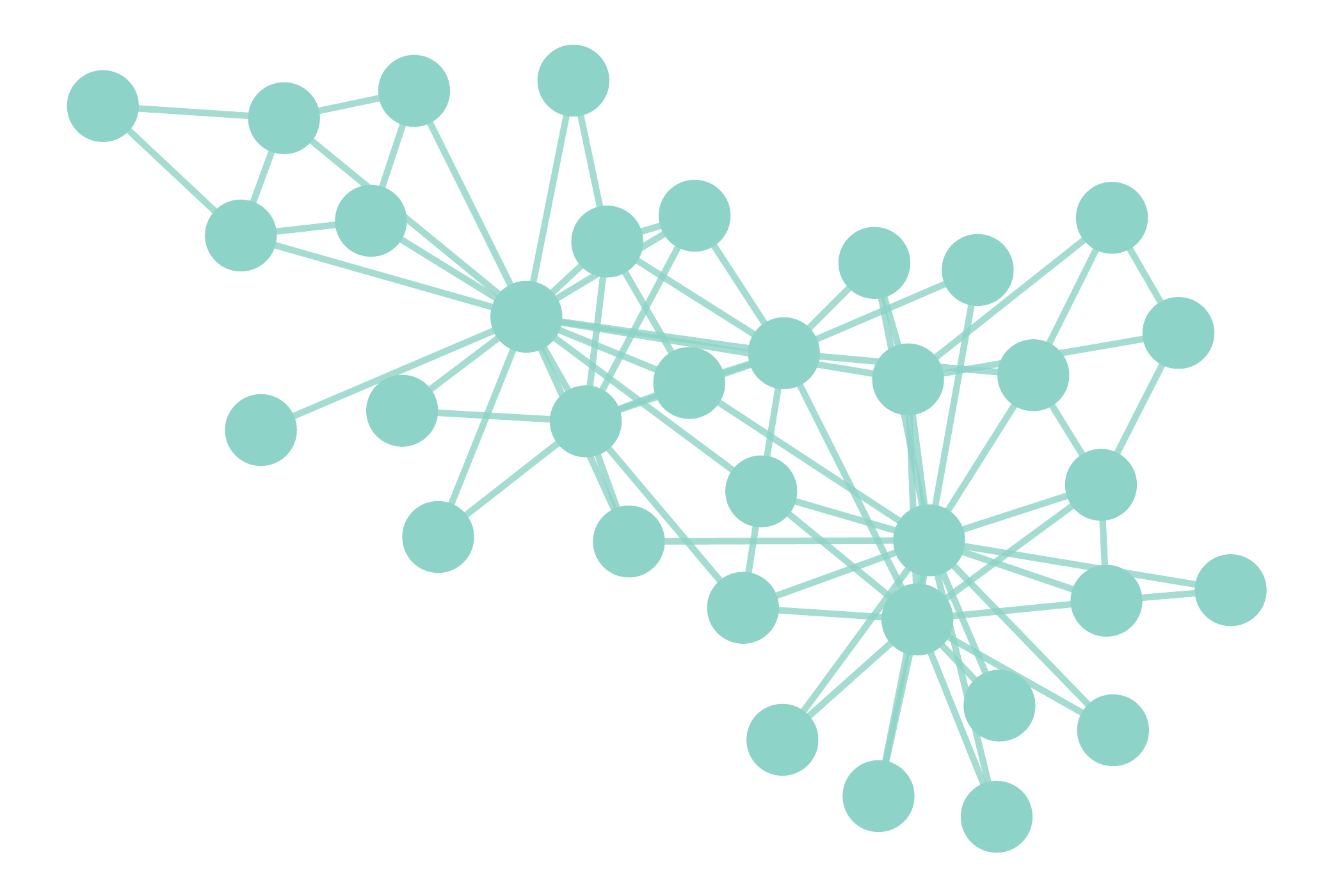}};
    \node at (-.4,-.7) {\includegraphics[width=.4\columnwidth]{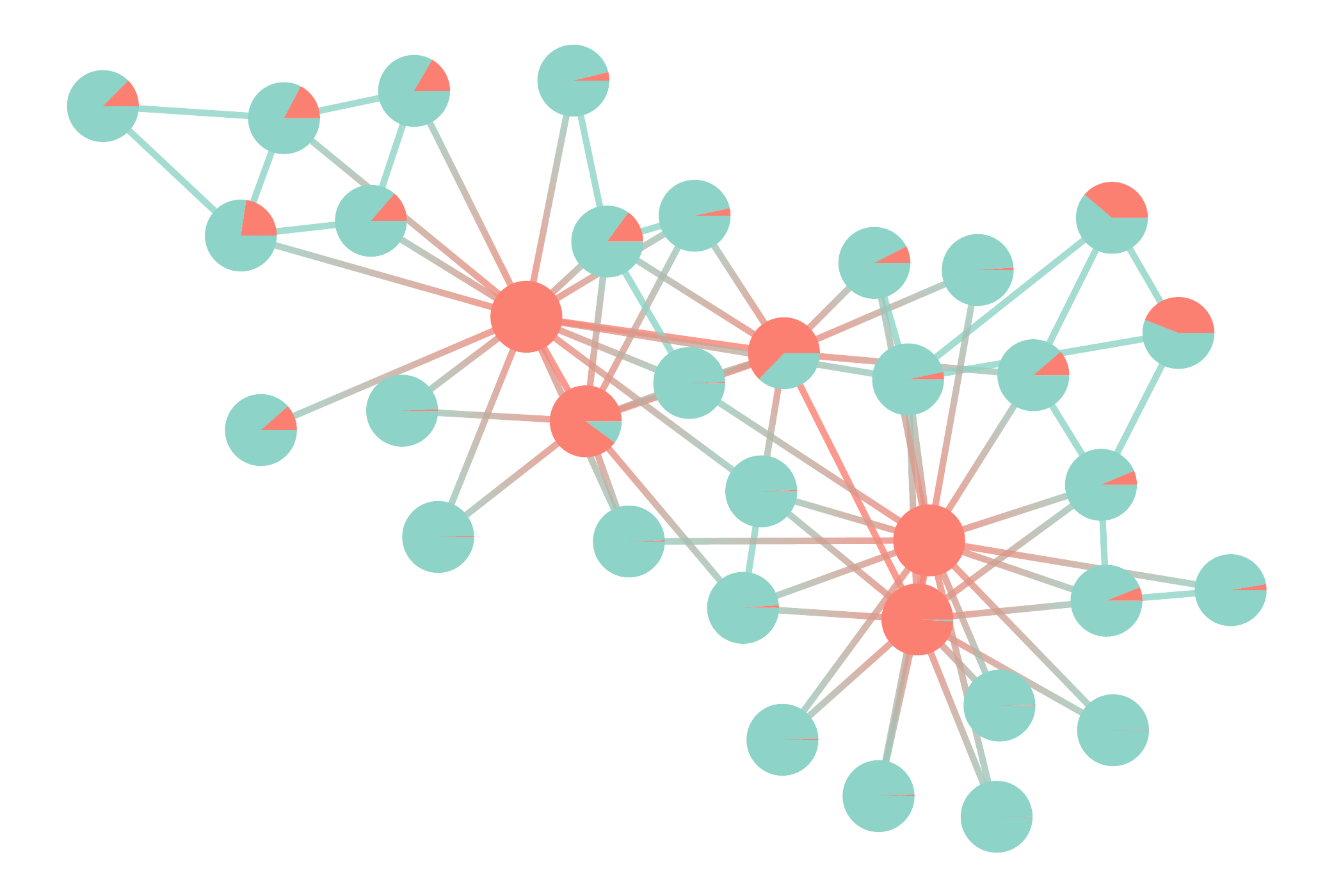}};
    \node at (2.4,-.1) {\includegraphics[width=.4\columnwidth]{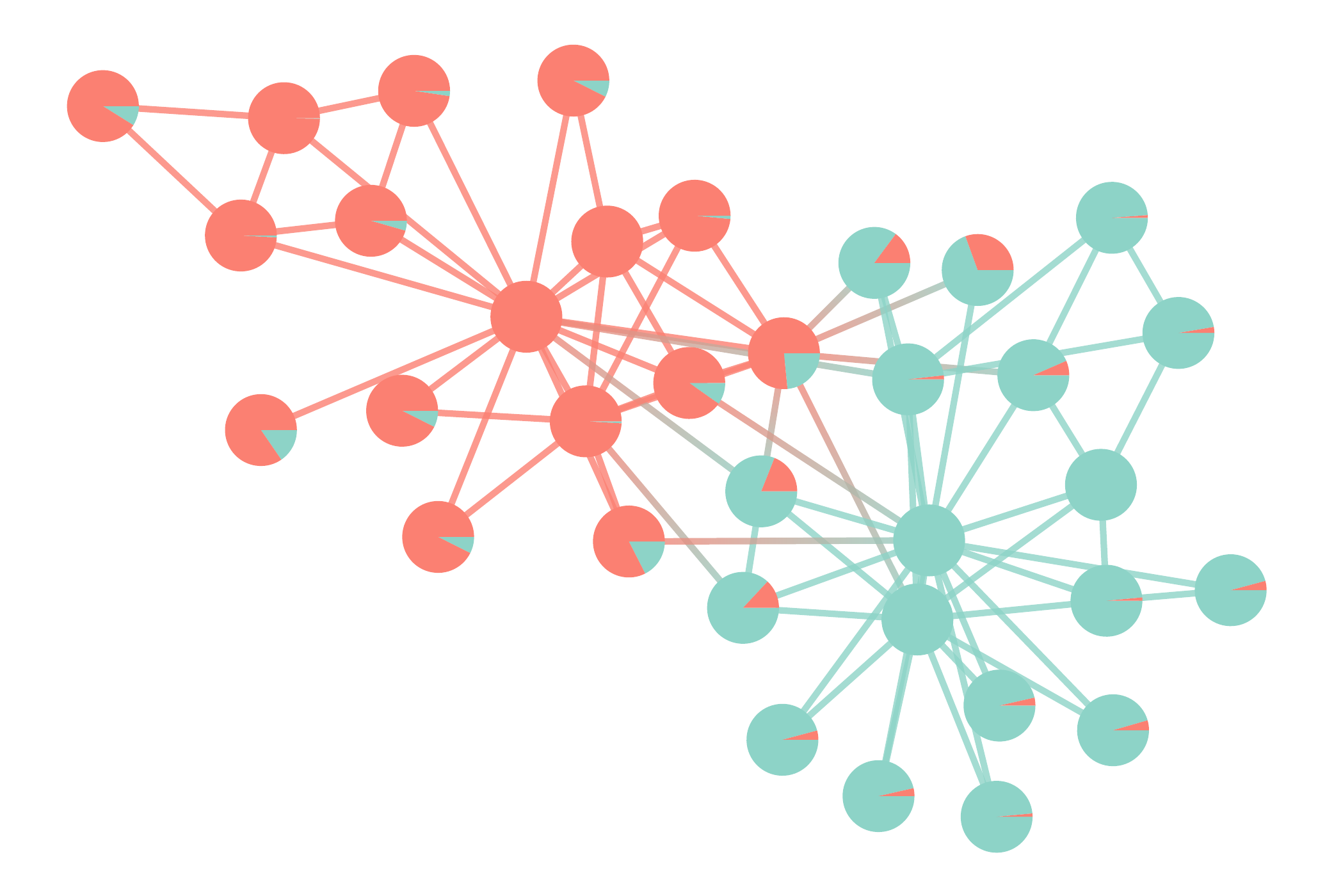}};

    \draw[->] (-1.8,1.3) -- (-2.6,1.2);
    \draw[->] (-1.7,-.7) -- (-2.4,-1.05);
    \draw[->] (2.3,-.7) .. controls (0,-3.2) and (-.2,-1) .. (-1.,-2.);
  \end{tikzpicture} &
  \begin{tikzpicture}
    \node at (0.,0) {\includegraphics[width=\columnwidth]{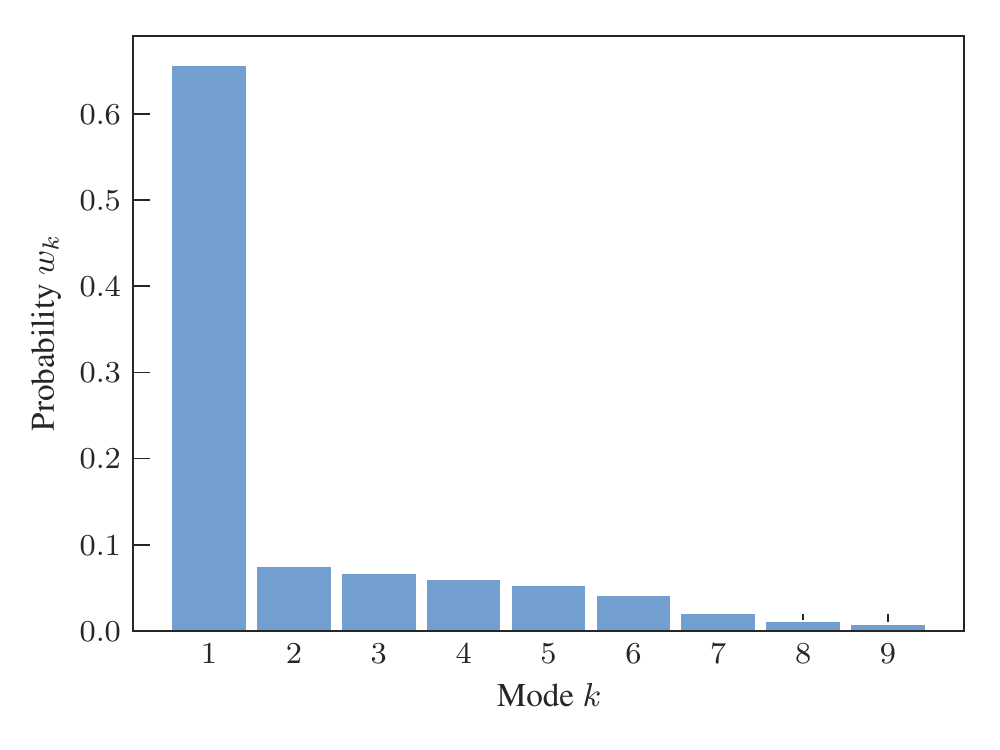}};
    \node at (-.55,1.5) {\includegraphics[width=.4\columnwidth]{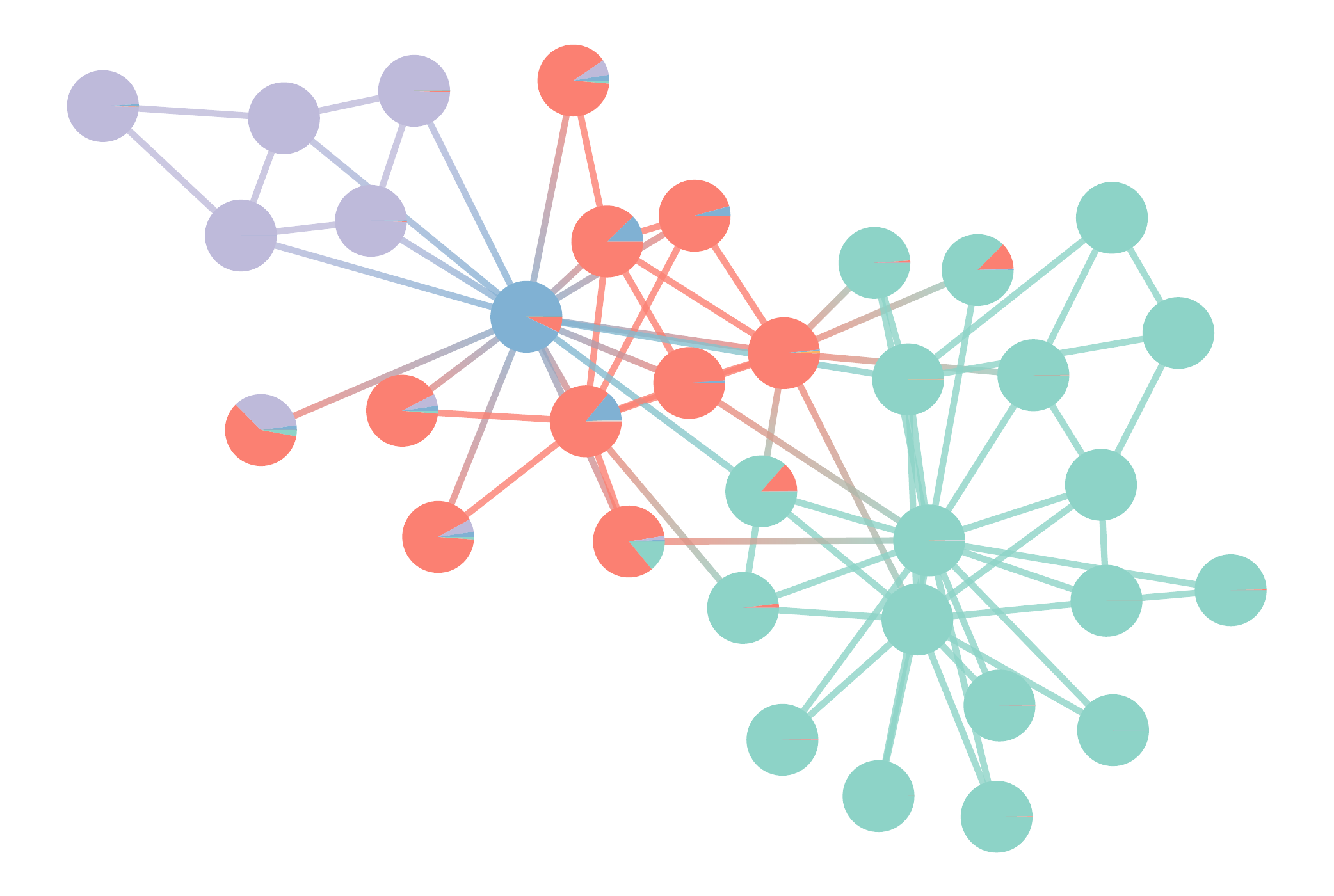}};
    \node at (-.4,-.5) {\includegraphics[width=.4\columnwidth]{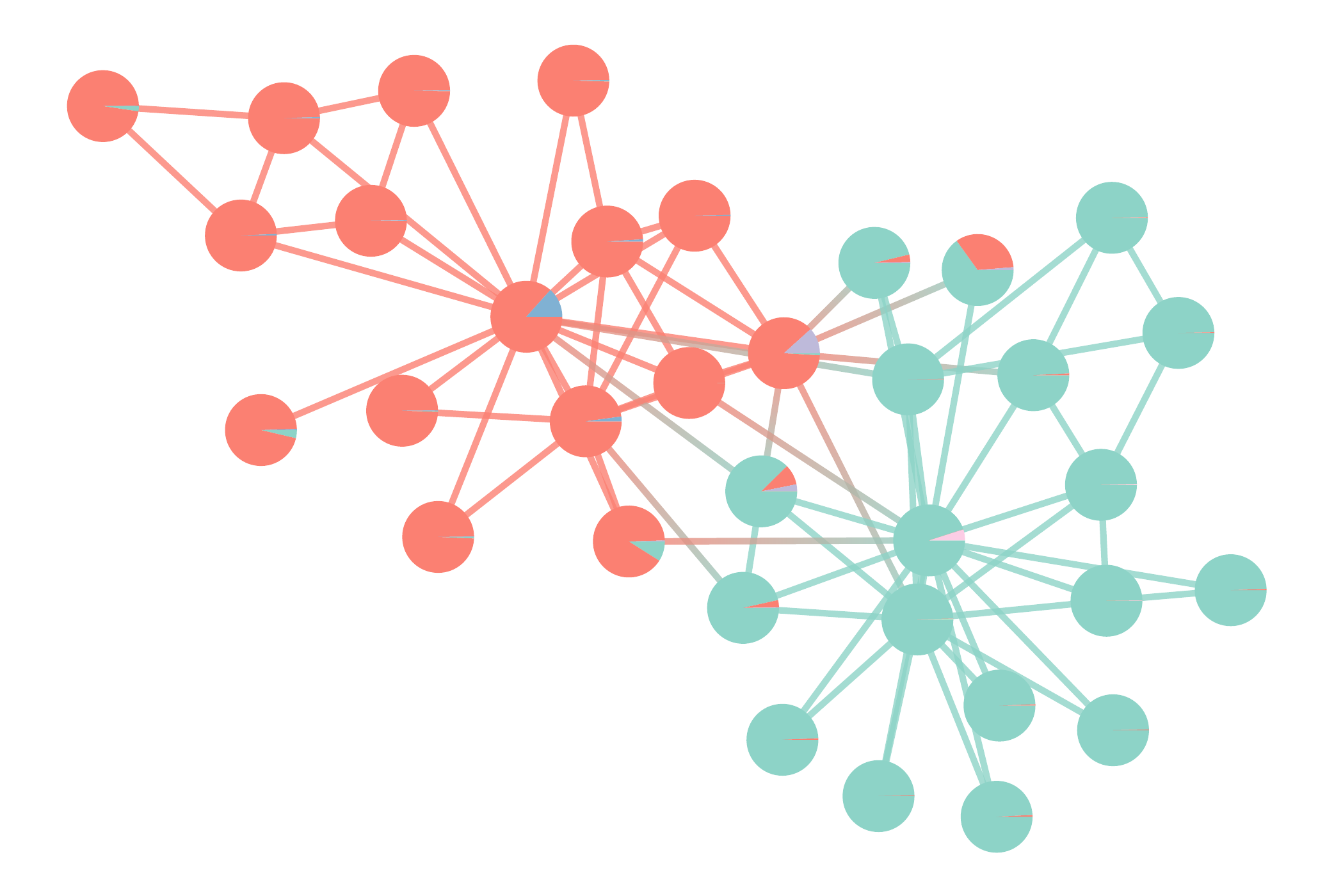}};
    \node at (2.4,-.16) {\includegraphics[width=.4\columnwidth]{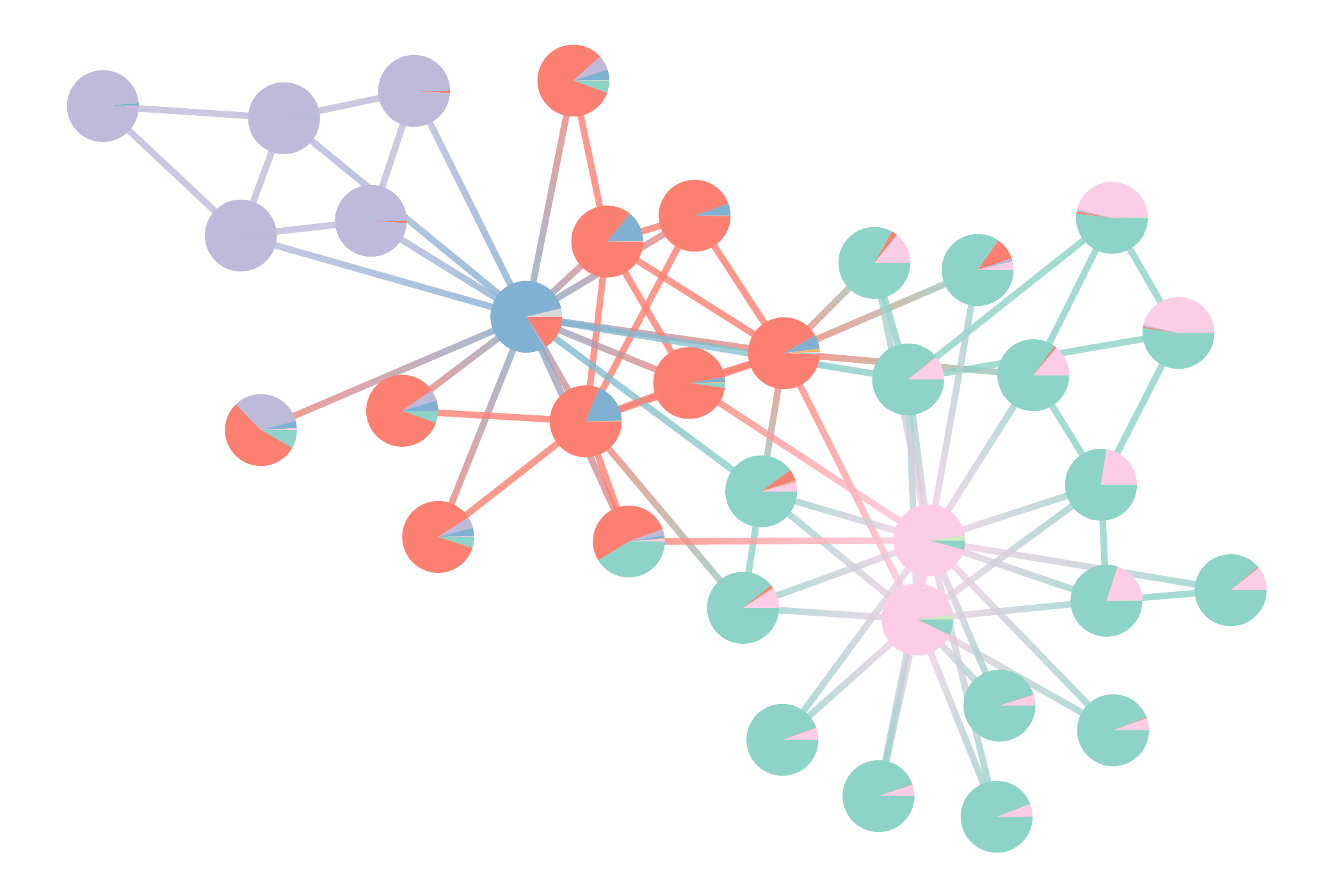}};

    \draw[->] (-1.4,1.3) -- (-2.1,1.1);
    \draw[->] (-1.2,-.9) -- (-1.8,-1.55);
    \draw[->] (2.3,-.7) .. controls (0,-2.85) and (-.2,-.7) .. (-1.08,-1.65);
  \end{tikzpicture}
  \end{tabular}
  \caption{Inferred partition modes from $M=10^5$ samples the posterior
  distribution obtained with the the Poisson DC-SBM (left) and latent
  Poisson DC-SBM (right) for the karate club network. The insets show
  the modes as indicated by the arrows, with the marginal distributions
  shown as pie diagrams on the nodes of the network.\label{fig:karate_modes}}
\end{figure*}

Differently from MAP and VI, the MOC estimation combines the properties
of all modes into a ``Frankenstein's monster,'' where local portions of
the final inferred partition correspond to different modes. As a result,
the resulting point estimate has a very low posterior probability, and
hence is a misleading representation of the population --- a classic
estimation failure of multimodal distributions.

The RMI estimate behaves differently, and corresponds to a typical
partition of mode (d), which has an overall plausibility of
$w_4=0.132$. We can understand this choice by inspecting its
composition, and noticing that the more plausible modes (a) to (c)
correspond to partitions where groups of (d) are merged
together. Because of this, the RMI similarity sees this partition as the
``center'' composed of the building blocks required to obtain the other
ones via simple operations. But by no means it is the most likely
explanation of the data according to the model, and given that it is a
division into a larger number of groups, it is more likely to be an
overfit, in view of the existence of simpler modes (a) to (c).

\subsection{Evaluating model consistency}

\begin{figure}
  \begin{tikzpicture}
    \node at (0,0) {\includegraphics[width=\columnwidth]{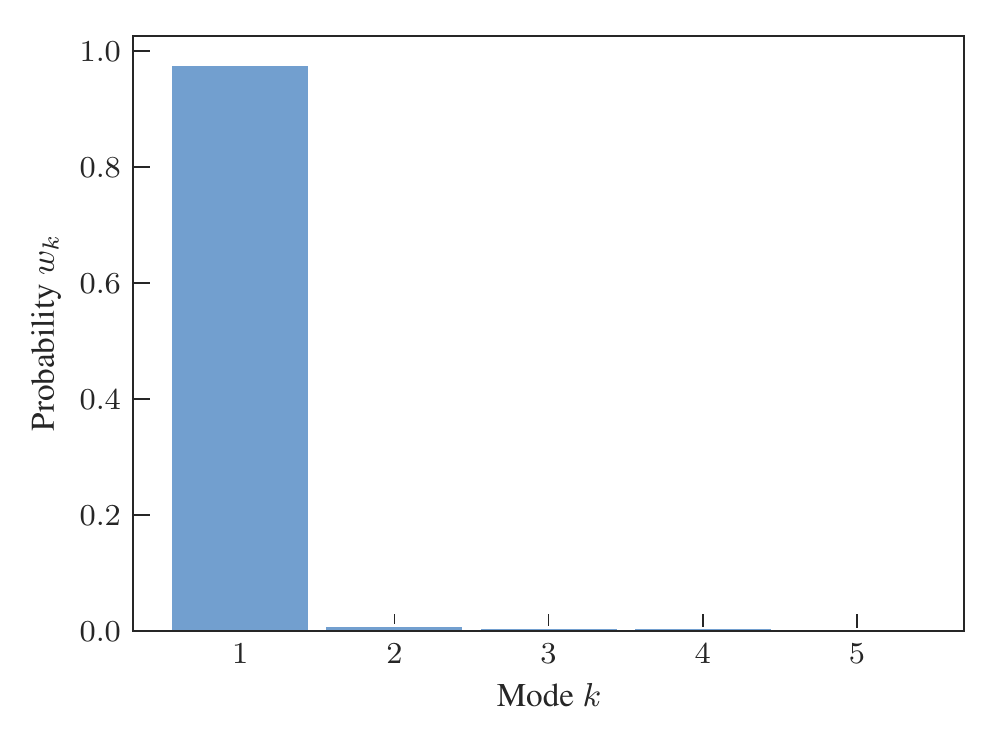}};
    \node at (0, 1.3) {\includegraphics[width=.38\columnwidth]{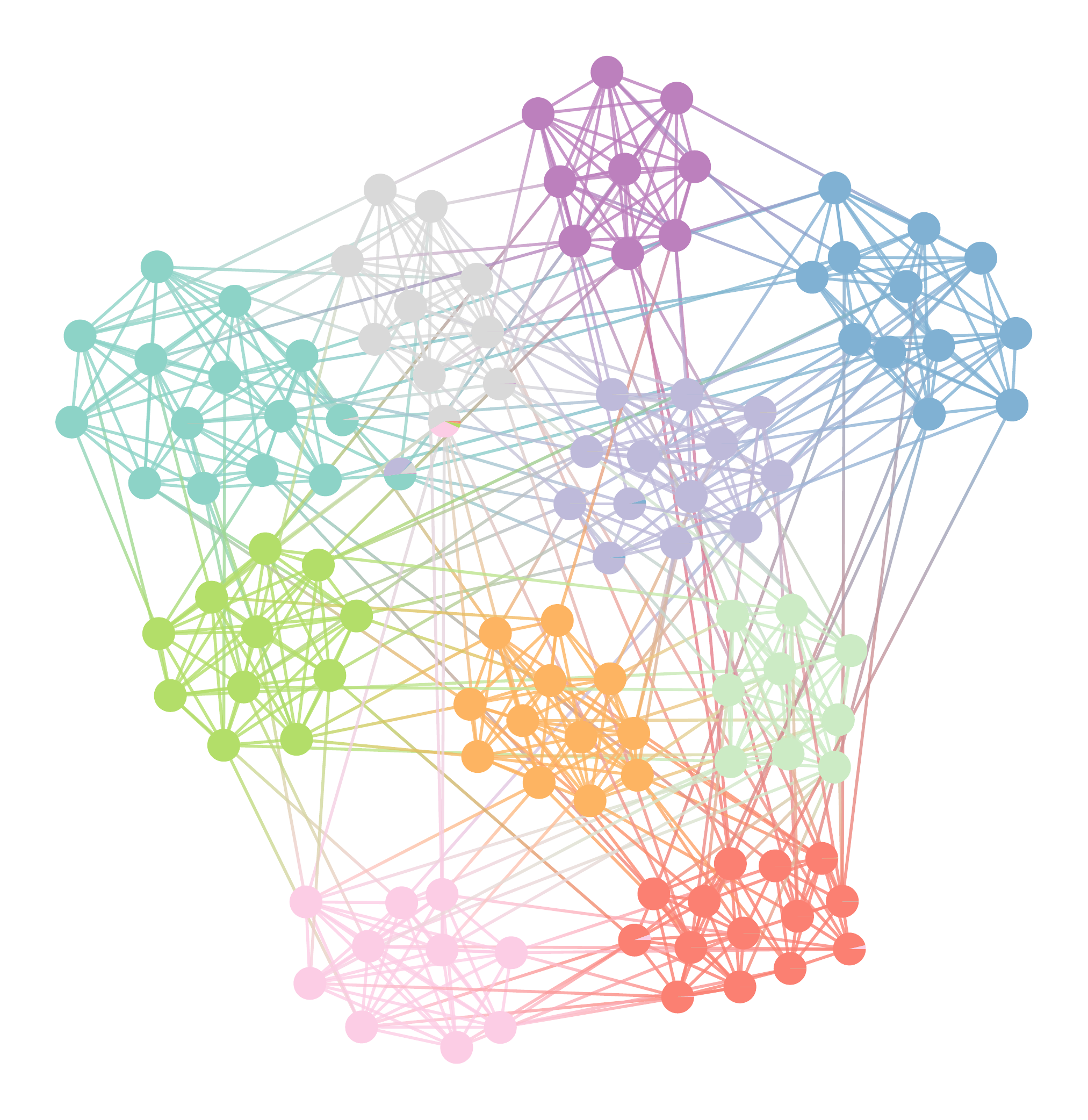}};
    \node at (2.5, -.55) {\includegraphics[width=.38\columnwidth]{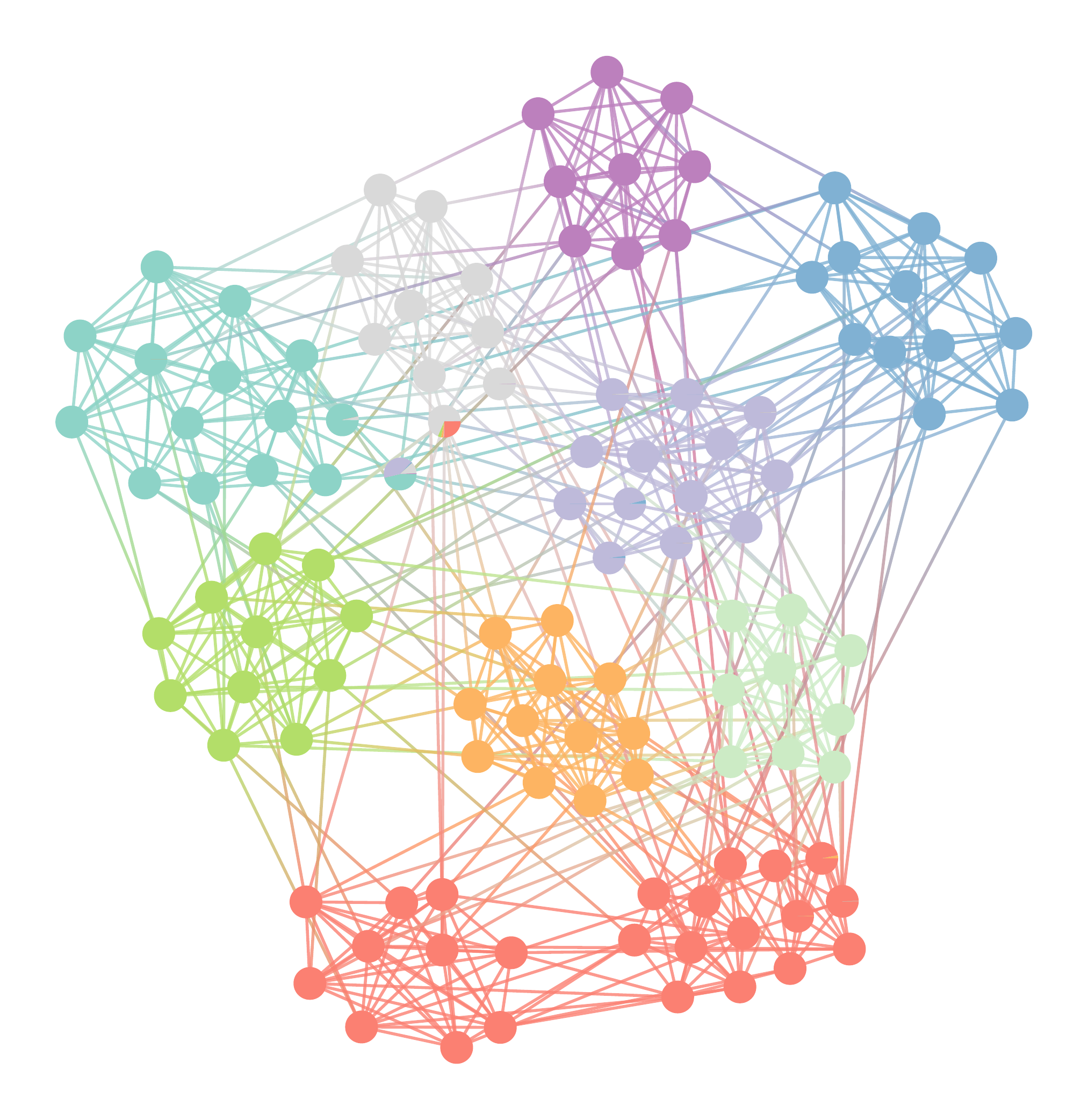}};

    \draw[->] (-1,.3) -- (-1.45,0);
    \draw[->] (1.1,-1) -- (-.8,-2.05);
  \end{tikzpicture}
  \caption{Inferred partition modes from $M=10^5$
  samples the posterior distribution obtained with the Poisson
  DC-SBM for the American college football network. The insets
  show the modes as indicated by the arrows, with the marginal distributions
  shown as pie diagrams on the nodes of the network.\label{fig:football_modes}}
\end{figure}

\begin{figure*}
  \begin{tabular}{ccc}
    \multirow[b]{2}{*}[14em]{\includegraphics[width=.9\columnwidth]{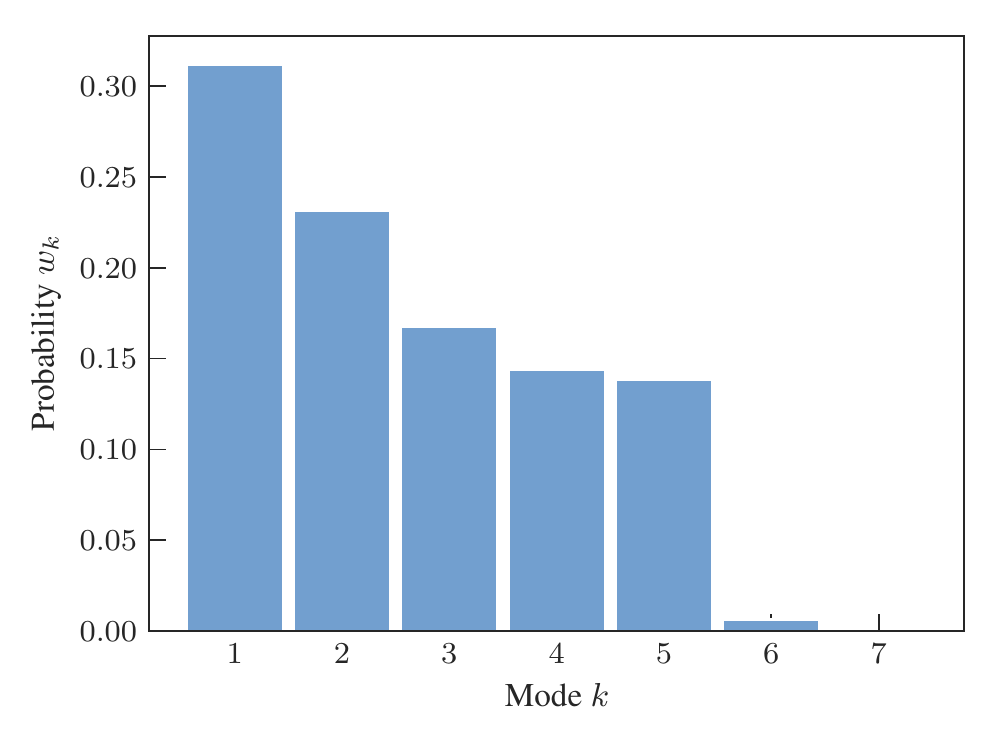}} &
    \includegraphics[width=.55\columnwidth, trim=.8cm .8cm .8cm .2cm]{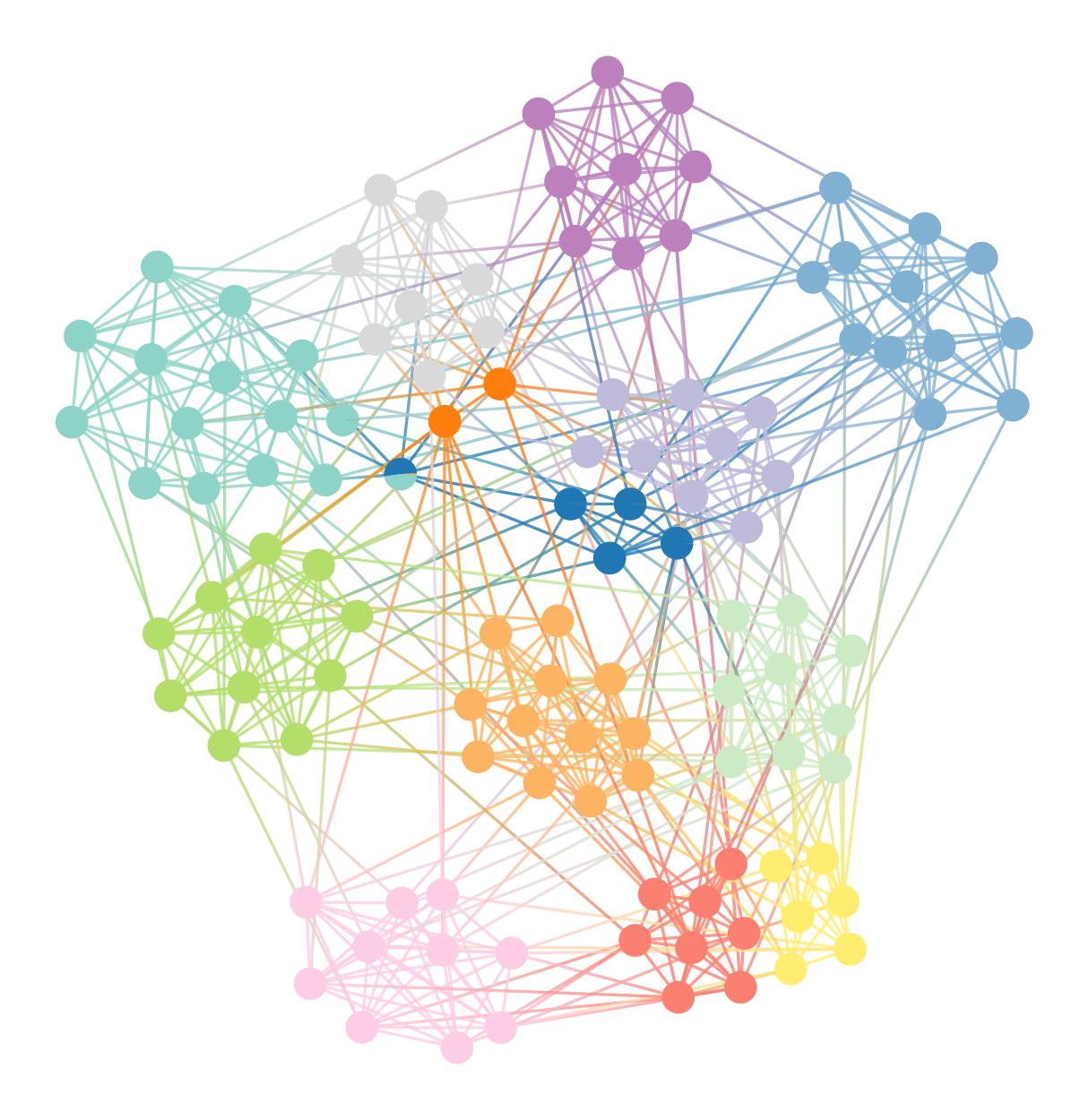}
      &
    \includegraphics[width=.55\columnwidth, trim=.8cm .8cm .8cm .2cm]{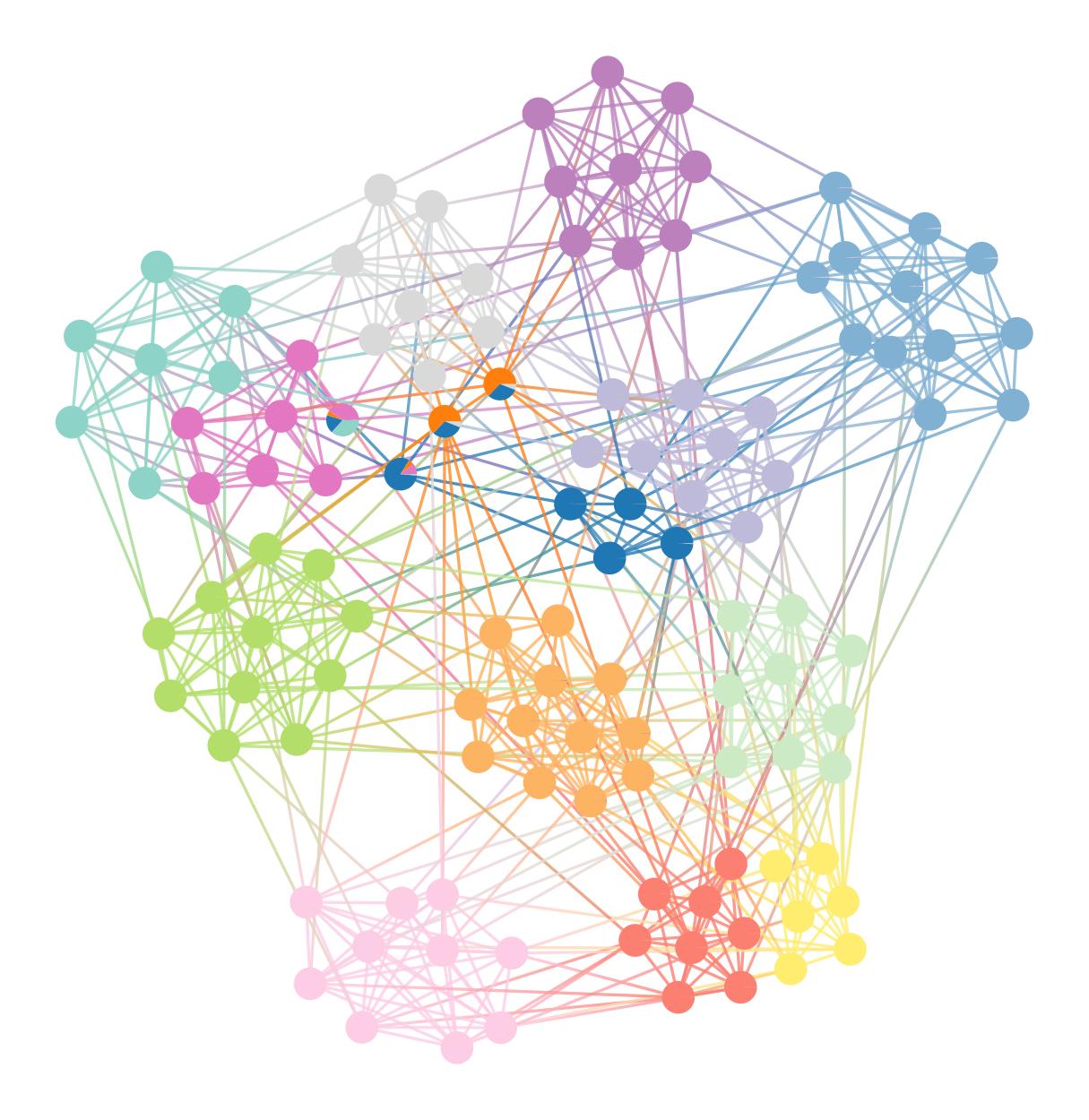}\\
    &
    (a) $k=1$ & (b) $k=2$
  \end{tabular}
  \caption{Inferred partition modes from $M=10^5$ samples of the latent Poisson DC-SBM
    posterior distribution for the American college football
    network. The left panel shows the mode fractions $w_k$, and the
    right panel the two largest modes, with the marginal distributions
    shown as pie diagrams on the nodes of the network.\label{fig:football-multi-modes}}
\end{figure*}

The full characterization of the posterior distribution with our
approach gives us the opportunity to assess the quality of fit between
model and data. Indeed, if the model was an excellent fit, e.g. if the
data were in fact generated by the SBM, we should expect a single mode
in the posterior distribution that is centered in the true
partition~\cite{decelle_asymptotic_2011} (although the broadness of the
mode, represented by the variance of the marginal distribution on the
nodes, will depend on how easily detectable the true partition
is). Therefore, the fact alone we observe multiple modes is an
indication of some degree of mismatch, with the model offering multiple
explanations for the data. Since our analysis allows us to inspect each
individual explanation, and ascribe to it a plausibility, this can be
used to make a more precise evaluation of the fit.

Inspecting the modes observed for the political books network in
Fig.~\ref{fig:polbooks_modes}, we notice that the four largest modes
amount approximately to different combinations of the same five groups
that appear in the fourth mode (Fig.~\ref{fig:polbooks_modes}d) ---
although the remaining modes deviate from this pattern. This is
reminiscent of a situation considered by Riolo and
Newman~\cite{riolo_consistency_2020}, who have applied RMI estimation
for artificial networks where none of the posterior samples matches the
true division, which is only uncovered by the RMI consensus. In
particular, in their scenario, the consensus exposed ``building
blocks,'' i.e. groups of nodes that tend to be clustered together,
although the building blocks themselves always appear merged together
into bigger groups. The situation where the partitions exhibit clear
shared building blocks that always appear merged together, but in
different combinations, begs the question as to why does the posterior
distribution fail to concentrate on the isolated building blocks in the
first place. One possibility is that the building blocks do not
correspond to the same kind of communities that the inference approach
is trying to uncover, e.g. in the case of the SBM these should be nodes
that have the same probability of connection to the rest of the
network. This would be a case of model mismatch, and hence it would be
difficult to interpret what the building blocks actually mean. Another
option, that we can address more directly, is that the model being used
underfits the data, i.e.  the model formulation fails to recognize the
available statistical evidence, resulting in the choice of simpler SBMs
with fewer groups, such that some ``true'' groups are merged together.
A common cause of underfitting is the use noninformative priors which
overly penalize larger numbers of groups, as was shown in
Ref.~\cite{peixoto_parsimonious_2013}. The use of hierarchical priors
solves this particular underfitting problem, as discussed in
Refs.~\cite{peixoto_hierarchical_2014,peixoto_nonparametric_2017}. Another
potential cause for underfitting is the use of Poisson formulations for
the SBM for networks with heterogeneous density, which assumes that the
observed simple graph is a possible realization of a multigraph model
that generates simple graphs with a very small
probability. Ref.~\cite{peixoto_latent_2020} introduced an alternative
SBM variation based on a simple but consequential modification of the
Poisson SBMs, where multigraphs are generated at a first stage, and the
multiedges are converted into simple edges, resulting in a Bernoulli
distribution obtained from the cumulative Poisson distribution. These
``latent Poisson'' SBMs also prevent underfitting, and in fact make the
posterior distribution concentrate on the correct answer for the
examples considered by Riolo and Newman~\cite{riolo_consistency_2020},
as shown in Ref.~\cite{peixoto_latent_2020}.

In Fig.~\ref{fig:polbooks-multi-modes} we show our method employed on
the posterior distribution of the political books network using the
latent Poisson DC-SBM with nested priors, which should be able to
correct the kinds of underfitting mentioned above. Indeed, the most
likely mode shows a more elaborate division of the network into $B=8$
groups, corresponding to particular subdivisions of the same
liberal-neutral-conservative groups seen previously. However, these
subdivisions are not quite the same as those seen in
Fig.~\ref{fig:polbooks_modes} for the Poisson SBM. Therefore, in this
example it would be futile to search for these uncovered groups in the
posterior distribution of the Poisson DC-SBM, even if we search for
overlaps between partitions. However, despite the more detailed division
of the network, the latent Poisson SBM is far from being a perfect fit
for this network, as we still observe $K=11$ modes, corresponding mostly
to different divisions of the ``conservative'' books. When comparing the
structure of the different modes, we see that these are not simple
combinations of the same subdivisions, but rather different
rearrangements. This seems to point to a kind of structure in the
network that is not fully captured by the strict division of the nodes
in discrete categories, at least not in the manner assumed by the SBM.

In Fig.~\ref{fig:karate_modes} we compare also the inferences obtained
with both SBM models for the karate club network considered
previously. The posterior distribution obtained with the Poisson DC-SBM
is very heterogeneous, with $K=30$ modes. It has as most plausible mode
one composed of a single partition into a single group (implying that
the degree sequence alone is enough to explain the network, and no
community structure is needed). The second most likely mode corresponds
to leader-follower partitions, largely dividing the nodes according to
degree (despite the degree correction). The putative division of this
network into two assortative communities comes only as the ninth most
likely mode. With such an extreme heterogeneity between partitions,
finding a consensus between them seems particularly futile, thus
explaining the obtained point estimates in
Fig.~\ref{fig:karate-centers}, in particular the odd behavior of the RMI
estimate that tries to assemble all diverging modes into a single
partition.  On the other hand, with the latent Poisson SBM the posterior
distribution changes drastically, as is shown in right panel of
Fig.~\ref{fig:karate_modes}. In this case the dominating mode
corresponds to partitions that, while not fully identical to the
accepted division, are more compatible with it, as they only further
divide one of the communities into two extra groups. The commonly
accepted division itself comes as a typical partition of the second most
likely mode. Overall, the posterior distribution becomes more
homogeneous, with only $K=9$ modes identified, and with most of the
posterior probability assigned to the first few.

It is important to observe that the heterogeneity of the posterior
distribution by itself cannot be used as a criterion in the decision of
which model is a better fit. Indeed, a typical behavior encountered in
statistical inference is the ``bias-variance
trade-off''~\cite{geman_neural_1992}, where a more accurate
representation of the data comes at the cost of increased variance in the
set of answers. We illustrate this with a network of American football
games~\cite{girvan_community_2002} shown in
Fig.~\ref{fig:football_modes}. The Poisson DC-SBM yields a very simple
posterior distribution, strongly concentrated on a typical partition
into $B=10$ groups. On the other hand, as seen in
Fig.~\ref{fig:football-multi-modes}, the latent Poisson DC-SBM yields a
more heterogeneous posterior distribution with $K=7$ modes, typically
uncovering a larger number of groups. It would be wrong to conclude that
the Poisson SBM provides a better fit only because it concentrates on a
single answer, if that single answer happens to be underfitting. But
from this analysis alone, it is not possible to say if the latent
Poisson SBM is not overfitting either. To make the final decision, we
need compute the total evidence for each model, as we will consider in
Sec.~\ref{sec:evidence}. This computation takes the heterogeneity of the
posterior distribution into consideration, but combined with the model
plausibility.

Before we proceed with model selection, we first show how the methods
constructed so far can be generalized for hierarchical partitions, which
form the basis of generically better-fitting models of community
structure in networks~\cite{peixoto_nonparametric_2017}.

\section{Hierarchical partitions}\label{sec:hierarchical}
\begin{figure*}
  \begin{tabular}{cccc}
    \multirow[b]{2}{*}[12.5em]{\includegraphics[width=.5\columnwidth]{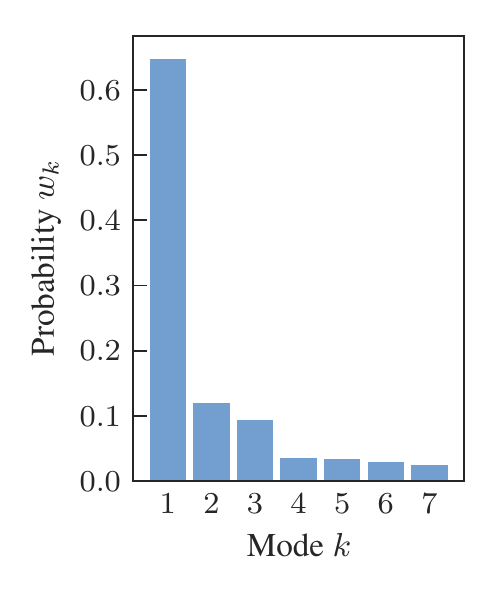}} &
    \includegraphics[width=.5\columnwidth, trim=.8cm .8cm .8cm .2cm]{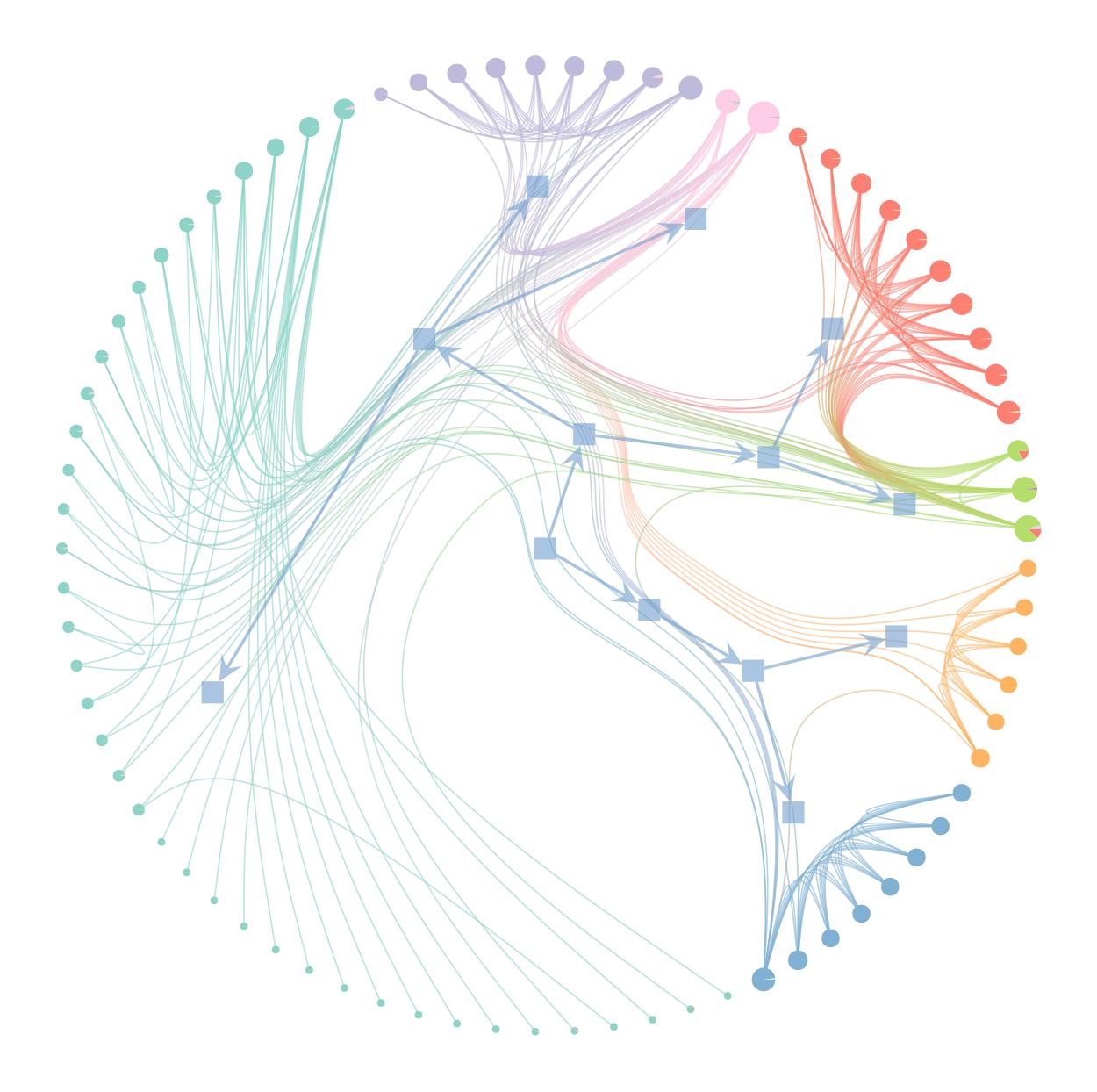}
    &
    \includegraphics[width=.5\columnwidth, trim=.8cm .8cm .8cm .2cm]{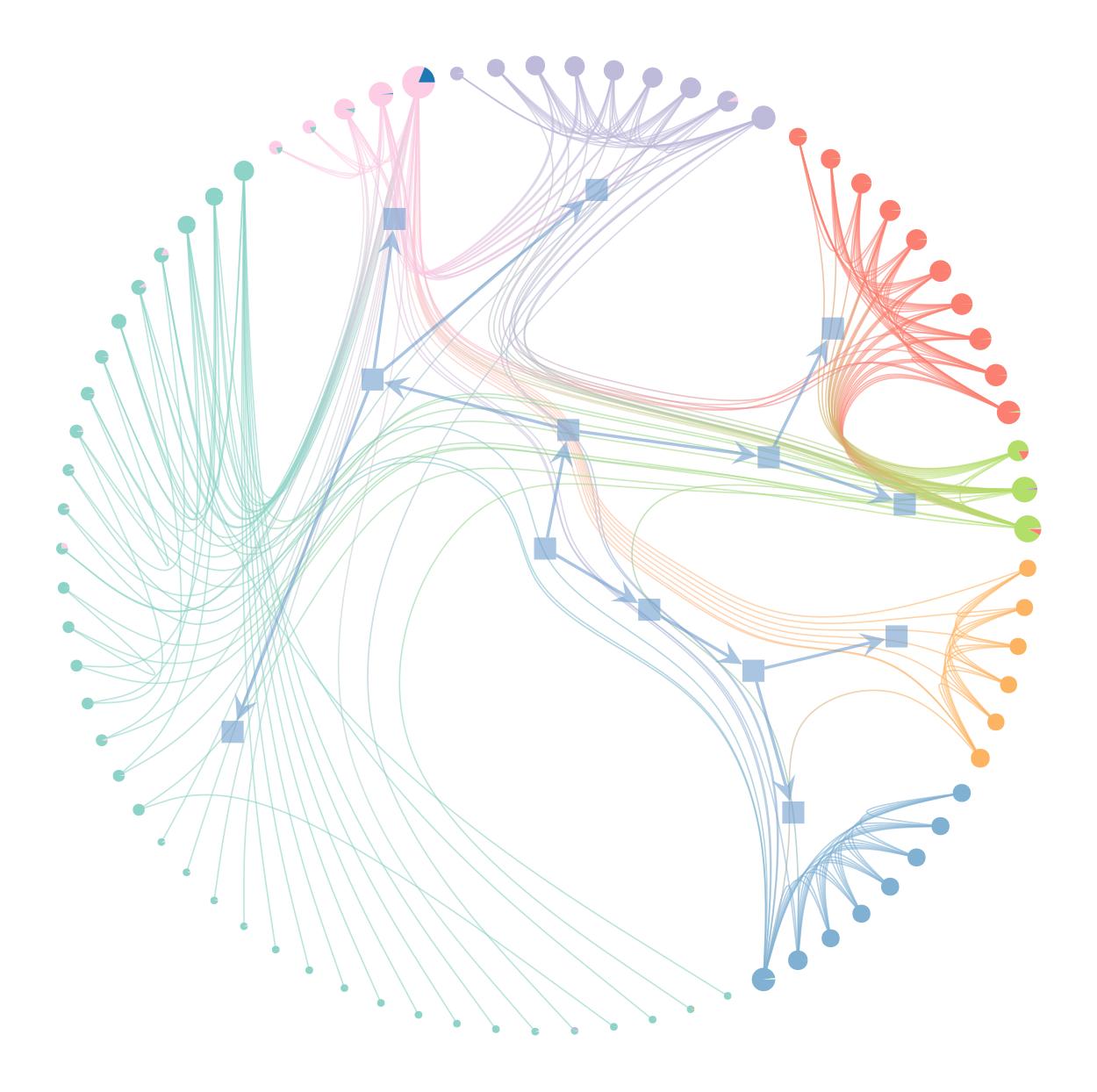}
    &
    \includegraphics[width=.5\columnwidth, trim=.8cm .8cm .8cm .2cm]{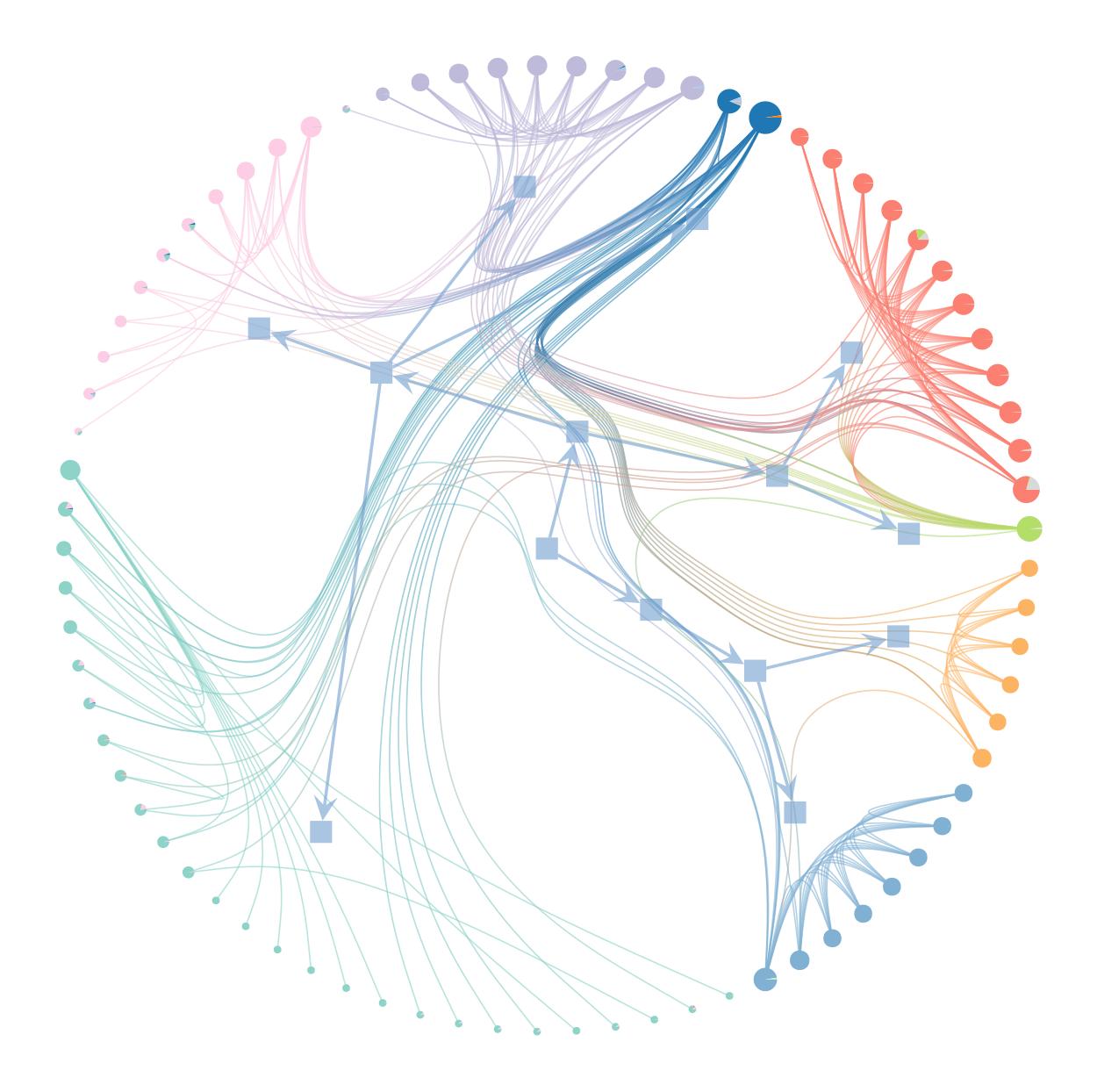}\\
    &
    (a) $k=1$ & (b) $k=2$ & (c) $k=3$
  \end{tabular}

  \caption{Inferred hierarchical partition modes from $M=10^5$ samples
    of the hierarchical latent Poisson DC-SBM posterior distribution for
    the co-occurrence network of characters of the Les Misérables
    novel. The left panel shows the mode fractions $w_k$, and the right
    panel the three largest modes, with the marginal distributions
    shown as pie diagrams on the nodes of the network.\label{fig:lesmis-modes}}
\end{figure*}

An important extension of SBM formulations is one where the choice of
priors is replaced by a nested sequence of priors and hyperpriors, where
groups of nodes are also clustered in their own meta-groups, associated
with a coarse-grained version of the network described via its own
smaller SBM, and so on recursively, resulting in a nested version of the
model~\cite{peixoto_hierarchical_2014,peixoto_nonparametric_2017}. This
hierarchical formulation recovers the usual SBMs when the hierarchy has
only a single level, and also introduces many useful properties,
including a dramatically reduced tendency to underfit large
networks~\cite{peixoto_hierarchical_2014,peixoto_nonparametric_2017} as
well a simultaneous description of the network structure at several
scales of resolution. This model variant takes as parameters a
hierarchical partition $\bar{\bb} = \{\bb_1,\dots,\bb_L\}$, where
 $b_i^{(l)}$ is the group membership of node $i$ in level $l$, and
each group label in level $l$ is a node in the above level $l+1$, which
results in the number of nodes in level $l$ being the number of groups
in the level below, $N_l=B_{l-1}$, except for the first level,
$N_1=N$. For this model, we have a posterior distribution over
hierarchical partitions given by
\begin{equation}
  \pi(\bar\bb) = \frac{P(\A|\bar\bb)P(\bar\bb)}{P(\A)}.
\end{equation}
Like in the non-hierarchical case, this posterior distribution is
invariant to label permutations, i.e.
\begin{equation}
  \pi(\bar\bb) = \pi(\bar\cc)
\end{equation}
if $\bar\bb$ and $\bar\cc$ are identical up a relabelling of the
groups. However in the hierarchical scenario the group relabellings that
keep the posterior distribution invariant must keep the same partitions
when projected at the lower levels. In other words, the invariant
permutation of the labels in level $l$ affects the nodes in level
$l+1$. More specifically, if we consider a bijection $\mu(r)$ for labels
at level $l$, such that $b_i^l(r) = \mu(c_i^l(r))$, then we must change
the membership in level $l+1$ to $b_{\mu(i)}^{l+1}=c_i^{l+1}$. If two
hierarchical partitions $\bar\bb$ and $\bar\cc$ are identical up to this
kind of transformation, we denote this with the indicator function
\begin{equation}
  [\bar\bb\sim\bar\cc] = 1,
\end{equation}
or $[\bar\bb\sim\bar\cc] = 0$ otherwise. Based on this, we can
generalize the random label model considered before to model
hierarchical partitions sampled from the posterior distribution. We
first assume that the labels at all levels are sampled independently as
\begin{equation}
  P_{\text{MF}}(\bar\cc|\bar{\bm p}) = \prod_{l=1}^{L}P_{\text{MF}}(\cc_l|\bm p_l),
\end{equation}
with
\begin{equation}
  P_{\text{MF}}(\cc_l|\bm p_l) = \prod_ip_i^l(c_i^l),
\end{equation}
where $p_i^l(r)$ is the probability that node $i$ in level $l$ belongs
to group $r$. After sampling a partition $\bar\cc$, we then obtain a
final partition $\bar\bb$ by choosing uniformly among all label
permutations, yielding
\begin{equation}\label{eq:hrlabel}
  P(\bar\bb|\bar{\bm p}) = \sum_{\bar\cc}P(\bar\bb|\bar\cc)P_{\text{MF}}(\bar\cc|\bar{\bm p}),
\end{equation}
where
\begin{equation}
  P(\bar\bb|\bar\cc) = \frac{[\bar\bb\sim\bar\cc]}{\prod_lq(\cc_l)!}.
\end{equation}
If we now consider $M$ sampled hierarchical partitions $\{\bar\bb\} =
\{\bar\bb^{(1)},\dots,\bar\bb^{(M)}\}$, the posterior distribution of the
hidden relabelled hierarchical partitions $\{\bar\cc\}$ is given by
\begin{multline}
  P(\{\bar\cc\}|\{\bar\bb\}, B_l)\propto
  \left(\prod_{m=1}^{M}[\bar\bb^{(m)}\sim\bar\cc^{(m)}]\right)\times\\
  \prod_l\prod_i\frac{(B_l-1)!}{(M+B_l-1)!}\prod_r n_i^{(l)}(r)!,
\end{multline}
where $n_i^{(l)}(r)=\sum_{m=1}^M\delta_{b_i^l,r}$ is how often node $i$
in level $l$ has group label $r$ in all samples. Similarly to before, if
we consider the conditional probability of a single partition
relabelling $\cc_l^{(m)}$, but marginalized over the upper levels
$l'>l$, we obtain
\begin{multline}
  P(\cc_l^{(m)} | \{\bar\bb\}, \{\bar\cc^{(m'\ne m)}\}, \{\cc_{l'< l}^{(m)}\}) \\
  \begin{aligned}
    &\propto
    \sum_{\cc_{l+1}^{(m)},\dots,\cc_{L}^{(m)}} P(\{\bar\cc\}|\{\bar\bb\})\\
    &\propto \prod_i\prod_r \left[{n_i'}^l(r)+1\right]^{\delta_{c_i^{l,m},r}},
  \end{aligned}
\end{multline}
where ${n'}^l_i(r)$ are the label counts excluding $\cc_l^{(m)}$. Just
like in the non-hierarchical case, we can write
\begin{equation}\label{eq:hmarg}
  \ln P(\cc_l^{(m)} | \{\bar\bb\}, \{\bar\cc^{(m'\ne m)}\}, \{\cc_{l'< l}^{(m)}\}) = \sum_{r}w_{r,\mu(r)},
\end{equation}
up to an unimportant additive constant, where
\begin{equation}
  w_{rs}=\sum_i\delta_{b_i^l,r}\ln\left[{n'}_i^l(s)+1\right],
\end{equation}
and $\mu(r)$ is the bijection that matches the groups labels between
$\cc_l^{(m)}$ and $\bb_l^{(m)}$. Therefore we can find the maximum of
Eq.~\ref{eq:hmarg} once more by solving the maximum weight bipartite
matching problem with weights given by $w_{rs}$. This leads to an
overall algorithm entirely analogous to the non-hierarchical case,
where, starting from some configuration, we remove a sample $m$ from the
ensemble, and add it again, choosing its labels according to the
maximization of Eq.~\ref{eq:hmarg}, starting from level $l=1$ and going
up until $l=L$, and stopping if such moves no longer increase the
posterior probability. Doing a relabel for every sample once takes time
$O[M\sum_l(N_l+ B_l^3)]$, where $N_l$ and $B_l$ are the typical number
of nodes and groups at level $l$. Typically, the number of groups
decreases exponentially with the hierarchical level,
$N_l=O(N/\sigma^{l-1})$ with $\sigma > 1$, so that we have $L = O(\log
N)$, and thus $\sum_lN_l = O(N)$, the entire running time for a single
``sweep'' over all samples is then simply $O[M(N + B^3)]$, where $B$ is
the number of labels in the first hierarchical level.

The mixed random label model of Sec.~\ref{sec:modes} can also be
generalized in a straightforward manner for hierarchical partitions,
i.e.
\begin{equation}
  P(\bar\bb|\bar{\bm p}, \w) = \sum_k P(\bar\bb|\bar{\bm p}, k)P(k|\w)
\end{equation}
where inside a mode $k$ the partitions are sampled according to the
hierarchical random label model given by Eq.~\ref{eq:hrlabel}. The
inference algorithm from this point onward is exactly the same as in
the non-hierarchical case, where we need only to relabel the
hierarchical partitions according to Eq.~\ref{eq:hmarg} when we move
them between modes.

In Fig.~\ref{fig:lesmis-modes} we show the inferred modes for
hierarchical partitions sampled from the posterior distribution using
the nested latent Poisson DC-SBM for a co-occurrence network of
characters of the Les Misérables novel~\cite{knuth_stanford_1993}. As
this example shows, this algorithm allows us to summarize a multimodal
distribution of hierarchical partitions in a rather compact manner. In
this particular example we see that the distribution is fairly dominated
by one of the modes (shown in Fig.~\ref{fig:lesmis-modes}a), followed by
less probable alternatives.

\subsubsection{Comparing and finding consensus between hierarchical partitions}

If we infer the hierarchical random label model above for two
hierarchical partitions $\bar\x$ and $\bar\y$, it amounts to solving a
recursive maximum bipartite weighted matching problem on every level,
starting from $l=1$ to $l=L$, using as weights the contingency table at
each level $l$,
\begin{equation}
  m_{rs}^{(l)} = \sum_{i\in \mathcal{N}_{\x^l}\cap \mathcal{N}_{\y^l}}\delta_{x_i^l,r}\delta_{y_i^l,s},
\end{equation}
where $\mathcal{N}_{\x}$ is the set of nodes in partition $\x$ (as
upper level partitions might have a disjoint set of nodes), and
propagating the matched labels to the upper levels. This is equivalent
to maximizing the recursive overlap across all levels
\begin{equation}
  w(\bar\x,\bar\y) = \sum_l\sum_i\delta_{x_i^l,\mu_l(\hat y_i^l)},
\end{equation}
where at each level we need to incorporate the relabeling at the
lower levels via
\begin{equation}
  \hat y_i^l = y^l_{\mu_{l-1}(i)}
\end{equation}
where $\bm\mu_l$ is a label bijection at level $l$, with the boundary
condition $\mu_0(i)=i$. This leads us to the hierarchical maximum
overlap distance, defined as
\begin{equation}
  d(\bar\x,\bar\y) = \sum_lN_l - \underset{\bm\mu_l}{\operatorname{argmax}}\sum_i\delta_{x_i^l,\mu_l(\hat y_i^l)},
\end{equation}
where $N_l=\max(|\mathcal{N}_{\x^l}|, |\mathcal{N}_{\y^l}|)$. A version
of this distance that is normalized in the range $[0,1]$ can be obtained
by dividing it by the largest possible value,
\begin{equation}
  \frac{d(\bar\x,\bar\y)}{\sum_l N_l - 1}.
\end{equation}
It is important to note here that hierarchy levels with a single node,
$N_l=1$, always have a contribution of zero to the distance, therefore
this measure can be applied to infinite hierarchies with $L\to\infty$,
as long as any level is eventually grouped into a single group. For
hierarchies with a single level, $L=1$, we recover the maximum overlap
distance considered previously, except for the normalized version, which
is slightly different with $d(\x,\y)/(N-1)$. This is also a valid
normalization for the non-hierarchical distance, since we must always
have $d(\x,\y) < N$. The label matching at level $l$ of the hierarchy
can be done in time $O[(q(\x_l)+q(\y_l))E_m^l + N_l]$, using the sparse
version of the Kuhn–Munkres
algorithm~\cite{kuhn_hungarian_1955,munkres_algorithms_1957,ramshaw_minimum-cost_2012},
where $E_m^l \le q(\x_l)q(\y_l)$ is the the number of nonzero entries in
the contingency matrix $m_{rs}$. If we assume once more the typical case
with $N_l=O(N/\sigma^{l-1})$ and $L=O(\log N)$, so that $\sum_lN_l=O(N)$,
the overall computation can then be done in time
$O[(q(\x^1)+q(\y^1))E_m^1 + N]$.

Following the same steps as before, we can use the hierarchical maximum
overlap distance as an error function
$\epsilon(\bar\x,\bar\y)=d(\bar\x,\bar\y)$ to define a MOC estimator
over hierarchical partitions based on the minimization of the mean posterior
loss,
\begin{equation}
  \hat{\bar\bb} =
  \underset{\bar\bb}{\operatorname{argmin}}\;\sum_{\bar\bb'}\epsilon(\bar\bb,\bar\bb')P(\bar\bb'|\A).
\end{equation}
Substituting its definition leads us to a set of self-consistent
equations at each level $l$,
\begin{equation}
  \hat b_i^l = \underset{r}{\operatorname{argmax}}\;\hat\pi_i^l(r|\{\mu_{\bb}^l\}),
\end{equation}
with the marginal distributions obtained over the relabeled partitions,
\begin{equation}
  \hat\pi_i^l(r|\{\mu_{\bb}^l\}) = \sum_{\bar\bb}\delta_{\mu_{\bb}^l(\tilde b_i^l),r}P(\bar\bb|\A),
\end{equation}
where the relabeling is done in order to maximize the overlap with $\hat{\bar\bb}$,
\begin{equation}
  \bm\mu_{\bb}^l = \underset{\bm\mu}{\operatorname{argmax}} \sum_i\delta_{\hat b_i, \mu(\tilde b_i)}.
\end{equation}
and where once again we need to recursively incorporate the relabellings at the lower levels,
\begin{equation}
  \tilde b_i^l = b^l_{\mu_{l-1}(i)}.
\end{equation}
We can define an uncertainty $\sigma_{\hat{\bar\bb}} \in [0,1]$ for this
estimator by inspecting the marginal distributions computed along the
way,
\begin{equation}
  \sigma_{\hat{\bar\bb}} = 1 - \frac{1}{N-L}\sum_l\frac{N_l-1}{N_l}\sum_i\hat\pi_i(\hat{b}_i^l|\{\mu_{\bb}^l\}).
\end{equation}
In the above sum, we omit levels with $N_l=1$ since those always have a
trivial marginal distribution concentrated on a single group. In
practice we implement this estimator by sampling a set of $M$
hierarchical partitions $\{\bar\bb\}$ from the posterior distribution
and then performing the sequential maximizations starting from $l=1$ to
$l=L$,
\begin{align}
  \hat b^l_i &= \underset{r}{\operatorname{argmax}}\;\sum_m \delta_{\mu_m(\tilde b^{l,m}_i), r}\\
  \bm\mu_m^l &= \underset{\bm\mu}{\operatorname{argmax}} \sum_r\hat m^{(l,m)}_{r,\mu(r)},
\end{align}
where $m^{(l,m)}_{r,s}$ is the contingency table of level $l$ of sample
$m$ with $\hat\bb^l$. The final solution is obtained when repeating the
above maximization no longer changes the result. Like in the
non-hierarchical case, this algorithm yields a local optimum of the
optimization problem, but not necessarily a global one, therefore it
needs to be repeated multiple times with different initial conditions,
and the best result kept. Since it involves the relabelling over all $M$
hierarchical partitions, the overall algorithmic complexity of a single
iteration is $O(MNB + MB^3)$, assuming once more the typical case with
$N_l=O(N/\sigma^{l-1})$ and $L=O(\log N)$.

\section{Model selection and evidence approximation}\label{sec:evidence}

\begin{table*}
  \resizebox{\textwidth}{!}{
\begin{tabular}{l|S[table-format = 5.1]S[table-format = 5.1]S[table-format = 5.1]S[table-format = 5.1]|S[table-format = 5.1]S[table-format = 5.1]S[table-format = 5.1]S[table-format = 5.1]|S[table-format = 5.1]S[table-format = 5.1]S[table-format = 5.1]S[table-format = 5.1]}
\multirow{3}{0pt}{Data} & \multicolumn{4}{c|}{Poisson} & \multicolumn{4}{c|}{Latent Poisson} & \multicolumn{4}{c}{Single partition}\\
& \multicolumn{2}{c}{Non-nested} & \multicolumn{2}{c|}{Nested} & \multicolumn{2}{c}{Non-nested} & \multicolumn{2}{c|}{Nested} & \multicolumn{2}{c}{Non-nested} & \multicolumn{2}{c}{Nested} \\
& {NDC} & {DC} & {NDC} & {DC} & {NDC} & {DC} & {NDC} & {DC} & {NDC} & {DC} & {NDC} & {DC} \\\hline\hline
Karate club~\cite{zachary_information_1977} & 213.1& 220.3& {\cellcolor[HTML]{E0E0E0}} 212.6& 221.7& 174.0& 172.4& {\cellcolor[HTML]{C0C0C0}} 170.6& 171.6& {\cellcolor[HTML]{E0E0E0}} 215.3& 222.7& {\cellcolor[HTML]{E0E0E0}} 215.3& 222.7\\
Dolphins~\cite{lusseau_bottlenose_2003} & 522.4& 539.3& {\cellcolor[HTML]{E0E0E0}} 522.1& 540.1& 480.9& 483.6& {\cellcolor[HTML]{C0C0C0}} 477.6& 478.7& {\cellcolor[HTML]{E0E0E0}} 529.6& 544.1& {\cellcolor[HTML]{E0E0E0}} 529.6& 544.1\\
Les Misérables~\cite{knuth_stanford_1993} & 674.1& 680.1& {\cellcolor[HTML]{E0E0E0}} 667.5& 672.4& 513.7& 471.0& 454.6& {\cellcolor[HTML]{C0C0C0}} 402.7& {\cellcolor[HTML]{E0E0E0}} 688.7& 697.6& {\cellcolor[HTML]{E0E0E0}} 688.7& 697.6\\
Political books~\cite{krebs_political_nodate} & 1305.2& 1334.4& {\cellcolor[HTML]{E0E0E0}} 1288.8& 1330.8& 1188.2& 1178.6& {\cellcolor[HTML]{C0C0C0}} 1136.7& 1137.4& 1321.9& 1343.4& {\cellcolor[HTML]{E0E0E0}} 1317.4& 1343.4\\
American football~\cite{girvan_community_2002} & 1722.4& 1769.2& {\cellcolor[HTML]{E0E0E0}} 1709.7& 1755.7& 1427.7& 1505.8& {\cellcolor[HTML]{C0C0C0}} 1319.8& 1373.1& 1738.9& 1785.9& {\cellcolor[HTML]{E0E0E0}} 1733.5& 1780.6\\
Network scientist~\cite{newman_finding_2006} & 3871.5& 3869.5& {\cellcolor[HTML]{E0E0E0}} 3592.6& 3645.1& 3728.4& 3611.5& 3059.9& {\cellcolor[HTML]{C0C0C0}} 3043.6& 4007.8& 3982.2& {\cellcolor[HTML]{E0E0E0}} 3813.4& 3826.2\\
High school~\cite{harris_national_2009} & 4530.5& 4620.6& {\cellcolor[HTML]{E0E0E0}} 4482.8& 4592.3& 4378.1& 4421.7& {\cellcolor[HTML]{C0C0C0}} 4257.4& 4307.6& 4599.9& 4676.8& {\cellcolor[HTML]{E0E0E0}} 4585.9& 4668.2\\
\emph{C. elegans} neurons~\cite{white_structure_1986} & 6968.2& 7040.3& {\cellcolor[HTML]{E0E0E0}} 6812.7& 6943.0& 6492.3& 6485.7& {\cellcolor[HTML]{C0C0C0}} 6048.3& 6411.3& 7043.7& 7144.4& {\cellcolor[HTML]{E0E0E0}} 6959.5& 7091.3\\
E-mail~\cite{guimera_self-similar_2003} & 25020.5& 24845.5& {\cellcolor[HTML]{E0E0E0}} 24145.3& 24264.8& 24577.1& 24047.4& 23544.7& {\cellcolor[HTML]{C0C0C0}} 23002.0& 25617.1& 25311.2& 25163.8& {\cellcolor[HTML]{E0E0E0}} 25094.7\\
Political blogs~\cite{adamic_political_2005} & 51389.1& 50638.2& 50528.9& {\cellcolor[HTML]{E0E0E0}} 50138.0& 47787.8& 46380.7& 46065.2& {\cellcolor[HTML]{C0C0C0}} 45006.4& 51639.1& 51084.1& 51195.2& {\cellcolor[HTML]{E0E0E0}} 50892.7\\
\end{tabular}}
\caption{Description length (negative log-evidence) $\Sigma=-\ln P(\A)$
for several networks and SBM variations, with DC and NDC indicating
degree-correction and not, respectively. The shaded cells indicate the
smallest value for the each model class, with the dark grey indicating
the best fitting model overall. The ``single partition'' columns
correspond to the two-part description length $\Sigma=-\ln P(\A,\bb)$
obtained with the best-fitting partition of the Poisson
model. \label{table:model-selection}}
\end{table*}

If we are interested in comparing two models $\mathcal{M}_1$ and
$\mathcal{M}_2$ in their plausibility for generating some network $\A$,
we can do so by computing the ratio of their posterior probability given
the data,
\begin{equation}
  \frac{P(\mathcal{M}_1|\A)}{P(\mathcal{M}_2|\A)} =
  \frac{P(\A|\mathcal{M}_1) P(\mathcal{M}_1)}{P(\A|\mathcal{M}_2) P(\mathcal{M}_2)}.
\end{equation}
Therefore, if we are \emph{a priori} agnostic about either model with
$P(\mathcal{M}_1)=P(\mathcal{M}_2)$, this ratio will be determined by
the total probability of the data $P(\A|\mathcal{M})$ according to that
model. This quantity is called the evidence, and appears as a
normalization constant in the posterior distribution of
Eq.~\ref{eq:sbm_posterior}. For any particular choice of model, it is
obtained by summing the joint probability of data and partitions over
all possible partitions (we drop the explicit dependence on
$\mathcal{M}$ from now on, to unclutter the expressions),
\begin{equation}
  P(\A) = \sum_{\bb}P(\A,\bb).
\end{equation}
Unfortunately, the exact computation of this sum is intractable since
the number of partitions is too large in most cases of interest. It
also cannot be obtained directly from samples of the posterior
distribution, which makes its estimation from MCMC also very
challenging. To illustrate this, it is useful to write the logarithm of
the evidence in the following manner,
\begin{align}\label{eq:free-energy}
  \ln P(\A) &= \sum_{\bb}\pi(\bb)\ln P(\A,\bb) - \sum_{\bb}\pi(\bb)\ln \pi(\bb)\\
            &= \avg{\ln P(\A,\bb)} + H(b)
\end{align}
where
\begin{equation}
  \pi(\bb) = \frac{P(\A,\bb)}{\sum_{\bb'}P(\A,\bb')} = \frac{P(\A,\bb)}{P(\A)}
\end{equation}
is the posterior distribution of Eq.~\ref{eq:sbm_posterior},
and
\begin{equation}
  \avg{\ln P(\A,\bb)} = \sum_{\bb}\pi(\bb)\ln P(\A,\bb),
\end{equation}
is the mean joint log-probability computed over the posterior
distribution, and finally
\begin{align}
  H(b) = -\sum_{\bb}\pi(\bb)\ln \pi(\bb)
\end{align}
is the entropy of the posterior distribution. Eq.~\ref{eq:free-energy}
has the shape of a negative Gibbs free energy of a physical ensemble, if
we interpret $\avg{\ln P(\A,\bb)}$ as the mean negative ``energy'' over
the ensemble of partitions. It tells us that what contributes to the
evidence is not only the mean joint probability, but also the
multiplicity of solutions with similar probabilities, which is captured
by the posterior entropy. In this formulation, we see that while it is
possible to estimate $\avg{\ln P(\A,\bb)}$ from MCMC simply be averaging
$\ln P(\A,\bb)$ for sufficiently many samples, the same approach does
not work for the entropy term $H(b)$, since it would require the
computation of the log-posterior $\ln\pi(\bb)$ for every sample,
something that cannot be done without knowing the normalization constant
$P(\A)$, which is what we want to find in the first place. However, the
mixed random label model of Sec.~\ref{sec:modes} can be used to fit the
posterior distribution, allowing us to compute the entropy term via the
inferred model, and use and the rich information gained on its structure
to perform model selection. Let us recall that the mixed random label
model, when inferred from partitions sampled from $\pi(\bb)$, amounts to
an approximation given by
\begin{align}
  \pi(\bb) \approx \sum_{k,\cc}P(\bb|\cc)P(\cc|k)P(k)
\end{align}
where $P(k) = w_k$ determines the mode mixture and
\begin{equation}
  P(\cc|k) = \prod_i p_i^{(k)}(c_i),
\end{equation}
are the independent marginal distributions of mode $k$ and finally
\begin{equation}
  P(\bb|\cc) = \frac{[\bb\sim\cc]}{q(\bb)!}
\end{equation}
is the random relabeling of groups. In most cases we have investigated,
the inferred modes tend to be very well separated (otherwise they would
get merged together into a larger mode), such that we can assume
\begin{align}
  \pi(\bb) \approx \underset{k, \cc}{\operatorname{max}}\; P(\bb|\cc)P(\cc|k)P(k),
\end{align}
This means we can write the entropy as
\begin{align}
  H(b) \approx H(b,c,k) = H(b|c)+H(c|k)+H(k)
\end{align}
where
\begin{equation}
  H(k) = -\sum_k w_k \ln w_k
\end{equation}
is the entropy of the mode mixture distribution, and
\begin{align}
  H(c|k) & = -\sum_k w_k \sum_{\cc} P(\cc|k)\ln P(\cc|k)\\
  &= -\sum_k w_k \sum_i\sum_rp_i^{(k)}(r)\ln p_i^{(k)}(r)
\end{align}
is the entropy of mode $k$ and
\begin{align}
  H(b|c) &= -\sum_{\cc} P(\cc) \sum_{\bb}P(\bb|\cc)\ln P(\bb|\cc)\\
  &= \sum_{\cc} P(\cc) \ln q(\cc)! = \sum_{\bb} P(\bb) \ln q(\bb)!,
\end{align}
is the relabelling entropy. Putting it all together we have the
following approximation for the evidence according to the mixed random
label model,
\begin{multline}
  \ln P(\A) \approx \avg{\ln P(\A,\bb)} + \avg{\ln q(\bb)!}  -\sum_k w_k \ln w_k\\
  -\sum_k w_k \sum_i\sum_rp_i^{(k)}(r)\ln p_i^{(k)}(r).
\end{multline}
We can extend this for hierarchical partitions in an entirely analogous
way, which leads to
\begin{multline}
  \ln P(\A) \approx \avg{\ln P(\A,\bar\bb)} + \sum_l\avg{\ln q(\bb^l)!}\\
  -\sum_k w_k \ln w_k
  -\sum_k w_k \sum_l\sum_i\sum_rp_i^{(l,k)}(r)\ln p_i^{(l,k)}(r).
\end{multline}
The above quantities are then computed by sampling $M$ partitions from
the posterior distribution, using them (or a superset thereof) to
compute the first two means $\avg{\ln P(\A,\bb)}$ and $\avg{\ln
q(\bb)!}$, and then fit the mixed random label model, from which the
parameters $\bm{w}$ and $\bm p$ are obtained, and then computing the
remaining terms.

In Table~\ref{table:model-selection} we show the evidence obtained for
several SBM variants and datasets, including latent Poisson versions
(which require special considerations, see
Appendix~\ref{app:latent_multigraph}). Overall, we find that when
considering the Poisson SBMs, degree correction is only favored for
larger networks, corroborating a similar previous analysis based on a
less accurate calculation~\cite{peixoto_nonparametric_2017}. This
changes for latent Poisson models, where for some networks the balance
tips in favor of degree correction. Overall, we find more evidence for
the latent Poisson models for all networks considered, which is
unsurprising given that they are all simple graphs. Likewise, we always
find more evidence for the hierarchical SBMs, which further demonstrate
their more flexible nature.

\subsection{Bayesian evidence and the minimum description length (MDL) criterion}

In this section we explore briefly some direct connections between
Bayesian model selection and the minimum description length (MDL)
criterion based on information theory~\cite{grunwald_minimum_2007}. We
begin by pointing out the simple fact that the MAP point estimate given
by the single most likely partition yields a lower bound for the
evidence, i.e.
\begin{equation}\label{eq:evidence_bound}
P(\A) = \sum_{\bb}P(\A,\bb) \ge \underset{\bb}{\operatorname{max}}\;P(\A,\bb).
\end{equation}
This means that taking into account the full posterior distribution,
rather than only its maximum, almost always can be used to compress the
data, as we now show. We can see this by inspecting first the usual
``two-part'' description length,
\begin{align}
  \Sigma_1(\A,\bb) &= -\ln P(\A,\bb)\\
  &= -\ln P(\A|\bb)  -\ln P(\bb)
\end{align}
which corresponds to amount of information necessary to describe the
data if one first describes the partition $\bb$ and then, conditioned on
it, the network $\A$. Therefore, finding the most likely partition $\bb$
means finding the one that most compresses the network, according to
this particular two-part encoding. However, the full posterior
distribution gives us a more efficient ``one-part'' encoding where no
explicit description of the partition is necessary. Simply defining the
joint distribution $P(\A,\bb)$ means we can compute the marginal
probability $P(\A) = \sum_{\bb} P(\A,\bb)$, which yields directly a
description length
\begin{equation}
  \Sigma_2(\A) = -\ln P(\A).
\end{equation}
According to Eq.~\ref{eq:evidence_bound} we have
\begin{equation}
  \Sigma_2(\A) \le \underset{\bb}{\operatorname{min}}\;\Sigma_1(\A,\bb),
\end{equation}
which means that considering all possible partitions can only increase
the overall compression achievable. In Table~\ref{table:model-selection}
we can verify that this holds for all results obtained.

In a slightly more concrete setting, let us consider a transmitter who
wants to convey the network $\A$ to a receiver, who both know the joint
distribution $P(\A,\bb)$. According to the two-part code, the
transmitter first sends the partition $\bb$, for that using $-\log_2
P(\bb)$ bits, and then sends the final network using $-\log_2 P(\A|\bb)$
bits, using in total $\Sigma_1(\A,\bb)/\ln 2$ bits. In practice, this is
achieved, for example, by both sender and receiver sharing the same two
tables of optimal prefix codes derived from $P(\bb)$ and $P(\A|\bb)$. On
the other hand, using the second one-part code, both transmitter and
receiver share only a single table of optimal prefix codes derived
directly from the marginal distribution $P(\A)$, which means that only
$\Sigma_2(\A)/\ln 2 = -\log_2P(\A)$ bits need to be transmitted.  In
practice, it will be more difficult to construct the one-part code since
it involves marginalizing over a high-dimensional distribution, which is
intractable via brute force --- although our mixed random label model
can be used as a basis of an analytical approximation. However what is
important in our model selection context is only that such a code
exists, not its computational tractability.

\section{Conclusion}
\label{sec:conclusion}

We have shown how the random label model can be used to solve the group
identification problem in community detection, allowing us to compute
marginal distributions of group membership on the nodes in a unambiguous
way. This led us to the notion of maximum overlap distance as a general
way of comparing two network partitions, which we then used as a loss
function to obtain the consensus of a population of network
partitions. By investigating the behavior of different loss functions on
artificial and empirical ensembles of heterogeneous partitions, we have
demonstrated that they can yield inconsistent results, due precisely to
a lack of uniformity between divisions. We then developed a more
comprehensive characterization of the posterior distribution, based on a
mixed version of the random label model that is capable of describing
multimodal populations of partitions, where multiple consensuses exist
at the same time. This kind of structure corresponds to a ``multiple
truths'' phenomenon, where a model can yield diverging hypotheses for the
same data. We showed how our method provides a compact representation
for structured populations of network partitions, and allows us to
assess quality of fit and perform model selection. The latter was
achieved by using the multimodal fit of the posterior distribution as a
proxy for the computation of its entropy, which is a key but often
elusive ingredient in Bayesian model selection.

Although we have focused on community detection, the methods developed
here are applicable for any kind of clustering problem from which a
population of answers can be produced. They allow us to be more detailed
in our assessment of the consistency of results when applied to real or
artificial data. In particular, we no longer need to rely on ``point
estimates'' that can give a very misleading picture of high dimensional
and structured populations of partitions, even if they attempt to
assemble a consensus among them. We achieve this without losing
interpretability, as our method yields groupings of partitions that
share a local consensus, each telling a different version of how the
data might have been generated, and weighted according to the
statistical evidence available.

\bibliography{bib}

\begin{thebibliography}{54}%
\makeatletter
\providecommand \@ifxundefined [1]{%
 \@ifx{#1\undefined}
}%
\providecommand \@ifnum [1]{%
 \ifnum #1\expandafter \@firstoftwo
 \else \expandafter \@secondoftwo
 \fi
}%
\providecommand \@ifx [1]{%
 \ifx #1\expandafter \@firstoftwo
 \else \expandafter \@secondoftwo
 \fi
}%
\providecommand \natexlab [1]{#1}%
\providecommand \enquote  [1]{``#1''}%
\providecommand \bibnamefont  [1]{#1}%
\providecommand \bibfnamefont [1]{#1}%
\providecommand \citenamefont [1]{#1}%
\providecommand \href@noop [0]{\@secondoftwo}%
\providecommand \href [0]{\begingroup \@sanitize@url \@href}%
\providecommand \@href[1]{\@@startlink{#1}\@@href}%
\providecommand \@@href[1]{\endgroup#1\@@endlink}%
\providecommand \@sanitize@url [0]{\catcode `\\12\catcode `\$12\catcode
  `\&12\catcode `\#12\catcode `\^12\catcode `\_12\catcode `\%12\relax}%
\providecommand \@@startlink[1]{}%
\providecommand \@@endlink[0]{}%
\providecommand \url  [0]{\begingroup\@sanitize@url \@url }%
\providecommand \@url [1]{\endgroup\@href {#1}{\urlprefix }}%
\providecommand \urlprefix  [0]{URL }%
\providecommand \Eprint [0]{\href }%
\providecommand \doibase [0]{http://dx.doi.org/}%
\providecommand \selectlanguage [0]{\@gobble}%
\providecommand \bibinfo  [0]{\@secondoftwo}%
\providecommand \bibfield  [0]{\@secondoftwo}%
\providecommand \translation [1]{[#1]}%
\providecommand \BibitemOpen [0]{}%
\providecommand \bibitemStop [0]{}%
\providecommand \bibitemNoStop [0]{.\EOS\space}%
\providecommand \EOS [0]{\spacefactor3000\relax}%
\providecommand \BibitemShut  [1]{\csname bibitem#1\endcsname}%
\let\auto@bib@innerbib\@empty
\bibitem [{\citenamefont {Fortunato}(2010)}]{fortunato_community_2010}%
  \BibitemOpen
  \bibfield  {author} {\bibinfo {author} {\bibfnamefont {Santo}\ \bibnamefont
  {Fortunato}},\ }\bibfield  {title} {\enquote {\bibinfo {title} {Community
  detection in graphs},}\ }\href {\doibase 16/j.physrep.2009.11.002} {\bibfield
   {journal} {\bibinfo  {journal} {Physics Reports}\ }\textbf {\bibinfo
  {volume} {486}},\ \bibinfo {pages} {75--174} (\bibinfo {year}
  {2010})}\BibitemShut {NoStop}%
\bibitem [{\citenamefont {Fortunato}\ and\ \citenamefont
  {Hric}(2016)}]{fortunato_community_2016}%
  \BibitemOpen
  \bibfield  {author} {\bibinfo {author} {\bibfnamefont {Santo}\ \bibnamefont
  {Fortunato}}\ and\ \bibinfo {author} {\bibfnamefont {Darko}\ \bibnamefont
  {Hric}},\ }\bibfield  {title} {\enquote {\bibinfo {title} {Community
  detection in networks: {A} user guide},}\ }\href {\doibase
  10.1016/j.physrep.2016.09.002} {\bibfield  {journal} {\bibinfo  {journal}
  {Physics Reports}\ } (\bibinfo {year} {2016}),\
  10.1016/j.physrep.2016.09.002}\BibitemShut {NoStop}%
\bibitem [{\citenamefont {Good}\ \emph {et~al.}(2010)\citenamefont {Good},
  \citenamefont {de~Montjoye},\ and\ \citenamefont
  {Clauset}}]{good_performance_2010}%
  \BibitemOpen
  \bibfield  {author} {\bibinfo {author} {\bibfnamefont {Benjamin~H.}\
  \bibnamefont {Good}}, \bibinfo {author} {\bibfnamefont {Yves-Alexandre}\
  \bibnamefont {de~Montjoye}}, \ and\ \bibinfo {author} {\bibfnamefont {Aaron}\
  \bibnamefont {Clauset}},\ }\bibfield  {title} {\enquote {\bibinfo {title}
  {Performance of modularity maximization in practical contexts},}\ }\href
  {\doibase 10.1103/PhysRevE.81.046106} {\bibfield  {journal} {\bibinfo
  {journal} {Physical Review E}\ }\textbf {\bibinfo {volume} {81}},\ \bibinfo
  {pages} {046106} (\bibinfo {year} {2010})}\BibitemShut {NoStop}%
\bibitem [{\citenamefont {Brandes}\ \emph {et~al.}(2006)\citenamefont
  {Brandes}, \citenamefont {Delling}, \citenamefont {Gaertler}, \citenamefont
  {Görke}, \citenamefont {Hoefer}, \citenamefont {Nikoloski},\ and\
  \citenamefont {Wagner}}]{brandes_modularity-np-completeness_2006}%
  \BibitemOpen
  \bibfield  {author} {\bibinfo {author} {\bibfnamefont {U.}~\bibnamefont
  {Brandes}}, \bibinfo {author} {\bibfnamefont {D.}~\bibnamefont {Delling}},
  \bibinfo {author} {\bibfnamefont {M.}~\bibnamefont {Gaertler}}, \bibinfo
  {author} {\bibfnamefont {R.}~\bibnamefont {Görke}}, \bibinfo {author}
  {\bibfnamefont {M.}~\bibnamefont {Hoefer}}, \bibinfo {author} {\bibfnamefont
  {Z.}~\bibnamefont {Nikoloski}}, \ and\ \bibinfo {author} {\bibfnamefont
  {D.}~\bibnamefont {Wagner}},\ }\bibfield  {title} {\enquote {\bibinfo {title}
  {On modularity-np-completeness and beyond},}\ }\href@noop {} {\bibfield
  {journal} {\bibinfo  {journal} {ITI Wagner, Faculty of Informatics,
  Universität Karlsruhe (TH), Tech. Rep}\ }\textbf {\bibinfo {volume} {19}},\
  \bibinfo {pages} {2006} (\bibinfo {year} {2006})}\BibitemShut {NoStop}%
\bibitem [{\citenamefont {Decelle}\ \emph {et~al.}(2011)\citenamefont
  {Decelle}, \citenamefont {Krzakala}, \citenamefont {Moore},\ and\
  \citenamefont {Zdeborová}}]{decelle_asymptotic_2011}%
  \BibitemOpen
  \bibfield  {author} {\bibinfo {author} {\bibfnamefont {Aurelien}\
  \bibnamefont {Decelle}}, \bibinfo {author} {\bibfnamefont {Florent}\
  \bibnamefont {Krzakala}}, \bibinfo {author} {\bibfnamefont {Cristopher}\
  \bibnamefont {Moore}}, \ and\ \bibinfo {author} {\bibfnamefont {Lenka}\
  \bibnamefont {Zdeborová}},\ }\bibfield  {title} {\enquote {\bibinfo {title}
  {Asymptotic analysis of the stochastic block model for modular networks and
  its algorithmic applications},}\ }\href {\doibase 10.1103/PhysRevE.84.066106}
  {\bibfield  {journal} {\bibinfo  {journal} {Physical Review E}\ }\textbf
  {\bibinfo {volume} {84}},\ \bibinfo {pages} {066106} (\bibinfo {year}
  {2011})}\BibitemShut {NoStop}%
\bibitem [{\citenamefont {Guimerà}\ and\ \citenamefont
  {Sales-Pardo}(2009)}]{guimera_missing_2009}%
  \BibitemOpen
  \bibfield  {author} {\bibinfo {author} {\bibfnamefont {Roger}\ \bibnamefont
  {Guimerà}}\ and\ \bibinfo {author} {\bibfnamefont {Marta}\ \bibnamefont
  {Sales-Pardo}},\ }\bibfield  {title} {\enquote {\bibinfo {title} {Missing and
  spurious interactions and the reconstruction of complex networks},}\ }\href
  {\doibase 10.1073/pnas.0908366106} {\bibfield  {journal} {\bibinfo  {journal}
  {Proceedings of the National Academy of Sciences}\ }\textbf {\bibinfo
  {volume} {106}},\ \bibinfo {pages} {22073 --22078} (\bibinfo {year}
  {2009})}\BibitemShut {NoStop}%
\bibitem [{\citenamefont {Clauset}\ \emph {et~al.}(2008)\citenamefont
  {Clauset}, \citenamefont {Moore},\ and\ \citenamefont
  {Newman}}]{clauset_hierarchical_2008}%
  \BibitemOpen
  \bibfield  {author} {\bibinfo {author} {\bibfnamefont {Aaron}\ \bibnamefont
  {Clauset}}, \bibinfo {author} {\bibfnamefont {Cristopher}\ \bibnamefont
  {Moore}}, \ and\ \bibinfo {author} {\bibfnamefont {M.~E.~J.}\ \bibnamefont
  {Newman}},\ }\bibfield  {title} {\enquote {\bibinfo {title} {Hierarchical
  structure and the prediction of missing links in networks},}\ }\href
  {\doibase 10.1038/nature06830} {\bibfield  {journal} {\bibinfo  {journal}
  {Nature}\ }\textbf {\bibinfo {volume} {453}},\ \bibinfo {pages} {98--101}
  (\bibinfo {year} {2008})}\BibitemShut {NoStop}%
\bibitem [{\citenamefont {Calatayud}\ \emph {et~al.}(2019)\citenamefont
  {Calatayud}, \citenamefont {Bernardo-Madrid}, \citenamefont {Neuman},
  \citenamefont {Rojas},\ and\ \citenamefont
  {Rosvall}}]{calatayud_exploring_2019}%
  \BibitemOpen
  \bibfield  {author} {\bibinfo {author} {\bibfnamefont {Joaquín}\
  \bibnamefont {Calatayud}}, \bibinfo {author} {\bibfnamefont {Rubén}\
  \bibnamefont {Bernardo-Madrid}}, \bibinfo {author} {\bibfnamefont {Magnus}\
  \bibnamefont {Neuman}}, \bibinfo {author} {\bibfnamefont {Alexis}\
  \bibnamefont {Rojas}}, \ and\ \bibinfo {author} {\bibfnamefont {Martin}\
  \bibnamefont {Rosvall}},\ }\bibfield  {title} {\enquote {\bibinfo {title}
  {Exploring the solution landscape enables more reliable network community
  detection},}\ }\href {\doibase 10.1103/PhysRevE.100.052308} {\bibfield
  {journal} {\bibinfo  {journal} {Physical Review E}\ }\textbf {\bibinfo
  {volume} {100}},\ \bibinfo {pages} {052308} (\bibinfo {year} {2019})},\
  \bibinfo {note} {publisher: American Physical Society}\BibitemShut {NoStop}%
\bibitem [{\citenamefont {Riolo}\ and\ \citenamefont
  {Newman}(2020)}]{riolo_consistency_2020}%
  \BibitemOpen
  \bibfield  {author} {\bibinfo {author} {\bibfnamefont {Maria~A.}\
  \bibnamefont {Riolo}}\ and\ \bibinfo {author} {\bibfnamefont {M.~E.~J.}\
  \bibnamefont {Newman}},\ }\bibfield  {title} {\enquote {\bibinfo {title}
  {Consistency of community structure in complex networks},}\ }\href {\doibase
  10.1103/PhysRevE.101.052306} {\bibfield  {journal} {\bibinfo  {journal}
  {Physical Review E}\ }\textbf {\bibinfo {volume} {101}},\ \bibinfo {pages}
  {052306} (\bibinfo {year} {2020})},\ \bibinfo {note} {publisher: American
  Physical Society}\BibitemShut {NoStop}%
\bibitem [{\citenamefont {Strehl}\ and\ \citenamefont
  {Ghosh}(2002)}]{strehl_cluster_2002}%
  \BibitemOpen
  \bibfield  {author} {\bibinfo {author} {\bibfnamefont {Alexander}\
  \bibnamefont {Strehl}}\ and\ \bibinfo {author} {\bibfnamefont {Joydeep}\
  \bibnamefont {Ghosh}},\ }\bibfield  {title} {\enquote {\bibinfo {title}
  {Cluster {Ensembles} --- {A} {Knowledge} {Reuse} {Framework} for {Combining}
  {Multiple} {Partitions}},}\ }\href
  {http://www.jmlr.org/papers/v3/strehl02a.html} {\bibfield  {journal}
  {\bibinfo  {journal} {Journal of Machine Learning Research}\ }\textbf
  {\bibinfo {volume} {3}},\ \bibinfo {pages} {583--617} (\bibinfo {year}
  {2002})}\BibitemShut {NoStop}%
\bibitem [{\citenamefont {Topchy}\ \emph {et~al.}(2005)\citenamefont {Topchy},
  \citenamefont {Jain},\ and\ \citenamefont {Punch}}]{topchy_clustering_2005}%
  \BibitemOpen
  \bibfield  {author} {\bibinfo {author} {\bibfnamefont {A.}~\bibnamefont
  {Topchy}}, \bibinfo {author} {\bibfnamefont {A.K.}\ \bibnamefont {Jain}}, \
  and\ \bibinfo {author} {\bibfnamefont {W.}~\bibnamefont {Punch}},\ }\bibfield
   {title} {\enquote {\bibinfo {title} {Clustering ensembles: models of
  consensus and weak partitions},}\ }\href {\doibase 10.1109/TPAMI.2005.237}
  {\bibfield  {journal} {\bibinfo  {journal} {IEEE Transactions on Pattern
  Analysis and Machine Intelligence}\ }\textbf {\bibinfo {volume} {27}},\
  \bibinfo {pages} {1866--1881} (\bibinfo {year} {2005})},\ \bibinfo {note}
  {conference Name: IEEE Transactions on Pattern Analysis and Machine
  Intelligence}\BibitemShut {NoStop}%
\bibitem [{\citenamefont {Goder}\ and\ \citenamefont
  {Filkov}(2008)}]{goder_consensus_2008}%
  \BibitemOpen
  \bibfield  {author} {\bibinfo {author} {\bibfnamefont {Andrey}\ \bibnamefont
  {Goder}}\ and\ \bibinfo {author} {\bibfnamefont {Vladimir}\ \bibnamefont
  {Filkov}},\ }\bibfield  {title} {\enquote {\bibinfo {title} {Consensus
  {Clustering} {Algorithms}: {Comparison} and {Refinement}},}\ }in\ \href
  {\doibase 10.1137/1.9781611972887.11} {\emph {\bibinfo {booktitle} {2008
  {Proceedings} of the {Workshop} on {Algorithm} {Engineering} and
  {Experiments} ({ALENEX})}}},\ \bibinfo {series and number} {Proceedings}\
  (\bibinfo  {publisher} {Society for Industrial and Applied Mathematics},\
  \bibinfo {year} {2008})\ pp.\ \bibinfo {pages} {109--117}\BibitemShut
  {NoStop}%
\bibitem [{\citenamefont {Lancichinetti}\ and\ \citenamefont
  {Fortunato}(2012)}]{lancichinetti_consensus_2012}%
  \BibitemOpen
  \bibfield  {author} {\bibinfo {author} {\bibfnamefont {Andrea}\ \bibnamefont
  {Lancichinetti}}\ and\ \bibinfo {author} {\bibfnamefont {Santo}\ \bibnamefont
  {Fortunato}},\ }\bibfield  {title} {{\selectlanguage {english}\enquote
  {\bibinfo {title} {Consensus clustering in complex networks},}\ }}\href
  {\doibase 10.1038/srep00336} {\bibfield  {journal} {\bibinfo  {journal}
  {Scientific Reports}\ }\textbf {\bibinfo {volume} {2}},\ \bibinfo {pages}
  {1--7} (\bibinfo {year} {2012})},\ \bibinfo {note} {number: 1 Publisher:
  Nature Publishing Group}\BibitemShut {NoStop}%
\bibitem [{\citenamefont {Zhang}\ and\ \citenamefont
  {Moore}(2014)}]{zhang_scalable_2014}%
  \BibitemOpen
  \bibfield  {author} {\bibinfo {author} {\bibfnamefont {Pan}\ \bibnamefont
  {Zhang}}\ and\ \bibinfo {author} {\bibfnamefont {Cristopher}\ \bibnamefont
  {Moore}},\ }\bibfield  {title} {{\selectlanguage {english}\enquote {\bibinfo
  {title} {Scalable detection of statistically significant communities and
  hierarchies, using message passing for modularity},}\ }}\href {\doibase
  10.1073/pnas.1409770111} {\bibfield  {journal} {\bibinfo  {journal}
  {Proceedings of the National Academy of Sciences}\ }\textbf {\bibinfo
  {volume} {111}},\ \bibinfo {pages} {18144--18149} (\bibinfo {year} {2014})},\
  \bibinfo {note} {publisher: National Academy of Sciences Section: Physical
  Sciences}\BibitemShut {NoStop}%
\bibitem [{\citenamefont {Tandon}\ \emph {et~al.}(2019)\citenamefont {Tandon},
  \citenamefont {Albeshri}, \citenamefont {Thayananthan}, \citenamefont
  {Alhalabi},\ and\ \citenamefont {Fortunato}}]{tandon_fast_2019}%
  \BibitemOpen
  \bibfield  {author} {\bibinfo {author} {\bibfnamefont {Aditya}\ \bibnamefont
  {Tandon}}, \bibinfo {author} {\bibfnamefont {Aiiad}\ \bibnamefont
  {Albeshri}}, \bibinfo {author} {\bibfnamefont {Vijey}\ \bibnamefont
  {Thayananthan}}, \bibinfo {author} {\bibfnamefont {Wadee}\ \bibnamefont
  {Alhalabi}}, \ and\ \bibinfo {author} {\bibfnamefont {Santo}\ \bibnamefont
  {Fortunato}},\ }\bibfield  {title} {\enquote {\bibinfo {title} {Fast
  consensus clustering in complex networks},}\ }\href {\doibase
  10.1103/PhysRevE.99.042301} {\bibfield  {journal} {\bibinfo  {journal}
  {Physical Review E}\ }\textbf {\bibinfo {volume} {99}},\ \bibinfo {pages}
  {042301} (\bibinfo {year} {2019})},\ \bibinfo {note} {publisher: American
  Physical Society}\BibitemShut {NoStop}%
\bibitem [{\citenamefont {Maaten}\ and\ \citenamefont
  {Hinton}(2008)}]{maaten_visualizing_2008}%
  \BibitemOpen
  \bibfield  {author} {\bibinfo {author} {\bibfnamefont {Laurens van~der}\
  \bibnamefont {Maaten}}\ and\ \bibinfo {author} {\bibfnamefont {Geoffrey}\
  \bibnamefont {Hinton}},\ }\bibfield  {title} {\enquote {\bibinfo {title}
  {Visualizing {Data} using t-{SNE}},}\ }\href
  {https://www.jmlr.org/papers/v9/vandermaaten08a.html} {\bibfield  {journal}
  {\bibinfo  {journal} {Journal of Machine Learning Research}\ }\textbf
  {\bibinfo {volume} {9}},\ \bibinfo {pages} {2579--2605} (\bibinfo {year}
  {2008})}\BibitemShut {NoStop}%
\bibitem [{\citenamefont {McInnes}\ \emph {et~al.}(2018)\citenamefont
  {McInnes}, \citenamefont {Healy},\ and\ \citenamefont
  {Melville}}]{mcinnes_umap_2018}%
  \BibitemOpen
  \bibfield  {author} {\bibinfo {author} {\bibfnamefont {Leland}\ \bibnamefont
  {McInnes}}, \bibinfo {author} {\bibfnamefont {John}\ \bibnamefont {Healy}}, \
  and\ \bibinfo {author} {\bibfnamefont {James}\ \bibnamefont {Melville}},\
  }\bibfield  {title} {\enquote {\bibinfo {title} {{UMAP}: {Uniform} {Manifold}
  {Approximation} and {Projection} for {Dimension} {Reduction}},}\ }\href
  {http://arxiv.org/abs/1802.03426} {\bibfield  {journal} {\bibinfo  {journal}
  {arXiv:1802.03426 [cs, stat]}\ } (\bibinfo {year} {2018})},\ \bibinfo {note}
  {arXiv: 1802.03426}\BibitemShut {NoStop}%
\bibitem [{\citenamefont {Peixoto}(2019)}]{peixoto_bayesian_2019}%
  \BibitemOpen
  \bibfield  {author} {\bibinfo {author} {\bibfnamefont {Tiago~P.}\
  \bibnamefont {Peixoto}},\ }\bibfield  {title} {{\selectlanguage
  {english}\enquote {\bibinfo {title} {Bayesian {Stochastic}
  {Blockmodeling}},}\ }}in\ \href {\doibase 10.1002/9781119483298.ch11}
  {{\selectlanguage {english}\emph {\bibinfo {booktitle} {Advances in {Network}
  {Clustering} and {Blockmodeling}}}}}\ (\bibinfo  {publisher} {John Wiley \&
  Sons, Ltd},\ \bibinfo {year} {2019})\ pp.\ \bibinfo {pages}
  {289--332}\BibitemShut {NoStop}%
\bibitem [{\citenamefont {Holland}\ \emph {et~al.}(1983)\citenamefont
  {Holland}, \citenamefont {Laskey},\ and\ \citenamefont
  {Leinhardt}}]{holland_stochastic_1983}%
  \BibitemOpen
  \bibfield  {author} {\bibinfo {author} {\bibfnamefont {Paul~W.}\ \bibnamefont
  {Holland}}, \bibinfo {author} {\bibfnamefont {Kathryn~Blackmond}\
  \bibnamefont {Laskey}}, \ and\ \bibinfo {author} {\bibfnamefont {Samuel}\
  \bibnamefont {Leinhardt}},\ }\bibfield  {title} {\enquote {\bibinfo {title}
  {Stochastic blockmodels: {First} steps},}\ }\href {\doibase
  16/0378-8733(83)90021-7} {\bibfield  {journal} {\bibinfo  {journal} {Social
  Networks}\ }\textbf {\bibinfo {volume} {5}},\ \bibinfo {pages} {109--137}
  (\bibinfo {year} {1983})}\BibitemShut {NoStop}%
\bibitem [{\citenamefont
  {Peixoto}(2014{\natexlab{a}})}]{peixoto_efficient_2014}%
  \BibitemOpen
  \bibfield  {author} {\bibinfo {author} {\bibfnamefont {Tiago~P.}\
  \bibnamefont {Peixoto}},\ }\bibfield  {title} {\enquote {\bibinfo {title}
  {Efficient {Monte} {Carlo} and greedy heuristic for the inference of
  stochastic block models},}\ }\href {\doibase 10.1103/PhysRevE.89.012804}
  {\bibfield  {journal} {\bibinfo  {journal} {Physical Review E}\ }\textbf
  {\bibinfo {volume} {89}},\ \bibinfo {pages} {012804} (\bibinfo {year}
  {2014}{\natexlab{a}})}\BibitemShut {NoStop}%
\bibitem [{\citenamefont {Riolo}\ \emph {et~al.}(2017)\citenamefont {Riolo},
  \citenamefont {Cantwell}, \citenamefont {Reinert},\ and\ \citenamefont
  {Newman}}]{riolo_efficient_2017}%
  \BibitemOpen
  \bibfield  {author} {\bibinfo {author} {\bibfnamefont {Maria~A.}\
  \bibnamefont {Riolo}}, \bibinfo {author} {\bibfnamefont {George~T.}\
  \bibnamefont {Cantwell}}, \bibinfo {author} {\bibfnamefont {Gesine}\
  \bibnamefont {Reinert}}, \ and\ \bibinfo {author} {\bibfnamefont {M.~E.~J.}\
  \bibnamefont {Newman}},\ }\bibfield  {title} {\enquote {\bibinfo {title}
  {Efficient method for estimating the number of communities in a network},}\
  }\href {\doibase 10.1103/PhysRevE.96.032310} {\bibfield  {journal} {\bibinfo
  {journal} {Physical Review E}\ }\textbf {\bibinfo {volume} {96}},\ \bibinfo
  {pages} {032310} (\bibinfo {year} {2017})}\BibitemShut {NoStop}%
\bibitem [{\citenamefont
  {Peixoto}(2020{\natexlab{a}})}]{peixoto_merge-split_2020}%
  \BibitemOpen
  \bibfield  {author} {\bibinfo {author} {\bibfnamefont {Tiago~P.}\
  \bibnamefont {Peixoto}},\ }\bibfield  {title} {\enquote {\bibinfo {title}
  {Merge-split {Markov} chain {Monte} {Carlo} for community detection},}\
  }\href {http://arxiv.org/abs/2003.07070} {\bibfield  {journal} {\bibinfo
  {journal} {arXiv:2003.07070 [physics, stat]}\ } (\bibinfo {year}
  {2020}{\natexlab{a}})},\ \bibinfo {note} {arXiv: 2003.07070}\BibitemShut
  {NoStop}%
\bibitem [{\citenamefont {Krebs}()}]{krebs_political_nodate}%
  \BibitemOpen
  \bibfield  {author} {\bibinfo {author} {\bibfnamefont {V}~\bibnamefont
  {Krebs}},\ }\bibfield  {title} {\enquote {\bibinfo {title} {Political {Books}
  {Network}},}\ }\href@noop {} {\bibinfo  {journal} {unpublished, retrieved
  from Mark Newman's website:
  \url{http://www-personal.umich.edu/{\textasciitilde}mejn/netdata/}}\
  }\BibitemShut {NoStop}%
\bibitem [{\citenamefont {Karrer}\ and\ \citenamefont
  {Newman}(2011)}]{karrer_stochastic_2011}%
  \BibitemOpen
\bibfield  {journal} {  }\bibfield  {author} {\bibinfo {author} {\bibfnamefont
  {Brian}\ \bibnamefont {Karrer}}\ and\ \bibinfo {author} {\bibfnamefont
  {M.~E.~J.}\ \bibnamefont {Newman}},\ }\bibfield  {title} {\enquote {\bibinfo
  {title} {Stochastic blockmodels and community structure in networks},}\
  }\href {\doibase 10.1103/PhysRevE.83.016107} {\bibfield  {journal} {\bibinfo
  {journal} {Physical Review E}\ }\textbf {\bibinfo {volume} {83}},\ \bibinfo
  {pages} {016107} (\bibinfo {year} {2011})}\BibitemShut {NoStop}%
\bibitem [{\citenamefont {Peixoto}(2017)}]{peixoto_nonparametric_2017}%
  \BibitemOpen
  \bibfield  {author} {\bibinfo {author} {\bibfnamefont {Tiago~P.}\
  \bibnamefont {Peixoto}},\ }\bibfield  {title} {\enquote {\bibinfo {title}
  {Nonparametric {Bayesian} inference of the microcanonical stochastic block
  model},}\ }\href {\doibase 10.1103/PhysRevE.95.012317} {\bibfield  {journal}
  {\bibinfo  {journal} {Physical Review E}\ }\textbf {\bibinfo {volume} {95}},\
  \bibinfo {pages} {012317} (\bibinfo {year} {2017})}\BibitemShut {NoStop}%
\bibitem [{\citenamefont {Ramshaw}\ and\ \citenamefont
  {Tarjan}(2012)}]{ramshaw_minimum-cost_2012}%
  \BibitemOpen
  \bibfield  {author} {\bibinfo {author} {\bibfnamefont {Lyle}\ \bibnamefont
  {Ramshaw}}\ and\ \bibinfo {author} {\bibfnamefont {Robert~E.}\ \bibnamefont
  {Tarjan}},\ }\bibfield  {title} {\enquote {\bibinfo {title} {On minimum-cost
  assignments in unbalanced bipartite graphs},}\ }\href@noop {} {\bibfield
  {journal} {\bibinfo  {journal} {HP Labs, Palo Alto, CA, USA, Tech. Rep.
  HPL-2012-40R1}\ } (\bibinfo {year} {2012})}\BibitemShut {NoStop}%
\bibitem [{\citenamefont {Kuhn}(1955)}]{kuhn_hungarian_1955}%
  \BibitemOpen
  \bibfield  {author} {\bibinfo {author} {\bibfnamefont {H.~W.}\ \bibnamefont
  {Kuhn}},\ }\bibfield  {title} {{\selectlanguage {english}\enquote {\bibinfo
  {title} {The {Hungarian} method for the assignment problem},}\ }}\href
  {\doibase 10.1002/nav.3800020109} {\bibfield  {journal} {\bibinfo  {journal}
  {Naval Research Logistics Quarterly}\ }\textbf {\bibinfo {volume} {2}},\
  \bibinfo {pages} {83--97} (\bibinfo {year} {1955})},\ \bibinfo {note}
  {publisher: John Wiley \& Sons, Ltd}\BibitemShut {NoStop}%
\bibitem [{\citenamefont {Munkres}(1957)}]{munkres_algorithms_1957}%
  \BibitemOpen
  \bibfield  {author} {\bibinfo {author} {\bibfnamefont {James}\ \bibnamefont
  {Munkres}},\ }\bibfield  {title} {{\selectlanguage {english}\enquote
  {\bibinfo {title} {Algorithms for the {Assignment} and {Transportation}
  {Problems}},}\ }}\href {\doibase 10.1137/0105003} {\bibfield  {journal}
  {\bibinfo  {journal} {Journal of the Society for Industrial and Applied
  Mathematics}\ }\textbf {\bibinfo {volume} {5}},\ \bibinfo {pages} {32--38}
  (\bibinfo {year} {1957})}\BibitemShut {NoStop}%
\bibitem [{\citenamefont
  {Peixoto}(2014{\natexlab{b}})}]{peixoto_graph-tool_2014}%
  \BibitemOpen
  \bibfield  {author} {\bibinfo {author} {\bibfnamefont {Tiago~P.}\
  \bibnamefont {Peixoto}},\ }\bibfield  {title} {\enquote {\bibinfo {title}
  {The \texttt{graph-tool} python library},}\ }\href {\doibase
  10.6084/m9.figshare.1164194} {\bibfield  {journal} {\bibinfo  {journal}
  {figshare}\ } (\bibinfo {year} {2014}{\natexlab{b}}),\
  10.6084/m9.figshare.1164194},\ \bibinfo {note} {available at
  \url{https://graph-tool.skewed.de}.}\BibitemShut {Stop}%
\bibitem [{\citenamefont {Meilă}\ and\ \citenamefont
  {Heckerman}(2001)}]{meila_experimental_2001}%
  \BibitemOpen
  \bibfield  {author} {\bibinfo {author} {\bibfnamefont {Marina}\ \bibnamefont
  {Meilă}}\ and\ \bibinfo {author} {\bibfnamefont {David}\ \bibnamefont
  {Heckerman}},\ }\bibfield  {title} {{\selectlanguage {english}\enquote
  {\bibinfo {title} {An {Experimental} {Comparison} of {Model}-{Based}
  {Clustering} {Methods}},}\ }}\href {\doibase 10.1023/A:1007648401407}
  {\bibfield  {journal} {\bibinfo  {journal} {Machine Learning}\ }\textbf
  {\bibinfo {volume} {42}},\ \bibinfo {pages} {9--29} (\bibinfo {year}
  {2001})}\BibitemShut {NoStop}%
\bibitem [{\citenamefont {Meilǎ}(2005)}]{meila_comparing_2005}%
  \BibitemOpen
  \bibfield  {author} {\bibinfo {author} {\bibfnamefont {Marina}\ \bibnamefont
  {Meilǎ}},\ }\bibfield  {title} {\enquote {\bibinfo {title} {Comparing
  clusterings: an axiomatic view},}\ }in\ \href@noop {} {\emph {\bibinfo
  {booktitle} {Proceedings of the 22nd international conference on {Machine}
  learning}}}\ (\bibinfo {year} {2005})\ pp.\ \bibinfo {pages}
  {577--584}\BibitemShut {NoStop}%
\bibitem [{\citenamefont {Meilă}(2007)}]{meila_comparing_2007}%
  \BibitemOpen
  \bibfield  {author} {\bibinfo {author} {\bibfnamefont {Marina}\ \bibnamefont
  {Meilă}},\ }\bibfield  {title} {\enquote {\bibinfo {title} {Comparing
  clusterings—an information based distance},}\ }\href {\doibase
  10.1016/j.jmva.2006.11.013} {\bibfield  {journal} {\bibinfo  {journal}
  {Journal of Multivariate Analysis}\ }\textbf {\bibinfo {volume} {98}},\
  \bibinfo {pages} {873--895} (\bibinfo {year} {2007})}\BibitemShut {NoStop}%
\bibitem [{\citenamefont {Meilă}(2003)}]{meila_comparing_2003}%
  \BibitemOpen
  \bibfield  {author} {\bibinfo {author} {\bibfnamefont {Marina}\ \bibnamefont
  {Meilă}},\ }\bibfield  {title} {\enquote {\bibinfo {title} {Comparing
  {Clusterings} by the {Variation} of {Information}},}\ }in\ \href
  {http://link.springer.com/chapter/10.1007/978-3-540-45167-9_14} {\emph
  {\bibinfo {booktitle} {Learning {Theory} and {Kernel} {Machines}}}},\
  \bibinfo {series and number} {\bibinfo {series} {Lecture {Notes} in
  {Computer} {Science}}\ No.\ \bibinfo {number} {2777}},\ \bibinfo {editor}
  {edited by\ \bibinfo {editor} {\bibfnamefont {Bernhard}\ \bibnamefont
  {Schölkopf}}\ and\ \bibinfo {editor} {\bibfnamefont {Manfred~K.}\
  \bibnamefont {Warmuth}}}\ (\bibinfo  {publisher} {Springer Berlin
  Heidelberg},\ \bibinfo {year} {2003})\ pp.\ \bibinfo {pages}
  {173--187}\BibitemShut {NoStop}%
\bibitem [{\citenamefont {Newman}\ \emph {et~al.}(2020)\citenamefont {Newman},
  \citenamefont {Cantwell},\ and\ \citenamefont
  {Young}}]{newman_improved_2020}%
  \BibitemOpen
  \bibfield  {author} {\bibinfo {author} {\bibfnamefont {M.~E.~J.}\
  \bibnamefont {Newman}}, \bibinfo {author} {\bibfnamefont {George~T.}\
  \bibnamefont {Cantwell}}, \ and\ \bibinfo {author} {\bibfnamefont
  {Jean-Gabriel}\ \bibnamefont {Young}},\ }\bibfield  {title} {\enquote
  {\bibinfo {title} {Improved mutual information measure for clustering,
  classification, and community detection},}\ }\href {\doibase
  10.1103/PhysRevE.101.042304} {\bibfield  {journal} {\bibinfo  {journal}
  {Physical Review E}\ }\textbf {\bibinfo {volume} {101}},\ \bibinfo {pages}
  {042304} (\bibinfo {year} {2020})},\ \bibinfo {note} {publisher: American
  Physical Society}\BibitemShut {NoStop}%
\bibitem [{\citenamefont {Zachary}(1977)}]{zachary_information_1977}%
  \BibitemOpen
  \bibfield  {author} {\bibinfo {author} {\bibfnamefont {Wayne~W.}\
  \bibnamefont {Zachary}},\ }\bibfield  {title} {\enquote {\bibinfo {title} {An
  {Information} {Flow} {Model} for {Conflict} and {Fission} in {Small}
  {Groups}},}\ }\href {http://www.jstor.org/stable/3629752} {\bibfield
  {journal} {\bibinfo  {journal} {Journal of Anthropological Research}\
  }\textbf {\bibinfo {volume} {33}},\ \bibinfo {pages} {452--473} (\bibinfo
  {year} {1977})}\BibitemShut {NoStop}%
\bibitem [{\citenamefont {Peixoto}(2013)}]{peixoto_parsimonious_2013}%
  \BibitemOpen
  \bibfield  {author} {\bibinfo {author} {\bibfnamefont {Tiago~P.}\
  \bibnamefont {Peixoto}},\ }\bibfield  {title} {\enquote {\bibinfo {title}
  {Parsimonious {Module} {Inference} in {Large} {Networks}},}\ }\href {\doibase
  10.1103/PhysRevLett.110.148701} {\bibfield  {journal} {\bibinfo  {journal}
  {Physical Review Letters}\ }\textbf {\bibinfo {volume} {110}},\ \bibinfo
  {pages} {148701} (\bibinfo {year} {2013})}\BibitemShut {NoStop}%
\bibitem [{\citenamefont
  {Peixoto}(2014{\natexlab{c}})}]{peixoto_hierarchical_2014}%
  \BibitemOpen
  \bibfield  {author} {\bibinfo {author} {\bibfnamefont {Tiago~P.}\
  \bibnamefont {Peixoto}},\ }\bibfield  {title} {\enquote {\bibinfo {title}
  {Hierarchical {Block} {Structures} and {High}-{Resolution} {Model}
  {Selection} in {Large} {Networks}},}\ }\href {\doibase
  10.1103/PhysRevX.4.011047} {\bibfield  {journal} {\bibinfo  {journal}
  {Physical Review X}\ }\textbf {\bibinfo {volume} {4}},\ \bibinfo {pages}
  {011047} (\bibinfo {year} {2014}{\natexlab{c}})}\BibitemShut {NoStop}%
\bibitem [{\citenamefont {Peixoto}(2020{\natexlab{b}})}]{peixoto_latent_2020}%
  \BibitemOpen
  \bibfield  {author} {\bibinfo {author} {\bibfnamefont {Tiago~P.}\
  \bibnamefont {Peixoto}},\ }\bibfield  {title} {\enquote {\bibinfo {title}
  {Latent {Poisson} models for networks with heterogeneous density},}\ }\href
  {http://arxiv.org/abs/2002.07803} {\bibfield  {journal} {\bibinfo  {journal}
  {arXiv:2002.07803 [physics, stat]}\ } (\bibinfo {year}
  {2020}{\natexlab{b}})},\ \bibinfo {note} {arXiv: 2002.07803}\BibitemShut
  {NoStop}%
\bibitem [{\citenamefont {Geman}\ \emph {et~al.}(1992)\citenamefont {Geman},
  \citenamefont {Bienenstock},\ and\ \citenamefont
  {Doursat}}]{geman_neural_1992}%
  \BibitemOpen
  \bibfield  {author} {\bibinfo {author} {\bibfnamefont {Stuart}\ \bibnamefont
  {Geman}}, \bibinfo {author} {\bibfnamefont {Elie}\ \bibnamefont
  {Bienenstock}}, \ and\ \bibinfo {author} {\bibfnamefont {René}\ \bibnamefont
  {Doursat}},\ }\bibfield  {title} {\enquote {\bibinfo {title} {Neural
  {Networks} and the {Bias}/{Variance} {Dilemma}},}\ }\href {\doibase
  10.1162/neco.1992.4.1.1} {\bibfield  {journal} {\bibinfo  {journal} {Neural
  Computation}\ }\textbf {\bibinfo {volume} {4}},\ \bibinfo {pages} {1--58}
  (\bibinfo {year} {1992})},\ \bibinfo {note} {publisher: MIT
  Press}\BibitemShut {NoStop}%
\bibitem [{\citenamefont {Girvan}\ and\ \citenamefont
  {Newman}(2002)}]{girvan_community_2002}%
  \BibitemOpen
  \bibfield  {author} {\bibinfo {author} {\bibfnamefont {M.}~\bibnamefont
  {Girvan}}\ and\ \bibinfo {author} {\bibfnamefont {M.~E.~J.}\ \bibnamefont
  {Newman}},\ }\bibfield  {title} {\enquote {\bibinfo {title} {Community
  structure in social and biological networks},}\ }\href {\doibase
  10.1073/pnas.122653799} {\bibfield  {journal} {\bibinfo  {journal}
  {Proceedings of the National Academy of Sciences}\ }\textbf {\bibinfo
  {volume} {99}},\ \bibinfo {pages} {7821 --7826} (\bibinfo {year}
  {2002})}\BibitemShut {NoStop}%
\bibitem [{\citenamefont {Knuth}(1993)}]{knuth_stanford_1993}%
  \BibitemOpen
  \bibfield  {author} {\bibinfo {author} {\bibfnamefont {Donald~E.}\
  \bibnamefont {Knuth}},\ }\href@noop {} {\emph {\bibinfo {title} {The
  {Stanford} {GraphBase}: {A} {Platform} for {Combinatorial} {Computing}}}},\
  \bibinfo {edition} {1st}\ ed.\ (\bibinfo  {publisher} {Addison-Wesley
  Professional},\ \bibinfo {address} {New York, N.Y. : Reading, Mass},\
  \bibinfo {year} {1993})\BibitemShut {NoStop}%
\bibitem [{\citenamefont {Lusseau}\ \emph {et~al.}(2003)\citenamefont
  {Lusseau}, \citenamefont {Schneider}, \citenamefont {Boisseau}, \citenamefont
  {Haase}, \citenamefont {Slooten},\ and\ \citenamefont
  {Dawson}}]{lusseau_bottlenose_2003}%
  \BibitemOpen
  \bibfield  {author} {\bibinfo {author} {\bibfnamefont {David}\ \bibnamefont
  {Lusseau}}, \bibinfo {author} {\bibfnamefont {Karsten}\ \bibnamefont
  {Schneider}}, \bibinfo {author} {\bibfnamefont {Oliver~J.}\ \bibnamefont
  {Boisseau}}, \bibinfo {author} {\bibfnamefont {Patti}\ \bibnamefont {Haase}},
  \bibinfo {author} {\bibfnamefont {Elisabeth}\ \bibnamefont {Slooten}}, \ and\
  \bibinfo {author} {\bibfnamefont {Steve~M.}\ \bibnamefont {Dawson}},\
  }\bibfield  {title} {{\selectlanguage {english}\enquote {\bibinfo {title}
  {The bottlenose dolphin community of {Doubtful} {Sound} features a large
  proportion of long-lasting associations},}\ }}\href {\doibase
  10.1007/s00265-003-0651-y} {\bibfield  {journal} {\bibinfo  {journal}
  {Behavioral Ecology and Sociobiology}\ }\textbf {\bibinfo {volume} {54}},\
  \bibinfo {pages} {396--405} (\bibinfo {year} {2003})}\BibitemShut {NoStop}%
\bibitem [{\citenamefont {Newman}(2006)}]{newman_finding_2006}%
  \BibitemOpen
  \bibfield  {author} {\bibinfo {author} {\bibfnamefont {M.~E.~J.}\
  \bibnamefont {Newman}},\ }\bibfield  {title} {\enquote {\bibinfo {title}
  {Finding community structure in networks using the eigenvectors of
  matrices},}\ }\href {\doibase 10.1103/PhysRevE.74.036104} {\bibfield
  {journal} {\bibinfo  {journal} {Physical Review E}\ }\textbf {\bibinfo
  {volume} {74}},\ \bibinfo {pages} {036104} (\bibinfo {year}
  {2006})}\BibitemShut {NoStop}%
\bibitem [{\citenamefont {Harris}\ \emph {et~al.}(2009)\citenamefont {Harris},
  \citenamefont {Halpern}, \citenamefont {Whitsel}, \citenamefont {Hussey},
  \citenamefont {Tabor}, \citenamefont {Entzel},\ and\ \citenamefont
  {Udry}}]{harris_national_2009}%
  \BibitemOpen
  \bibfield  {author} {\bibinfo {author} {\bibfnamefont {Kathleen~Mullan}\
  \bibnamefont {Harris}}, \bibinfo {author} {\bibfnamefont {Carolyn~T.}\
  \bibnamefont {Halpern}}, \bibinfo {author} {\bibfnamefont {Eric}\
  \bibnamefont {Whitsel}}, \bibinfo {author} {\bibfnamefont {Jon}\ \bibnamefont
  {Hussey}}, \bibinfo {author} {\bibfnamefont {Joyce}\ \bibnamefont {Tabor}},
  \bibinfo {author} {\bibfnamefont {Pamela}\ \bibnamefont {Entzel}}, \ and\
  \bibinfo {author} {\bibfnamefont {J.~Richard}\ \bibnamefont {Udry}},\
  }\bibfield  {title} {\enquote {\bibinfo {title} {The national longitudinal
  study of adolescent to adult health: {Research} design},}\ }\href@noop {}
  {\bibfield  {journal} {\bibinfo  {journal} {See http://www. cpc. unc.
  edu/projects/addhealth/design (accessed 9 April 2015)}\ } (\bibinfo {year}
  {2009})}\BibitemShut {NoStop}%
\bibitem [{\citenamefont {White}\ \emph {et~al.}(1986)\citenamefont {White},
  \citenamefont {Southgate}, \citenamefont {Thomson},\ and\ \citenamefont
  {Brenner}}]{white_structure_1986}%
  \BibitemOpen
  \bibfield  {author} {\bibinfo {author} {\bibfnamefont {J.~G.}\ \bibnamefont
  {White}}, \bibinfo {author} {\bibfnamefont {E.}~\bibnamefont {Southgate}},
  \bibinfo {author} {\bibfnamefont {J.~N.}\ \bibnamefont {Thomson}}, \ and\
  \bibinfo {author} {\bibfnamefont {S.}~\bibnamefont {Brenner}},\ }\bibfield
  {title} {{\selectlanguage {english}\enquote {\bibinfo {title} {The structure
  of the nervous system of the nematode {Caenorhabditis} elegans},}\
  }}\href@noop {} {\bibfield  {journal} {\bibinfo  {journal} {Philosophical
  Transactions of the Royal Society of London. Series B, Biological Sciences}\
  }\textbf {\bibinfo {volume} {314}},\ \bibinfo {pages} {1--340} (\bibinfo
  {year} {1986})}\BibitemShut {NoStop}%
\bibitem [{\citenamefont {Guimerà}\ \emph {et~al.}(2003)\citenamefont
  {Guimerà}, \citenamefont {Danon}, \citenamefont {Díaz-Guilera},
  \citenamefont {Giralt},\ and\ \citenamefont
  {Arenas}}]{guimera_self-similar_2003}%
  \BibitemOpen
  \bibfield  {author} {\bibinfo {author} {\bibfnamefont {R.}~\bibnamefont
  {Guimerà}}, \bibinfo {author} {\bibfnamefont {L.}~\bibnamefont {Danon}},
  \bibinfo {author} {\bibfnamefont {A.}~\bibnamefont {Díaz-Guilera}}, \bibinfo
  {author} {\bibfnamefont {F.}~\bibnamefont {Giralt}}, \ and\ \bibinfo {author}
  {\bibfnamefont {A.}~\bibnamefont {Arenas}},\ }\bibfield  {title} {\enquote
  {\bibinfo {title} {Self-similar community structure in a network of human
  interactions},}\ }\href {\doibase 10.1103/PhysRevE.68.065103} {\bibfield
  {journal} {\bibinfo  {journal} {Physical Review E}\ }\textbf {\bibinfo
  {volume} {68}},\ \bibinfo {pages} {065103} (\bibinfo {year}
  {2003})}\BibitemShut {NoStop}%
\bibitem [{\citenamefont {Adamic}\ and\ \citenamefont
  {Glance}(2005)}]{adamic_political_2005}%
  \BibitemOpen
  \bibfield  {author} {\bibinfo {author} {\bibfnamefont {Lada~A.}\ \bibnamefont
  {Adamic}}\ and\ \bibinfo {author} {\bibfnamefont {Natalie}\ \bibnamefont
  {Glance}},\ }\bibfield  {title} {\enquote {\bibinfo {title} {The political
  blogosphere and the 2004 {U}.{S}. election: divided they blog},}\ }in\ \href
  {\doibase 10.1145/1134271.1134277} {\emph {\bibinfo {booktitle} {Proceedings
  of the 3rd international workshop on {Link} discovery}}},\ \bibinfo {series
  and number} {{LinkKDD} '05}\ (\bibinfo  {publisher} {ACM},\ \bibinfo
  {address} {New York, NY, USA},\ \bibinfo {year} {2005})\ pp.\ \bibinfo
  {pages} {36--43}\BibitemShut {NoStop}%
\bibitem [{\citenamefont {Grünwald}(2007)}]{grunwald_minimum_2007}%
  \BibitemOpen
  \bibfield  {author} {\bibinfo {author} {\bibfnamefont {Peter~D.}\
  \bibnamefont {Grünwald}},\ }\href@noop {} {\emph {\bibinfo {title} {The
  {Minimum} {Description} {Length} {Principle}}}}\ (\bibinfo  {publisher} {The
  MIT Press},\ \bibinfo {year} {2007})\BibitemShut {NoStop}%
\bibitem [{\citenamefont {Gates}\ \emph {et~al.}(2019)\citenamefont {Gates},
  \citenamefont {Wood}, \citenamefont {Hetrick},\ and\ \citenamefont
  {Ahn}}]{gates_element-centric_2019}%
  \BibitemOpen
  \bibfield  {author} {\bibinfo {author} {\bibfnamefont {Alexander~J.}\
  \bibnamefont {Gates}}, \bibinfo {author} {\bibfnamefont {Ian~B.}\
  \bibnamefont {Wood}}, \bibinfo {author} {\bibfnamefont {William~P.}\
  \bibnamefont {Hetrick}}, \ and\ \bibinfo {author} {\bibfnamefont {Yong-Yeol}\
  \bibnamefont {Ahn}},\ }\bibfield  {title} {{\selectlanguage {english}\enquote
  {\bibinfo {title} {Element-centric clustering comparison unifies overlaps and
  hierarchy},}\ }}\href {\doibase 10.1038/s41598-019-44892-y} {\bibfield
  {journal} {\bibinfo  {journal} {Scientific Reports}\ }\textbf {\bibinfo
  {volume} {9}},\ \bibinfo {pages} {1--13} (\bibinfo {year} {2019})},\ \bibinfo
  {note} {number: 1 Publisher: Nature Publishing Group}\BibitemShut {NoStop}%
\bibitem [{\citenamefont {Zhang}\ \emph {et~al.}(2016)\citenamefont {Zhang},
  \citenamefont {Moore},\ and\ \citenamefont {Newman}}]{zhang_community_2016}%
  \BibitemOpen
  \bibfield  {author} {\bibinfo {author} {\bibfnamefont {Pan}\ \bibnamefont
  {Zhang}}, \bibinfo {author} {\bibfnamefont {Cristopher}\ \bibnamefont
  {Moore}}, \ and\ \bibinfo {author} {\bibfnamefont {M.~E.~J.}\ \bibnamefont
  {Newman}},\ }\bibfield  {title} {\enquote {\bibinfo {title} {Community
  detection in networks with unequal groups},}\ }\href {\doibase
  10.1103/PhysRevE.93.012303} {\bibfield  {journal} {\bibinfo  {journal}
  {Physical Review E}\ }\textbf {\bibinfo {volume} {93}},\ \bibinfo {pages}
  {012303} (\bibinfo {year} {2016})}\BibitemShut {NoStop}%
\bibitem [{\citenamefont {Rezaei}\ and\ \citenamefont
  {Fränti}(2016)}]{rezaei_set_2016}%
  \BibitemOpen
  \bibfield  {author} {\bibinfo {author} {\bibfnamefont {Mohammad}\
  \bibnamefont {Rezaei}}\ and\ \bibinfo {author} {\bibfnamefont {Pasi}\
  \bibnamefont {Fränti}},\ }\bibfield  {title} {\enquote {\bibinfo {title}
  {Set {Matching} {Measures} for {External} {Cluster} {Validity}},}\ }\href
  {\doibase 10.1109/TKDE.2016.2551240} {\bibfield  {journal} {\bibinfo
  {journal} {IEEE Transactions on Knowledge and Data Engineering}\ }\textbf
  {\bibinfo {volume} {28}},\ \bibinfo {pages} {2173--2186} (\bibinfo {year}
  {2016})},\ \bibinfo {note} {conference Name: IEEE Transactions on Knowledge
  and Data Engineering}\BibitemShut {NoStop}%
\bibitem [{\citenamefont {Amigó}\ \emph {et~al.}(2009)\citenamefont {Amigó},
  \citenamefont {Gonzalo}, \citenamefont {Artiles},\ and\ \citenamefont
  {Verdejo}}]{amigo_comparison_2009}%
  \BibitemOpen
  \bibfield  {author} {\bibinfo {author} {\bibfnamefont {Enrique}\ \bibnamefont
  {Amigó}}, \bibinfo {author} {\bibfnamefont {Julio}\ \bibnamefont {Gonzalo}},
  \bibinfo {author} {\bibfnamefont {Javier}\ \bibnamefont {Artiles}}, \ and\
  \bibinfo {author} {\bibfnamefont {Felisa}\ \bibnamefont {Verdejo}},\
  }\bibfield  {title} {{\selectlanguage {english}\enquote {\bibinfo {title} {A
  comparison of extrinsic clustering evaluation metrics based on formal
  constraints},}\ }}\href {\doibase 10.1007/s10791-008-9066-8} {\bibfield
  {journal} {\bibinfo  {journal} {Information Retrieval}\ }\textbf {\bibinfo
  {volume} {12}},\ \bibinfo {pages} {461--486} (\bibinfo {year}
  {2009})}\BibitemShut {NoStop}%
\bibitem [{\citenamefont {Kunegis}(2013)}]{kunegis_konect:_2013}%
  \BibitemOpen
  \bibfield  {author} {\bibinfo {author} {\bibfnamefont {Jérôme}\
  \bibnamefont {Kunegis}},\ }\bibfield  {title} {\enquote {\bibinfo {title}
  {{KONECT}: {The} {Koblenz} {Network} {Collection}},}\ }in\ \href {\doibase
  10.1145/2487788.2488173} {\emph {\bibinfo {booktitle} {Proceedings of the
  {22Nd} {International} {Conference} on {World} {Wide} {Web}}}},\ \bibinfo
  {series and number} {{WWW} '13 {Companion}}\ (\bibinfo  {publisher} {ACM},\
  \bibinfo {address} {New York, NY, USA},\ \bibinfo {year} {2013})\ pp.\
  \bibinfo {pages} {1343--1350}\BibitemShut {NoStop}%
\bibitem [{\citenamefont {Ghasemian}\ \emph {et~al.}(2019)\citenamefont
  {Ghasemian}, \citenamefont {Hosseinmardi},\ and\ \citenamefont
  {Clauset}}]{ghasemian_evaluating_2019}%
  \BibitemOpen
  \bibfield  {author} {\bibinfo {author} {\bibfnamefont {Amir}\ \bibnamefont
  {Ghasemian}}, \bibinfo {author} {\bibfnamefont {Homa}\ \bibnamefont
  {Hosseinmardi}}, \ and\ \bibinfo {author} {\bibfnamefont {Aaron}\
  \bibnamefont {Clauset}},\ }\bibfield  {title} {\enquote {\bibinfo {title}
  {Evaluating {Overfit} and {Underfit} in {Models} of {Network} {Community}
  {Structure}},}\ }\href {\doibase 10.1109/TKDE.2019.2911585} {\bibfield
  {journal} {\bibinfo  {journal} {IEEE Transactions on Knowledge and Data
  Engineering}\ ,\ \bibinfo {pages} {1--1}} (\bibinfo {year}
  {2019})}\BibitemShut {NoStop}%
\end{thebibliography}%

\appendix

\section{Properties of the maximum overlap distance}\label{app:max-overlap}

In Sec.~\ref{sec:max-overlap} of the main text we considered the maximum
overlap distance, which corresponds to the minimal classification error,
i.e. the smallest possible number of nodes with an incorrect group
placement in a partition $\y$ if another partition $\x$ is assumed to be
the correct one. It is defined as
\begin{equation}
  d(\x, \y) = N - \underset{\bm\mu}{\operatorname{max}}\;\sum_i\delta_{\mu(x_i),y_i}.
\end{equation}
This measure has been considered before in
Refs.~\cite{meila_experimental_2001, meila_comparing_2005,
meila_comparing_2007}, and here we review some of its useful properties.

\begin{figure}
  \includegraphics[width=\columnwidth]{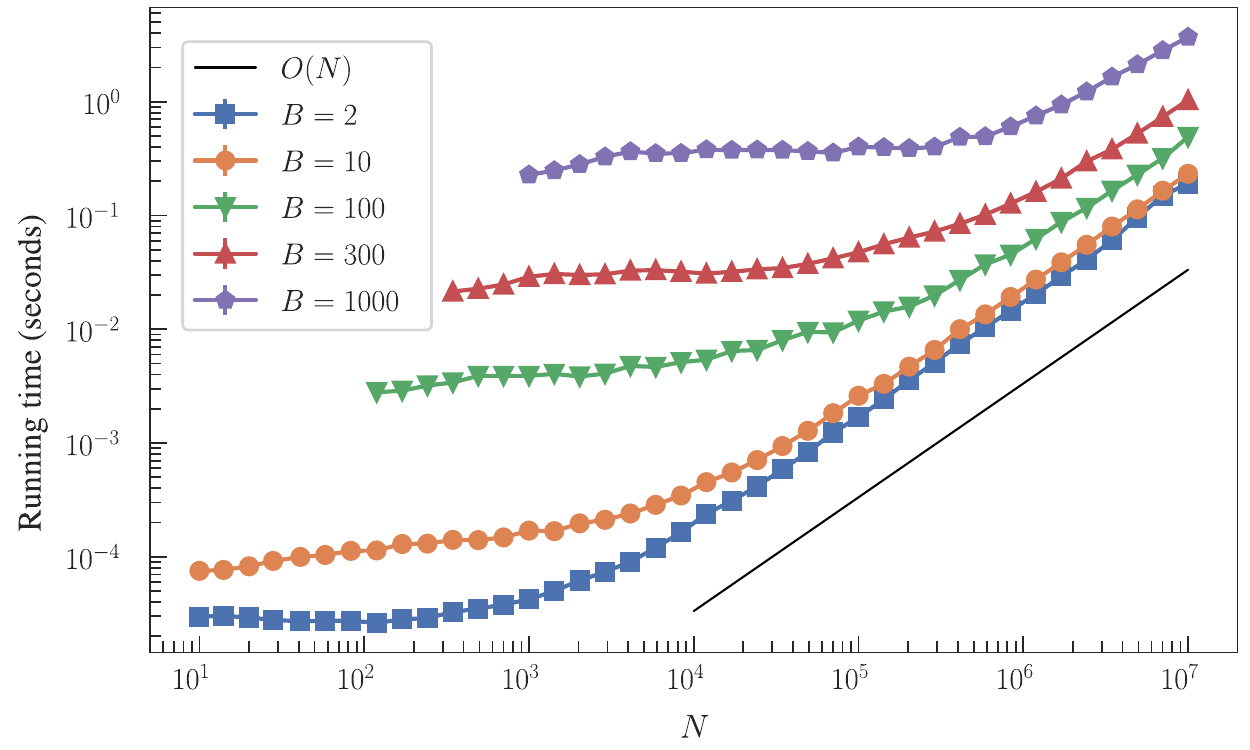}

  \caption{Time required to compute $d(\x,\y)$ for $\x$ and $\y$
  both randomly sampled with $q(\x)=q(\y)=B$ groups, as shown in the legend,
  as a function of $N$, averaged over 100 samples, using an Intel
  i9-9980HK CPU. The solid line shows a $O(N)$ slope.
  \label{fig:overlap_computation}}
\end{figure}

\begin{enumerate}
\item \textbf{Simple interpretation.} Since it quantifies the
  classification error, it is easy to intuitively understand what the
  distance is conveying. In particular its normalized version $d(\x,
  \y)/N$ yields values in the range $[0,1]$ which can be interpreted as
  fractions of differing nodes, and hence allows the direct comparison
  between results obtained for partitions of different sizes and numbers
  of groups.

\item \textbf{Behaves well for unbalanced partitions.} The distance
  $d(\x,\y)$ behaves as one would expect even when the partitions have
  very different number of groups, or the number of groups approaches
  $N$ for either $\x$ or $\y$, unlike alternatives such as mutual
  information~\cite{gates_element-centric_2019}. More specifically, if
  we simply increase the number of groups of either partition being
  compared, this does not spuriously introduce small values of
  $d(\x,\y)$. We see this by noticing that if $q(\x)=B$ and $q(\y)=N$,
  the maximum overlap is always $\omega(\x, \y) = B$, since each group
  in $\x$ can be trivially matched with any of the single-node groups in
  $\y$, yielding
  \begin{equation}
    d(\x,\y) = N - B,
  \end{equation} which leads to the maximum normalized distance
  $d(\x,\y)/N \to 1$ as $N \gg B$.

\item \textbf{Simple asymptotic behavior for uncorrelated partitions.}
  Suppose partitions $\x$ and $\y$ are sampled independently and
  uniformly from the set of all possible partitions into $q(\x)$ and
  $q(\y)$ labelled groups, respectively. In this case, as $N\gg 1$, the
  contingency table will tend to the uniform one with
  $m_{rs}=N/[q(\x)q(\y)]$, which results in the asymptotic normalized
  distance given by
  \begin{equation}
    \lim_{N\to\infty}\frac{d(\x,\y)}{N} = \frac{1}{\max(q(\x),q(\y))}.
  \end{equation}
  Although it is not a substitute for a proper hypothesis
  test (which would need to account for finite values of $N$), this
  asymptotic value gives a rule of thumb of how to interpret the
  distance between two partitions as a strength of statistical
  correlation.

\item \textbf{Defines a metric space.} The distance
  $d(\x,\y)$ is a proper metric, since it fulfills the properties of
  identity $d(\x,\x)=0$, non-negativity $d(\x,\y)\ge 0$, symmetry
  $d(\x,\y) = d(\y, \x)$, and most notably, triangle inequality
  $d(\x,\z) \le d(\x,\y) + d(\y,\z)$ (we offer a simple proof of this in
  Appendix~\ref{app:triangle}). This makes this notion of distance
  well-defined, unambiguous, and conforming to intuition.

\item\textbf{Information-theoretic interpretation.} The maximum overlap
  has a direct information theoretic interpretation, due to its
  connection to the random label generative model exposed
  earlier. According to the model of Eq.~\ref{eq:random_label_marginal}
  the joint probability of observing two partitions $\{\cc\} = \{\x,
  \y\}$, up to an arbitrary relabelling of the groups, is given by
  \begin{equation}
    P(\x,\y) = \frac{2^{\omega(\x, \y)}}{[B(B+1)]^N}.
  \end{equation}
  This means that any two partitions have a joint description
  length
  \begin{align}
    \Sigma(\x,\y) &= -\log_2  P(\x,\y) \\
    & = N\log_2[B(B+1)] - \omega(\x, \y),
  \end{align}
  which measures the amount of information (in bits)
  necessary to describe both partitions. The above quantity is
  proportional to the negative value of the maximum overlap
  $\omega(\x,\y)$, and hence is proportional to $d(\x,\y)$. (Note that
  this is not the most efficient encoding scheme based on the maximum
  overlap, we consider an alternative in Appendix~\ref{app:encoding}.)

\item\textbf{Efficient computation.} As discussed previously,
  computing the maximum overlap involves solving an instance of the
  maximum bipartite weighted matching problem, with weights given by the
  contingency table, $w_{rs}=m_{rs}$ (see Fig.~\ref{fig:matching}),
  which can be done using the Kuhn–Munkres
  algorithm~\cite{kuhn_hungarian_1955,munkres_algorithms_1957}. In its
  sparse version, the running time is bound by $O[(q(\x)+q(\y))E_m]$,
  with $E_m \le q(\x)q(\y)$ being the number of nonzero entries in the
  contingency matrix
  $m_{rs}$~\cite{ramshaw_minimum-cost_2012}. Combining this with the
  work required to build the contingency table itself, the computation
  of $d(\x,\y)$ is bound by $O[(q(\x)+q(\y))E_m + N]$.  Therefore the
  running time will depend on whether we expect the number of labels and
  the density of the contingency table to be much smaller or comparable
  to $N$. In the former case, the maximum matching algorithm takes a
  comparatively negligible time, and the linear term dominates, yielding
  a running time $O(N)$. Otherwise, if we have $q(\x)=O(N)$ or
  $q(\y)=O(N)$, then $E_m = O(N)$, and hence the running time will be
  quadratic, $O(N^2)$. However, the latter scenario is atypical when $N$
  is very large, so therefore we most often encounter the linear regime
  allowing for very fast computations (see
  Fig.~\ref{fig:overlap_computation}).
\end{enumerate}

The maximum overlap distance has been used before in situations where
the labeling is unambiguous or the number of labels is so small that
exhaustive iteration over label permutations is feasible
(e.g.~\cite{decelle_asymptotic_2011, zhang_community_2016}), but, to the
best of our knowledge, rarely in combination with the maximum bipartite
weighted matching algorithm as outlined above (with an exception being
Ref.~\cite{rezaei_set_2016} that employed it when comparing with other
metrics) which makes it usable in general settings. Instead, more focus
has been given to measures such as mutual information (and its several
variants)~\cite{amigo_comparison_2009} or variation of information
(VI)~\cite{meila_comparing_2003}, which are based on the contingency
table without requiring us to obtain a label matching. As pointed out by
Meilă~\cite{meila_comparing_2003}, it is not meaningful to talk about
the ``best'' way of comparing partitions without any context, since such
a task must be unavoidably tied with our ultimate objective. Therefore,
a different set of axiomatic conditions might prefer another
dissimilarity function, and indeed it can be proven that no single
function can simultaneously fulfil some elementary set of
axioms~\cite{meila_comparing_2005}.  In particular, since the maximum
overlap distance is based only on the number of nodes correctly
classified, it ignores the nodes that do not match, and hence does not
exploit any potential regularity with which the labels are
\emph{mismatched}. Other functions such as variation of information,
might provide alternatives which can be used to highlight different
properties of partition ensembles. Nevertheless, few other dissimilarity
functions share the same ease of interpretation with the maximum overlap
distance, while possessing its other useful formal properties, such as
natural normalization, information-theoretical interpretation, and the
fact it defines a metric space.

Among the alternative partition similarities and dissimilarities, the
recently introduced reduced mutual information
(RMI)~\cite{newman_improved_2020} deserves particular mention. This is
because, like the maximum overlap distance, it is related to a joint
description length of two partitions, which in the case of RMI involves
encoding the full contingency table. This means that both similarities
can be compared to each other in their own terms, and the most
appropriate measure must yield the shortest description length. We
perform a succinct comparison between RMI and an overlap-based encoding
in Appendix~\ref{app:encoding}. We will also consider both RMI and VI
more closely in the following section.

\section{Maximum overlap distance obeys triangle inequality}\label{app:triangle}

Here we show that the maximum overlap distance of
Eq.~\ref{eq:overlap_dist} obeys triangle inequality, i.e.
\begin{equation}\label{eq:triangle}
  d(\x,\z) \le d(\x,\y) + d(\y,\z),
\end{equation}
for any set of labelled partitions $\x$, $\y$, and $\z$. Let us consider
the maximum overlap
\begin{equation}
  \omega(\x,\y) = N - d(\x,\y) = \underset{\bm\mu}{\operatorname{max}}\;\sum_i\delta_{\mu(x_i),y_i}.
\end{equation}
Now for an arbitrary choice of $\x$, $\y$ and $\z$ let us consider the
sum
\begin{equation}
  \omega(\x, \y) +  \omega(\y, \z).
\end{equation}
The maximum value either term in the above sum can take is $N$,
corresponding to partitions that are identical up to relabeling,
i.e. $[\x\sim\y]=1$ or $[\y\sim\z]=1$. If we condition on one of the terms
taking its maximum value $N$, the remaining term can take a value at
most $\omega(\x,\z)$, either via the first term with $\omega(\x,\y) = \omega(\x,\z)$ if
$[\y\sim\z]=1$ or via the second term with $\omega(\y,\z) = \omega(\x,\z)$ with
$[\x\sim\y]=1$. This means we can write
\begin{equation}
  \omega(\x, \y) +  \omega(\y, \z) \le N + \omega(\x,\z).
\end{equation}
Substituting $\omega(\x,\y) = N - d(\x,\y)$ and rearranging gives us
Eq.~\ref{eq:triangle}.

\section{Encoding partitions based on overlap}\label{app:encoding}

As described in the main text, the random label model yields a description
length for a pair of partitions given by
\begin{align}
  \Sigma(\x,\y) &= -\ln  P(\x,\y) \\
  & = N\ln[B(B+1)] - \omega(\x, \y)\ln 2,
  \end{align}
Likewise, if we observe $\y$, and use to
describe partition $\x$, the additional amount of information we need
to convey is
\begin{align}
  \Sigma(\x|\y) &= -\ln  P(\x|\y) \\
  &= -\ln  P(\x,\y)/P(\y)\\
  & = N\ln(B+1) - \omega(\x, \y)\ln 2,
\end{align}
where we have used $P(\y)=1/B^N$ from
Eq.~\ref{eq:random_label_marginal}.  From this we can note that this
encoding is sub-optimal in the sense that even when the overlapping is
maximal with $\omega(\x,\y)=N$, the additional information needed to
encode $\x$ is $\Sigma(\x|\y) = N\ln[(B+1)/2]$ which is scales as $O(N)$
when $B > 1$.

Nevertheless we can develop a different encoding that is more efficient
at using the overlap information. We do so by incorporating it as an
explicit parameter as follows:
\begin{enumerate}
\item We sample an overlap value $\omega$ uniformly in the range $[1,N]$, such that
  \begin{equation}
    P(\omega) = \frac{1}{N}.
  \end{equation}
\item We chose a subset $V_{\omega}$ of the $N$ nodes of size $\omega$,
  uniformly with probability
  \begin{equation}
    P(V_{\omega}|\omega) = {N \choose \omega}^{-1}.
  \end{equation}
\item For the nodes in $V_{\omega}$ we sample a partition $\z$ with probability
  \begin{equation}
    P(\z|V_{\omega},\bm\gamma) = \prod_{i\in V_{\omega}}\gamma_{z_i},
  \end{equation}
  which leads to a marginal distribution
  \begin{align}
    P(\z|V_{\omega}) &= \int P(\z|V_{\omega},\bm\gamma)P(\bm\gamma)\;\dd\bm\gamma\\
    &= {\omega+B -1 \choose \omega}^{-1} \frac{\omega!}{\prod_rn_z(r)!},
  \end{align}
  where $n_z(r) = \sum_{i\in V_{\omega}}\delta_{z_i,r}$, assuming a
  uniform prior $P(\bm\gamma)=(B-1)!$.
\item For the remaining $N-\omega$ nodes not in $V_{\omega}$ we sample
      the values
  of partitions $\x$ and $\y$ analogously, i.e.
  \begin{align*}
    P(\x|V_{\omega}) &= {N - \omega + B - 1 \choose N - \omega}^{-1} \frac{(N-\omega)!}{\prod_rn_x(r)!},\\
    P(\y|V_{\omega}) &= {N - \omega + B - 1 \choose N - \omega}^{-1} \frac{(N-\omega)!}{\prod_rn_y(r)!},
  \end{align*}
  with $n_x(r) = \sum_{i\not\in V_{\omega}}\delta_{x_i,r}$ and $n_y(r) = \sum_{i\not\in V_{\omega}}\delta_{y_i,r}$.
\item For the nodes $i \in V_\omega$ we set $x_i=y_i=z_i$, and we choose a
      label bijection $\bm\mu$ uniformly at random from the set of size
      $B!$ and use to relabel either $\x$ or $\y$ arbitrarily.
\end{enumerate}
In the end, this model generates partitions $\x$ and $\y$ that have an
overlap at least $\omega$, although the actual overlap can be larger by
chance. The scheme above allows groups to be unpopulated in the final
partition, which is sub-optimal, but this can be neglected for our
current purpose. The final joint probability of this scheme is
\begin{multline}
  P(\x,\y,\z,V_{\omega},\omega,\bm\mu) = \\
  P(\x|V_{\omega})P(\y|V_{\omega})P(\z|V_{\omega})P(V_{\omega}|\omega)P(\omega)P(\bm\mu),
\end{multline}
which leads to a description length
\begin{multline}
  \Sigma(\x,\y,\z,V_{\omega},\omega,\bm\mu) = -\ln P(\x,\y,\z,V_{\omega},\omega,\bm\mu) =\\
  2\ln{N - \omega + B - 1 \choose N - \omega} + \ln{\omega + B - 1 \choose \omega} + {}\\
  \ln\frac{(N-\omega)!}{\prod_rn_x(r)!} +  \ln\frac{(N-\omega)!}{\prod_rn_y(r)!} +
  \ln \frac{\omega!}{\prod_rn_z(r)!} + {}\\ \ln{N\choose\omega} + \ln N + \ln B!.
\end{multline}
The minimum description length for $\x$ and $\y$ is given by
\begin{equation}
\Sigma(\x,\y) = \underset{\z,V_{\omega},\omega,\bm\mu}{\min} \Sigma(\x,\y,\z,V_{\omega},\omega,\bm\mu).
\end{equation}
which corresponds simply to finding the maximum overlap $\omega(\x,\y)$
and the corresponding label matching between $\x$ and $\y$ from which
$V_{\omega}$, $\z$ and $\mu$ can be derived. It is easy to see now that
if the overlap is maximal with $\omega=N$, the description length amounts to
\begin{multline}
  \Sigma(\x,\y) =  \ln{N + B - 1 \choose N} + \ln N! - \sum_r\ln n_y(r)! + {} \\ \ln N + \ln B!,
\end{multline}
where we have chosen $\y$ as the reference partition arbitrarily, but
without loss of generality.  Hence, if we subtract the necessary
information required to describe $\y$, given by
\begin{equation}
  -\ln P(\y) = \ln{N + B - 1 \choose N} + \ln N! - \sum_r\ln n_y(r)!
\end{equation}
we are left with negligible logarithmic terms
\begin{equation}
  \Sigma(\x|\y) = \ln N + \ln B!,
\end{equation}
meaning the additional information needed to describe $\x$ given $\y$ is
vanishingly small with respect to $N$, and hence the code is efficient
in this case.

\begin{figure}[h!]
  \begin{tabular}{c}
    \includegraphics[width=\columnwidth]{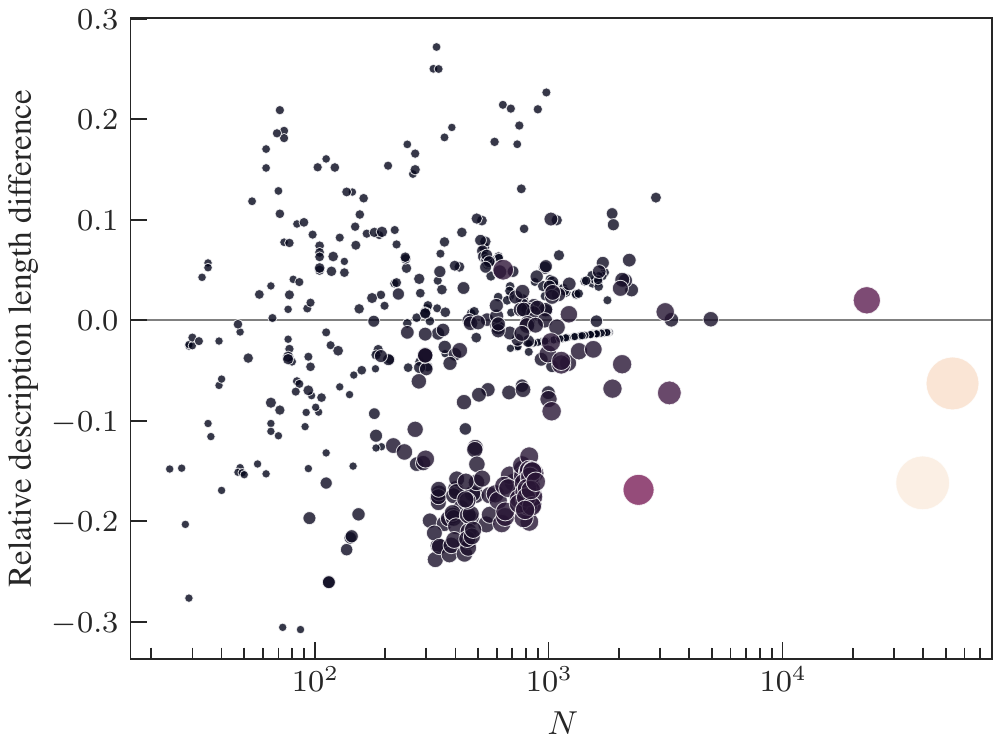}\\
    \includegraphics[width=\columnwidth]{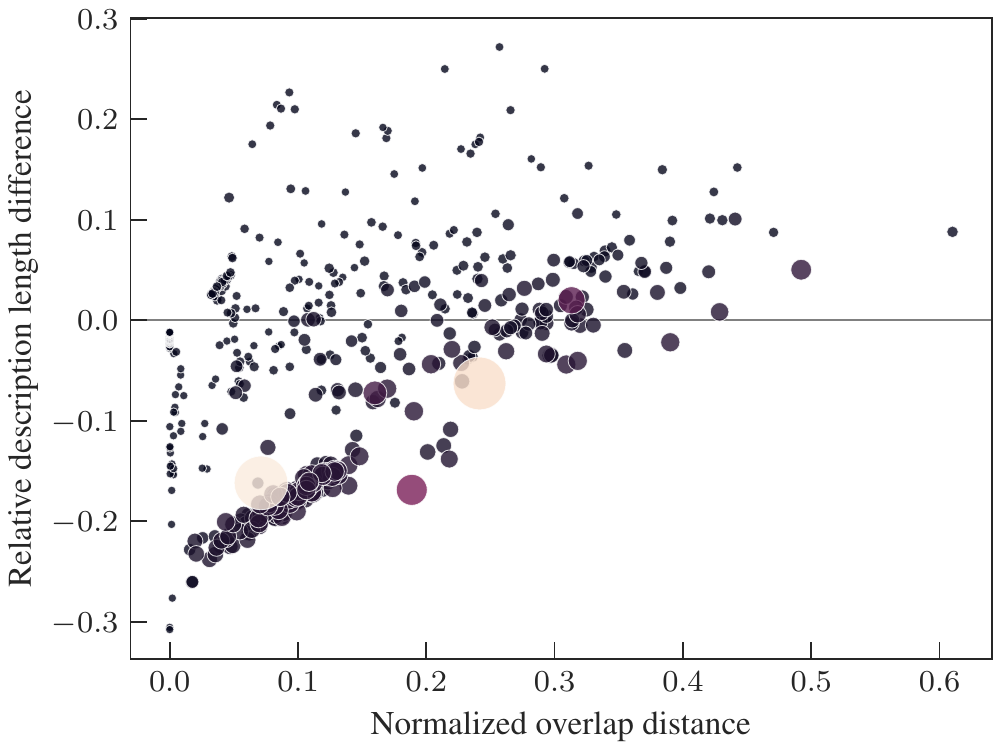}\\
    \includegraphics[width=\columnwidth]{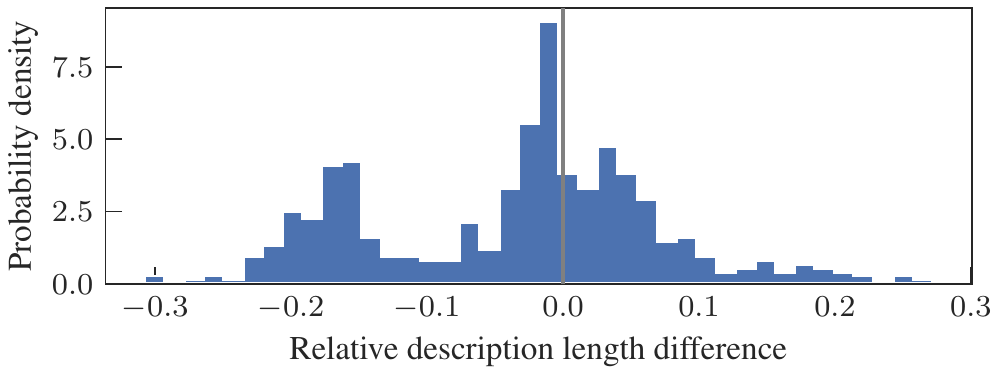}
  \end{tabular} \caption{(Top panel) Average relative description length
  difference $(\Sigma -
  \Sigma_{\text{RMI}})/\max(\Sigma,\Sigma_{\text{RMI}})$ between maximum
  overlap and RMI encodings for empirical networks with $N$ nodes,
  averaged over pairs of partitions independently sampled from the
  Poisson DC-SBM posterior distribution. The point size and color
  indicate the size of the network. (Middle panel) Like top panel, but
  to each network is computed the mean normalized overlap distance.
  (Bottom panel) Histogram of average relative description length
  differences over all empirical
  networks. \label{fig:partition-encoding}}
\end{figure}

It is instructive to compare the above scheme with the reduced mutual
information (RMI) encoding recently proposed in
Ref.~\cite{newman_improved_2020}. It corresponds to a three part scheme
where one encodes first partition $\y$, then the full contingency table
between both partitions $m_{rs}$, and finally the remaining partition
$\x$, leading to a description length
\begin{multline}
  \Sigma'_{\text{RMI}}(\x,\y) = \ln {N-1 \choose B_y + 1} + \ln {N-1 \choose B_x + 1} + {}\\
  \ln \frac{N!}{\prod_rn_y(r)!} + \sum_r\ln\frac{n_x(r)!}{\prod_sm_{rs}} + \ln\Omega(\bm{n}_x,\bm{n}_y),
\end{multline}
where $B_x$ and $B_y$ are the number of labels in partitions $\x$ and
$\y$ and $\Omega(\bm{n}_x,\bm{n}_y)$ is the number of possible
contingency tables with row and column sums given by $\bm{n}_x$ and
$\bm{n}_y$, which cannot be computed in closed form, but for which
approximations are available (see
Ref.~\cite{newman_improved_2020}). Note that the encoding above is not
symmetric, i.e. in general $\Sigma_{\text{RMI}}(\x,\y) \ne
\Sigma_{\text{RMI}}(\y,\x)$, as the overall description length will
depend on which partition is encoded first (although the relative
description length $\Sigma_{\text{RMI}}(\x) -
\Sigma_{\text{RMI}}(\x,\y)$ is always symmetric). Therefore the minimum
description length amounts to choosing the optimal partition to encode
first
\begin{equation}
  \Sigma_{\text{RMI}}(\x,\y) = \min\left[\Sigma'_{\text{RMI}}(\x,\y), \Sigma'_{\text{RMI}}(\y,\x)\right].
\end{equation}

In Fig.~\ref{fig:partition-encoding} we compare the compression of two
partitions sampled independently from the DC-SBM posterior distribution
of 571 empirical networks selected from the
Konect~\cite{kunegis_konect:_2013} and
CommunityFitNet~\cite{ghasemian_evaluating_2019} repositories. Overall,
we observe somewhat mixed results with, the overlap encoding providing a
better compression for around $61\%$ of the networks. As we might
expect, the overlap encoding tends to provide a better description if
the overlap between partitions is very high, such that a full
description of the non-matching nodes becomes superfluous. Otherwise,
for highly differing partitions, the RMI encoding is able to capture
similarities more efficiently.

\section{Comparison with dimensionality reduction}\label{app:dimred}

\begin{figure}
  \begin{tabular}{c}
    \includegraphics[width=\columnwidth, trim=2cm 0cm 1cm 3cm, clip]{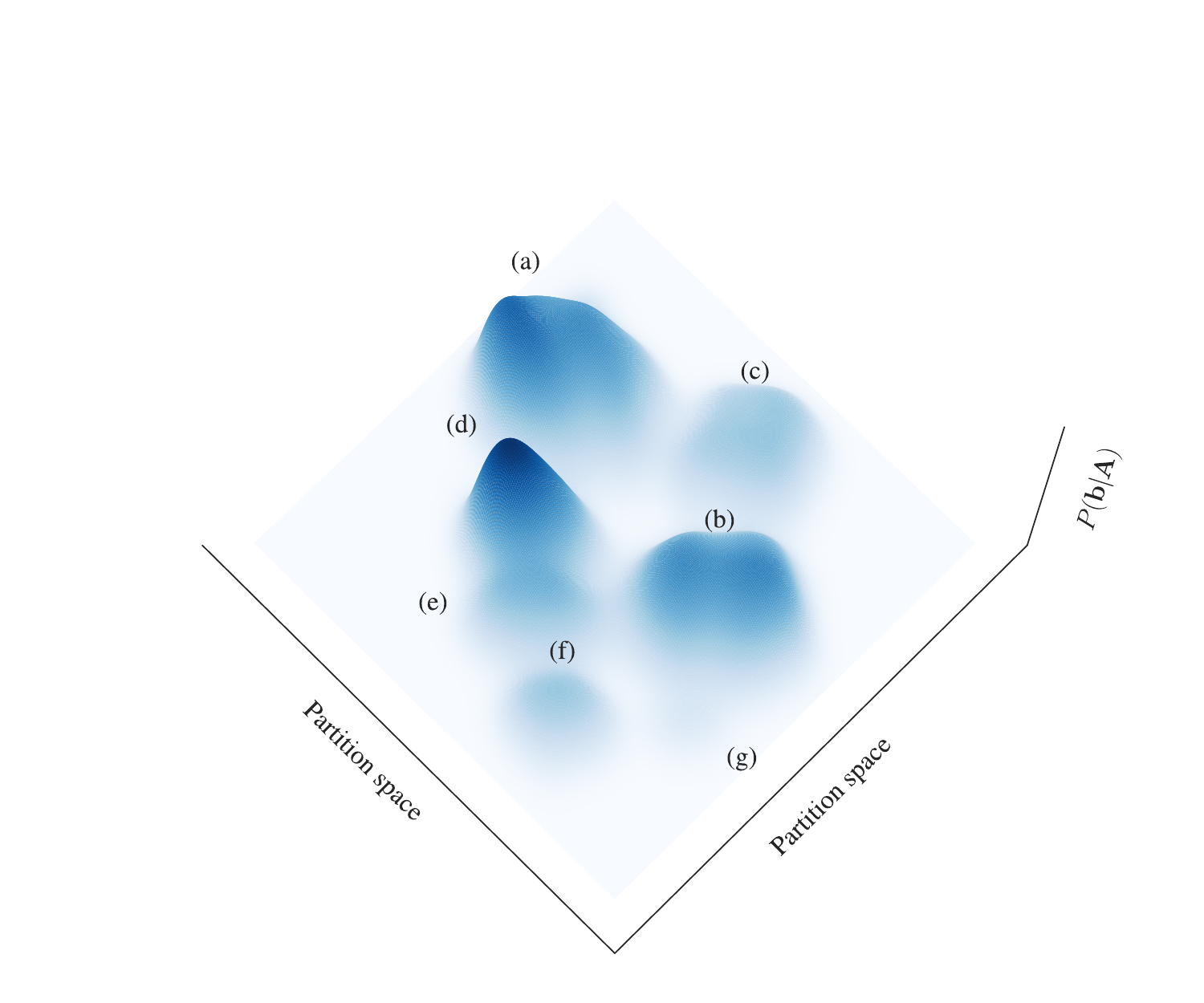}\\
    (a) Variation of information (VI)\\
    \includegraphics[width=\columnwidth, trim=2cm 0cm 1cm 3cm, clip]{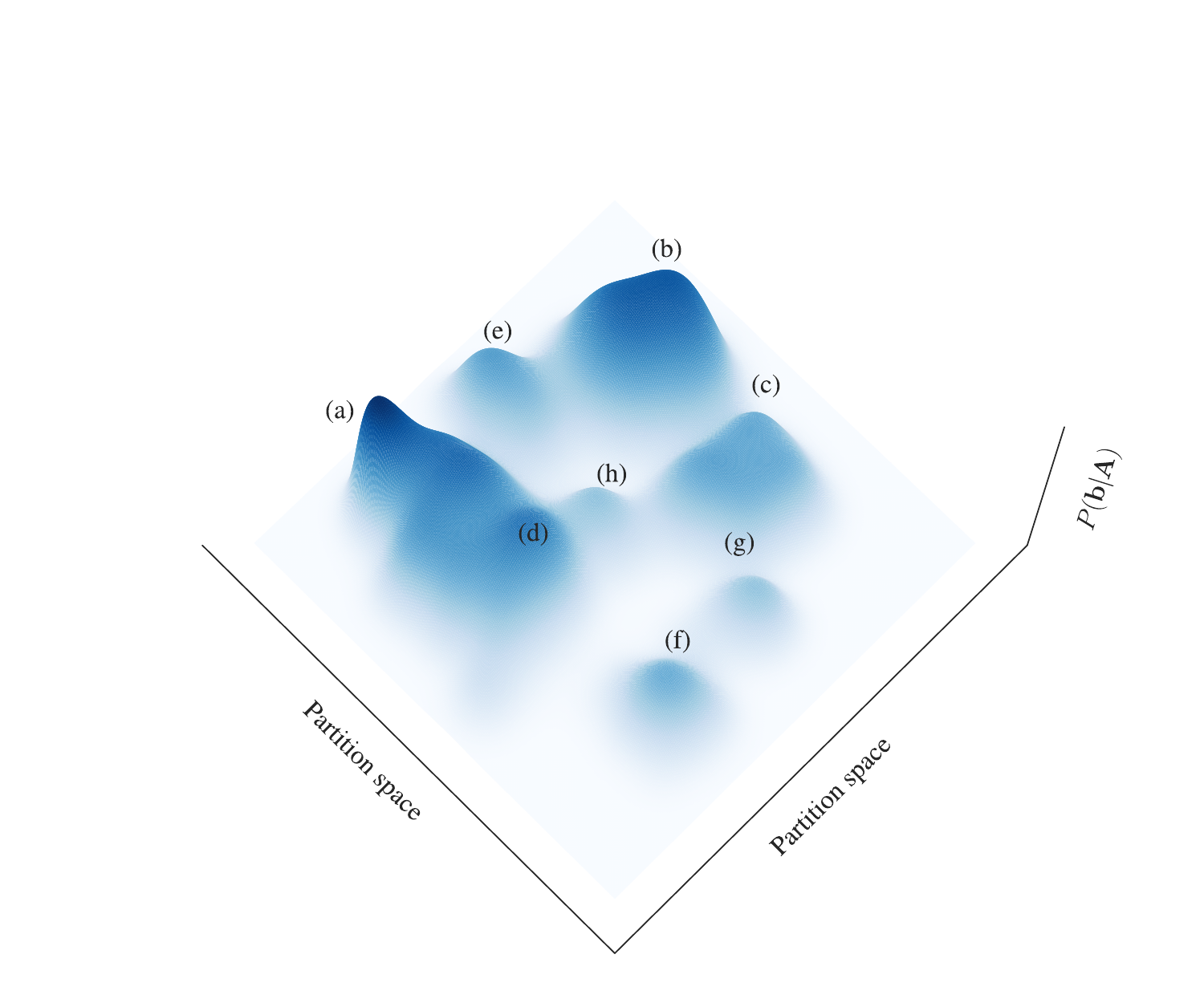}\\
    (b) Reduced mutual information (RMI)
  \end{tabular} \caption{Projection of the partition distribution in two
  dimensions according to the UMAP dimensionality reduction
  algorithm~\cite{mcinnes_umap_2018}, for the same data of
  Fig.~\ref{fig:polbooks-multi-modes}, using (a) the variation of
  information and (b) the (negative) reduced mutual information as
  dissimilarity functions. The labels indicate a correspondence of the
  modes with those found in Fig.~\ref{fig:polbooks-multi-modes}
  according to the majority of partitions.\label{fig:dimred}}
\end{figure}

The clustering algorithm presented in Sec.~\ref{sec:modes} of the main
text is based on a particular definition of what a mode is, according to
the random label model presented in Sec.~\ref{sec:random-label}. As has
been shown in Fig.~\ref{fig:polbooks-multi-modes}, there is an intimate
relationship between the clusters founds and the metric space of
partitions as defined by the maximum overlap distance, such that
dimensionality reduction algorithms like UMAP tend to identify the same
clusters. One may wonder, however, if this picture changes if we
consider another underlying metric space defined by a different distance
function. To give a glimpse into this question, in Fig.~\ref{fig:dimred}
are shown the results of dimensionality reduction using both the
variation and information and reduced mutual information\footnote{The
reduced mutual information is not a metric distance, since it does not
obey triangle inequality, hence it is not really suitable for use with
UMAP, which requires a true metric. Nevertheless, the results obtained
are robust even to this inconsistency.} functions, both of which make
use of the entire contingency table when comparing partitions.
As we can see, not only the overall the multimodal structure is
preserved, but also the composition of the modes is compatible with was
obtained in Fig.~\ref{fig:polbooks-multi-modes}, showing that the
existence of the clusters is not intrinsically tied with the modelling
choices made, but are in fact a property of the data that can be probed
in different ways. Naturally, the local shapes and relative positions of
the modes vary according to the distance used --- and in fact even
across different runs of the UMAP algorithm, since it is
nondeterministic.

We stress that the approach we present in the main text offers many
advantages over dimensionality reduction, namely: 1. We know from the
beginning what the identified modes mean, and is not something that
needs to be interpreted \emph{a posteriori}; 2. Clustering is performed
in a nonparametric manner, without having to decide on an embedding
dimension, or even the number of clusters that need to be
found. Dimensionality reduction, on the other hand, comprises only an
intermediary step that yields an input to a surrogate clustering
algorithm, like k-means, which is often parametric.

\section{Evidence for latent Poisson SBMs}\label{app:latent_multigraph}

The latent Poisson SBMs of Ref.~\cite{peixoto_latent_2020} are generative
models for simple graphs, where at first a multigraph $\G$ is
generated with probability
\begin{equation}
  P(\G|\bb)
\end{equation}
from a Poisson SBM, and then a simple graph is obtained by collapsing
the multiedges to simple edges with
\begin{equation}
  P(A_{ij}|\G) =
  \begin{cases}
    1 & \text{ if } i\neq j \text{ and } G_{ij} > 0,\\
    0 & \text{ otherwise.}
  \end{cases}
\end{equation}
The joint posterior distribution of partitions and latent multiedges is
then
\begin{equation}
  P(\bb,\G | \A) = \frac{P(\A|\G)P(\G|\bb)P(\bb)}{P(\A)},
\end{equation}
with evidence given by
\begin{equation}
  P(\A) = \sum_{\bb,\G}P(\A,\G,\bb).
\end{equation}
Because of the latent multiedges, we need to approximate the evidence in
a similar, but different manner. We write the log evidence as
\begin{align}
  \ln P(\A) &= \sum_{\bb,\G}\pi(\bb,\G)\ln P(\A,\G,\bb)\nonumber \\
            &\qquad - \sum_{\bb,\G}\pi(\bb,\G)\ln \pi(\G,\bb)\\
            &= \avg{\ln P(\A,\G,\bb)} + H(b,G)
\end{align}
where
\begin{equation}
  \pi(\G,\bb) = \frac{P(\A,\G,\bb)}{\sum_{\G',\bb'}P(\A,\G',\bb')}
\end{equation}
is the joint posterior distribution. For our approximation we assume the
factorization,
\begin{equation}
  \pi(\G,\bb) \approx \pi(\G)\pi(\bb),
\end{equation}
together with the ``mean-field'' over the latent multiedges,
\begin{equation}
  \pi(\G) = \prod_{i\le j}q_{ij}(G_{ij})
\end{equation}
with the marginals estimated via MCMC
\begin{equation}
  q_{ij}(x) = \sum_{\G,\bb}\delta_{G_{ij},x} \pi(\G,\bb),
\end{equation}
so that the latent edge entropy can be computed as
\begin{equation}
  H(G) = -\sum_{i\le j}\sum_{x}q_{ij}(x)\ln q_{ij}(x).
\end{equation}
From this we obtain the final approximation,
\begin{align}
  \ln P(\A) = \avg{\ln P(\A,\G,\bb)} + H(b) + H(G),
\end{align}
where $H(b)$ is computed using the mixed random label models as done in
the main text. The approximation for the hierarchical model follows
analogously.

\end{document}